\documentclass[12pt]{article}
\usepackage{jheppub}
\pdfoutput=1

\usepackage{amsmath,bbm,array,amsfonts,graphicx,wrapfig,lscape,float,mathtools,multirow,longtable}

\newcommand{\be}{\begin{equation}}
\newcommand{\ee}{\end{equation}}
\newcommand{\beq}{\begin{equation}}
\newcommand{\beql}[1]{\begin{equation}\label{#1}}
\newcommand{\eeq}{\end{equation}}
\newcommand{\ba}{\begin{array}}
\newcommand{\ea}{\end{array}}
\newcommand{\bea}{\begin{eqnarray}}
\newcommand{\beal}[1]{\begin{eqnarray}\label{#1}}
\newcommand{\eea}{\end{eqnarray}}
\newcommand{\ben}{\begin{enumerate}}
\newcommand{\een}{\end{enumerate}}
\newcommand{\bean}{\begin{eqnarray*}}
\newcommand{\eean}{\end{eqnarray*}}
\newcommand{\eref}[1]{(\ref{#1})}
\newcommand{\sref}[1]{\S\ref{#1}}
\newcommand{\tref}[1]{Table~\ref{#1}}
\newcommand{\nn}{\nonumber}

\newcommand{\fref}[1]{Figure \ref{#1}}
\newcommand{\btab}[1]{\begin{tabular}{#1}}
\newcommand{\etab}{\end{tabular}}

\def \Black {\pdfsetcolor{0 0 0 1}}

\newcommand{\comment}[1]{}

\newcommand{\qed}{\nobreak \ifvmode \relax \else
      \ifdim\lastskip<1.5em \hskip-\lastskip
      \hskip1.5em plus0em minus0.5em \fi \nobreak
      \vrule height0.75em width0.5em depth0.25em\fi}

\newcommand{\Tr}{\text{Tr}}

\DeclareMathOperator*{\residue}{Res}
\DeclareMathOperator*{\JK residue}{JK-Res}

\title{Elliptic Genera of 2d (0,2) Gauge Theories from Brane Brick Models}

\author[a,b]{Sebastian Franco,}
\author[c]{Dongwook Ghim,}
\author[c,d,e]{Sangmin Lee,}
\author[f]{Rak-Kyeong Seong}

\affiliation[a]{
Physics Department, The City College of the CUNY \\
160 Convent Avenue, New York, NY 10031, USA}

\affiliation[b]{The Graduate School and University Center, The City University of New York  \\
365 Fifth Avenue, New York NY 10016, USA }

\affiliation[c]{
Department of Physics and Astronomy, Seoul National University, Seoul 08826, Korea
}

\affiliation[d]{
Center for Theoretical Physics, Seoul National University, Seoul 08826, Korea
}

\affiliation[e]{
College of Liberal Studies, Seoul National University, Seoul 08826, Korea
}

\affiliation[f]{
Department of Physics and Astronomy, Uppsala University, SE-751 08 Uppsala, Sweden
}

\emailAdd{sfranco@ccny.cuny.edu}
\emailAdd{sg1841@snu.ac.kr}
\emailAdd{sangmin@snu.ac.kr}
\emailAdd{rakkyeongseong@gmail.com}

\preprint{
\begin{flushright}
CCNY-HEP-17-01 \\
SNUTP17-002\\
UUITP-05/17
\end{flushright}
}

\abstract{
We compute the elliptic genus of abelian $2d$ $(0,2)$ gauge theories corresponding to brane brick models. 
These theories are worldvolume theories on a single D1-brane probing a toric Calabi-Yau 4-fold singularity. We identify a match with the elliptic genus of the non-linear sigma model on the same Calabi-Yau background, which is computed using a new localization formula. The matching implies that the quantum effects do not drastically alter the correspondence between the geometry and the $2d$ $(0,2)$  gauge theory. In theories whose matter sector suffers from abelian gauge anomaly, we propose an ansatz for an anomaly cancelling term in the integral formula for the elliptic genus. We provide an example in which two brane brick models related to each other by Gadde-Gukov-Putrov triality give the same elliptic genus.
}

\begin{document}

\maketitle

\newpage
\section{Introduction}

Geometry has played a key role in the study of supersymmetric gauge theories and their dynamics. Comparing the moduli space of vacua has led to the discovery and verification of dualities. In three or higher dimensions, it is possible to examine how quantum effects modify the classical moduli space of vacua. Depending on the number of supersymmetries and other factors, the moduli space can be (partially) lifted or its geometry can significantly deviate from the classical one.

In two dimensions, under suitable conditions, gauge theories with classical moduli space of vacua may flow to a non-linear sigma model whose target space is the quantum corrected version of the classical moduli space. The pioneering work \cite{Witten:1993yc} on gauged linear sigma models (GLSM) shows that non-linear sigma models and Landau-Ginzburg (LG) theories can appear as different phases of the same gauge theory, thereby establishing a connection between the two. Non-abelian gauge theories may exhibit even richer structures with various phases of the quantum moduli space \cite{Hori:2006dk}.

In recent years, renewed interest in $2d$ $\mathcal{N}=(0,2)$ supersymmetric gauge theories has led to exciting discoveries. While gauged linear sigma models with $(0,2)$ SUSY have been studied for many years with heterotic model building in mind (see, e.g., \cite{Distler:1993mk,Silverstein:1994ih}), the study of $2d$ $(0,2)$ non-abelian gauge theories has been limited in its scope until recently. One of the most interesting recent breakthroughs is Gadde-Gukov-Putrov (GGP) triality \cite{Gadde:2013lxa,Gadde:2014ppa}, which identifies three seemingly unrelated quiver gauge theories that flow to the same superconformal field theory at low energies.

There are various ways to realize $2d$ $(0,2)$ gauge theories in string and M-theory \cite{Mohri:1997ef,GarciaCompean:1998kh,Gadde:2013sca,Franco:2015tna,Franco:2015tya,Franco:2016nwv,Franco:2016qxh,Franco:2016fxm,Benini:2012cz,Benini:2015bwz,Tatar:2015sga,Schafer-Nameki:2016cfr,Apruzzi:2016iac,Apruzzi:2016nfr}. We will focus on {\it brane brick models} \cite{Franco:2015tna,Franco:2015tya,Franco:2016nwv,Franco:2016qxh,Franco:2016fxm}, which arise from D1-branes probing non-compact toric Calabi-Yau 4-fold singularities (CY$_4$). A brane brick model is a type IIA brane configuration of D4-branes suspended between a NS5 brane that wraps a holomorphic surface $\Sigma$ given by the probed Calabi-Yau geometry. The general structure and construction of brane brick models was first spelled out in \cite{Franco:2015tna}. The connection between the CY$_4$ geometry and the gauge theory through brane brick models was elaborated in \cite{Franco:2015tya}. How triality is realized in terms brane brick models was explained in \cite{Franco:2016nwv}. Finally, in \cite{Franco:2016qxh} it was shown how the results of \cite{Franco:2015tna,Franco:2015tya,Franco:2016nwv} can be recast geometrically from a mirror CY$_4$ perspective. 

This paper addresses yet another aspect of brane brick models. Our main goal is to compute the elliptic genus of {\em abelian} brane brick models as a means to probe their infrared dynamics. Given the geometric origin of brane brick models, the most naive candidate for the infrared theory is a non-linear sigma model whose target space is the CY$_4$ associated with the gauge theory. The naive guess turns out to be correct. In a number of examples, we show that the gauge theory computation and the sigma model computation of the elliptic genus agree perfectly. To the extent the elliptic genus can differentiate theories, the infrared behavior of the gauge theory is the same as that of the sigma model.

There are several technical aspects of our computation that make the comparison between the gauge theory and the geometry non-trivial. For gauge theories, the supersymmetric localization for the elliptic genus was carried out in depth in \cite{Gadde:2013ftv,Benini:2013nda,Benini:2013xpa}; see also \cite{Benini:2016qnm} for a recent review and further references. Localization reduces the path integral to a finite dimensional contour integral over gauge fugacity variables, supplemented by the Jeffrey-Kirwan (JK) residue prescription \cite{JK-residue}. In simple examples, the contour integral can be performed explicitly and the result is a function of the modular parameter $\tau$ and flavor fugacity variables. 

For the non-linear sigma model we obtain a simple geometric formula by combining elements from related works in the literature \cite{Lerche:1988zy,Martelli:2006yb}. For any triangulation of the toric diagram of the CY$_4$, the geometric formula expresses the elliptic genus as a sum over tetrahedra in the triangulation. As expected, the sum is independent of the specifics of the triangulation.

The gauge theory computation is further complicated by the fact that, in some theories, the matter sector produces non-vanishing abelian gauge anomalies. Since the gauge theories that we consider have a clear string-theoretic origin, the anomaly should be cancelled through an interaction between open string modes and closed string modes. We have not been able to derive the precise anomaly cancelling mechanism from string theory. Instead, assuming the existence of a canceling mechanism, we have found an ansatz for the anomaly cancelling factor in the JK integral formula, which works for a large class of examples. Our ansatz is valid for theories in which the total number of chiral fields is greater than the number of Fermi fields in such a way that the anomaly polynomial can be written as a sum of squares with positive coefficients. 

As an application of these computations, we will check GGP triality for brane brick models. Brane brick models differ from the SQCD-like theories considered in the original papers on triality \cite{Gadde:2013lxa,Gadde:2014ppa} in that they correspond to quivers without flavor symmetry nodes. 
Triality has been proven in \cite{Gadde:2013lxa} by the use of an elliptic genus computation for SQCD-like theories and it is reasonable to expect that these calculations extend to more involved theories.
However, a systematic study of more complicated theories, such as quiver theories with only gauge nodes, has been lacking due to the increasing complexities of the required JK residue computations. For brane brick models, so far, triality has been shown to leave the CY$_4$ target space invariant by using the underlying geometry of the brane brick model construction \cite{Franco:2016nwv}.
In this paper, we will verify that brane brick models connected via triality share the same elliptic genus, as expected from the results in \cite{Franco:2016nwv}. This paper contains examples of the elliptic genus computation for simple brane brick models. 

The rest of this paper is organized as follows. 
In section \sref{sec:(0,2)review}, we briefly review $2d$ $(0,2)$ theories and brane brick models \cite{Franco:2015tna,Franco:2015tya,Franco:2016nwv} and set up our notation.
Section \sref{sec:JK-integral} reviews how to compute the elliptic genus from the gauge theory following \cite{Gadde:2013ftv,Benini:2013nda,Benini:2013xpa}.
In section \sref{sec:geo}, we propose a geometric formula that computes the elliptic genus from a triangulation of the toric diagram of the CY$_4$. 
In section \sref{sec:anomaly}, we discuss the general form of anomalies that are present in abelian brane brick models. We propose an ansatz for an anomaly cancelling factor that works in a large class of examples. 
In section \sref{sec:orbi}, we compute the elliptic genus of some orbifold models and find perfect agreement between the gauge theory and geometric computations.
In \sref{sec:non-orbi}, we confirm the agreement of the two computations in two non-orbifold models. In one of the examples, we also confirm the expectation that two gauge theories that are related by triality share the same elliptic genus. 
Section \sref{sec:discussion} concludes the paper with a discussion on future directions.

\section{Review of 2d (0,2) Gauge Theories and Brane Brick Models}
\label{sec:(0,2)review}

\subsection*{2d (0,2) Gauge Theories}\label{02notation}

We now briefly review basic aspects of $2d$ $(0,2)$ gauge theories to establish our notation. For more thorough reviews, we refer to \cite{Witten:1993yc,GarciaCompean:1998kh,Gadde:2013lxa}. 

There are three types of supermultiplets in $2d$ $(0,2)$ gauge theories. We will use superfield formalism with superspace coordinates $(x^\mu, \theta^+, \overline{\theta}^+ )$. All component fields are assumed to be complex-valued unless specified otherwise. The multiplets are:

\begin{itemize}
\item{\underline{Chiral multiplet $\Phi_i$}}

The physical component fields are a boson $\phi$ and a right-moving Fermion $\psi_+$. 

\item{\underline{Fermi multiplet $\Lambda_a$}} 

The only physical field is the left-moving fermion $\lambda_-$, which is a supersymmetry singlet in the free theory limit. Besides, the superfield contains an auxiliary field $G$ and a coupling to a holomorphic function $E(\Phi)$ of chiral superfields through a deformed chirality condition. 

\item{\underline{Vector multiplet $V_\alpha$}}

It contains the real gauge boson $v^\mu$, complex gaugini $\chi_{-}$, $\overline{\chi}_-$, and a real auxiliary field $D$. They couple to matter fields minimally through a supersymmetric completion of the gauge-covariant derivative.

\end{itemize}

\noindent
For each Fermi multiplet $\Lambda_a$, in addition to the holomorphic $E_a$-term mentioned above, it is possible to introduce another holomorphic term called $J^a(\Phi)$. 
The $(0,2)$ supersymmetry requires that $J$- and $E$-terms satisfy an overall constraint:
\begin{align}
\sum_a \mathrm{tr} \left[ E_a(\Phi_i) J^a(\Phi_i) \right]= 0 \,.
\end{align}

Integrating out the auxiliary fields $D_\alpha$, we obtain a familiar looking $D$-term potential (and its fermionic partner). For abelian theories, the potential takes the form
\begin{align}
V_D = \sum_\alpha \left( \sum_i q_{\alpha i} |\phi_i|^2 - t_\alpha \right)^2 \,,
\label{V_D}
\end{align}
where $t_\alpha$ are complexified Fayet-Iliopoulos (FI) parameters.
Integrating out the auxiliary fields $G_a$, we obtain what may be called an $F$-term potential, 
\begin{align}
V_F = \sum_a \left( \mathrm{tr}|E_a(\phi)|^2 +  \mathrm{tr}|J^a(\phi)|^2 \right)\,,
\label{V_JE}
\end{align}
as well as Yukawa-like interactions between scalars and pairs of fermions.

\subsection*{Brane Brick Models}\label{bbdef}

We can represent the $2d$ (0,2) quiver gauge theory that lives on the worldvolume of D1-branes probing a toric CY$_4$ by a brane brick model \cite{Franco:2015tna,Franco:2015tya,Franco:2016nwv,Franco:2016qxh}. When we T-dualize the D1-branes at the $\text{CY}_4$ singularity, we obtain a Type IIA brane configuration of D4-branes wrapping a 3-torus $T^3$ and suspended from an NS5-brane that wraps a holomorphic surface $\Sigma$ intersecting with $T^3$. This Type IIA brane configuration, which we call the brane brick model, is summarized in \tref{Brane brick-config}. The holomorphic surface $\Sigma$ encodes the geometry of the probed toric Calabi-Yau 4-fold and originates from the zero locus of the Newton polynomial of its toric diagram. 

\begin{table}[ht!!]
\centering
\begin{tabular}{l|cccccccccc}
\; & 0 & 1 & 2 & 3 & 4 & 5 & 6 & 7 & 8 & 9 \\
\hline
$\text{D4}$ & $\times$ & $\times$ & $\times$ & $\cdot$ & $\times$ & $\cdot$ & $\times$ & $\cdot$ & $\cdot$ & $\cdot$  \\
$\text{NS5}$ & $\times$ & $\times$ & \multicolumn{6}{c}{----------- \ $\Sigma$ \ ------------} & $\cdot$ & $\cdot$ \\
\end{tabular}
\caption{Brane brick model configuration of branes.}
\label{Brane brick-config}
\end{table}

The brane brick model encodes all the data needed to write down the full Lagrangian of the gauge theory. Moreover, it combines geometric and combinatorial data in a powerful way that enables us to analyze various properties of the gauge theory. 
Sometimes, it is more convenient to work with the periodic quiver, which is the graph dual of the brane brick model. Being graph dual, they contain exactly the same information. The dictionary between the brane brick model (or equivalently the periodic quiver) and the gauge theory is summarized in \tref{tbrick}.

\begin{table}[h]
\centering
\resizebox{\hsize}{!}{
\begin{tabular}{|l|l|l|}
\hline
{\bf Brane Brick Model} \ \ &  {\bf Gauge Theory} \ \ \ \ \ \ \  & {\bf Quiver diagram} \ \ \ 
\\
\hline\hline
Brick  & Gauge group & Node \\
\hline
Oriented face  & Bifundamental chiral field & Oriented (black) arrow 
\\
between bricks $i$ and $j$ & from node $i$ to node $j$  & from node $i$ to node $j$ \\
\hline
Unoriented square face  & Bifundamental Fermi field & Unoriented (red) line \\
between bricks $i$ and $j$ & between nodes $i$ and $j$ & between nodes $i$ and $j$  \\
\hline
Edge  & Interaction by $J$- or $E$-term & Plaquette encoding \\ 
& & a $J$- or an $E$-term \\
\hline
\end{tabular}
}
\caption{
Dictionary between brane brick models and $2d$ gauge theories.
\label{tbrick}
}
\end{table}

\subsection*{Non-Compact Target Space and its Regularization}\label{ncregularize}

We are dealing with theories whose target spaces are non-compact. Such theories may contain an infinite number of states along flat directions, and the elliptic genus may not be well-defined. In order to regulate this, we will use three of the four global $U(1)$ isometries of the toric $\text{CY}_4$ in order to refine the elliptic genera. The remaining $U(1)$ in the $\text{CY}_4$ is identified with the $R$-symmetry of the gauge theory. It cannot be used as a refinement since it does not commute with the supercharges. Instead, it will be used to saturate the fermionic zero mode from the decoupled $U(1)$ gaugino in the path integral; see section \sref{sec:JK-integral}.

\subsection*{A Comment on the Central Charge $c_R$ and the $R$-charge}

The right-moving central charge $c_R$ and the $R$-charge assignments of a $2d$ $(0,2)$ SCFT are closely related via a $c$-extremization \cite{Benini:2013cda}. A naive application of this principle, however, leads to $c_R=0$ for brane brick models. This cannot be true as long as the theory is a non-trivial unitary CFT. A similar breakdown of $c$-extremization has been reported in the theory of a free $(0,2)$ chiral multiplet in \cite{Benini:2013cda}. This failure is presumably due to the non-compactness of the corresponding target space. The non-compactness makes the vacuum non-normalizable and allows for an additional non-holomorphic current whose two-point function with the $R$-current might not vanish, violating an assumption of the extremization principle. A remedy to this breakdown will be the subject of a future investigation.

\section{Elliptic Genus from Gauge Theory}
\label{sec:JK-integral}

Recently, several groups \cite{Gadde:2013ftv,Benini:2013nda,Benini:2013xpa} independently derived a localization formula for computing the elliptic genus of a $2d$ $(0,2)$ gauge theory. The elliptic genus has become a powerful tool for studying the dynamics of these theories. For example, it has been recently used to verify GGP triality \cite{Gadde:2013lxa}. 
This section summarizes how to compute the elliptic genus following \cite{Benini:2013nda,Benini:2013xpa}.

The elliptic genus is defined by the trace over the Ramond-Ramond (R-R) sector, in which fermionic fields satisfy periodic boundary condition, as follows\footnote{One can use the NS-NS boundary condition to define the elliptic genus \cite{Gadde:2013ftv}, which is different from the R-R boundary condition we use here. For $2d$ $(2,2)$ theories, spectral flow can be used to compare the results from different boundary conditions. This is not the case for $2d$ $(0,2)$ theories. We will focus on the R-R boundary condition, which makes it easier to compare with the geometric formula in \sref{sec:geo}}:
\begin{align}
\label{esa1e1}
{\mathcal I}(q, x_{i})=\text{Tr}_\text{RR} (-1)^{F} q^{H_L} \overline{q}^{H_R} \prod_a x_{a}^{K_{a}}~,~
\end{align}
where  $K_i$ are the Cartan generators for the global flavor symmetry group. The parameter $q$ and the fugacities $x_i$ have logarithmic counterparts defined as $q=e^{2\pi i \tau}$ and $x_a = e^{2\pi i w_a}$. Given a charge vector $\rho$, we have $x^\rho = \prod_a x_i^{\rho^a}= e^{2\pi i \rho^a w_a}$.  
Note that $(0,2)$ supersymmetry ensures that the $\overline{q}$-dependence drops out of \eref{esa1e1}.

For a $2d$ $(0,2)$ GLSM, \eref{esa1e1} can be evaluated in terms of a contour integral of a meromorphic $(r,0)$-form $Z_\text{1-loop}$,
\beal{esa1e2}
{\mathcal I}(q, x_{i})= \frac{1}{(2\pi i)^{r}} \oint_C Z_{\text{1-loop}} ~.~
\eea
where $r$ is the rank of gauge group $G$.
$Z_\text{1-loop}$ is defined on the moduli space $\mathfrak{M}$ of flat connections of $G$ over $T^2$, where the contour $C$ in \eref{esa1e2} is an $r$-dimensional cycle in $\mathfrak{M}$. Each $2d$ $(0,2)$ multiplet contributes to the integrand $Z_\text{1-loop}= \prod Z_\text{multiplets}$. 
Let $\Phi$, $\Lambda$, $V$ denote chiral, Fermi, vector multiplets, respectively. The one-loop determinants are given by
\begin{align} \label{integrand-all}
Z_\Phi  = \prod_{\rho \in \mathcal{R}} i \frac{\eta(q)}{\theta_{1}(q,x^{\rho})} \,,
\quad
Z_\Lambda = \prod_{\rho \in \mathcal{R}}i \frac{\theta_{1}(q,x^{\rho})}{\eta(q)}\,, \quad
Z_{V}|_{G=U(1)^r} = \prod_{i=1}^{r} \frac{2 \pi \eta (q)^2}{i} du_{i} ~,~
\end{align}
where $\rho$ are the weights for the representation $\mathfrak{R}$ of the gauge and flavor groups in which the chiral and Fermi multiplets transform. Note that for the vector multiplet contribution with gauge group $G=U(1)^r$, we have $z_{i}=e^{2 \pi i u_{i}}$ with $i=1,\ldots,r$. The integral in \eref{esa1e2} is evaluated over $u_i$.
The definitions of the functions $\theta_1(q,y)$ and $\eta(q)$ in \eqref{integrand-all} are reviewed in appendix \ref{sec:theta-conv}.

The contour integral in \eref{esa1e2} is evaluated by following the Jeffery-Kirwan (JK) residue prescription \cite{JK-residue}. The physical motivation for the prescription is given in \cite{Benini:2013nda,Benini:2013xpa}. In the end, the prescription gives a formula for the elliptic genus in \eref{esa1e2}:
\begin{align}
\label{jkformula}
{\mathcal I}(q;x_{i})=\frac{1}{|W|} \sum_{u_{*} \in \mathfrak{M}_{\text{sing}}^{*}} \underset{u=u_{*}}{\text{JK-Res}} \left( \text{Q}|_{u_{*}},\eta \right) \ Z_\text{1-loop}(q,u,a_{i}) \,,
\end{align}
where $|W|$ is the order of the Weyl group $W$ of the gauge group $G$. In addition, $\text{Q}|_{u_{*}}$ is the set of charges labeling fields that give rise to the pole at $u=u_{*}$. Each charge is a normal covector of a singular hyperplane. $\eta$ is a generic charge covector that selects a set of poles $u_{*}$ that contribute to the JK residues in \eref{jkformula} depending on their covectors $\text{Q}|_{u_{*}}$. 

A pole is called \textit{non-degenerate} when it is determined by the intersection of exactly $r$ hyperplanes. Assuming a pole is at the origin ($u_*=0$), we can label these $r$ hyperplanes by their charge covectors $Q_{i} \in \mathbb{R}^r$, since the $i$-th hyperplane is defined as $Q_{i}(u)=0$. 
The JK residue for a non-degenerate pole at $u_{*}=0$ then takes the form
\beal{JKdefinition}
 \underset{u=0}{\text{JK-Res}} \left( \text{Q}|_{0},\eta \right) \frac{du_{1} \wedge \cdots \wedge du_{r}}{Q_{j_1}(u) \cdots Q_{j_r}(u) } = \left \{ \begin{array}{ll}
\frac{1}{|\text{det}(Q_{j_1} \cdots Q_{j_r})|} \quad & \textrm{if $\eta \in$ Cone($Q_{j_1}$,$\cdots$,$Q_{j_r}$)}\\
0 & \textrm{otherwise}\\
\end{array} \right.
~,~
\nn\\
\eea
where Cone($Q_{j_1}$,$\cdots$,$Q_{j_r}$) is a subspace of $\mathbb{R}^r$ spanned by $Q_{j_1}$,$\cdots$,$Q_{j_r}$ with positive coefficients. Let us make two important observations regarding the role of $\eta$. First, it determines which poles contribute to the index. Second, the final answer to the integral is independent of the choice of $\eta$. In other words, individual poles contributing to the index depend on the choice of $\eta$, but the final sum over all residues is independent of the choice.

One can also have a situation where $l > r$ hyperplanes intersect at $u_{*}=0$. We call the corresponding poles \textit{degenerate}. In appendix \ref{sdegpoles}, we present the so-called \textit{flag method} \cite{Benini:2013xpa} for resolving the JK residue for degenerate poles. The flag method generalizes the JK residue formula in \eref{JKdefinition} for any type of pole and arbitrary high rank $r$ of the gauge group. In fact, the computational complexity of the JK residue formula increases extremely fast with the rank $r$. For the present paper, this is not an issue since we focus on abelian theories with gauge group $U(1)^r$ for small values of $r$.

\subsection*{U(1) Decoupling and the Modified Elliptic Genus\label{U1decouple}}

Abelian gauge theories from brane brick models have gauge group $G=U(1)^r$, with all matter fields transforming in bifundamental or adjoint representations. 
As a result, the overall diagonal $U(1)$ decouples from the rest of the theory, leaving us with $U(1)^{r-1}$. This decoupling can be easily implemented by the following redefinition of gauge holonomy variables,
\begin{align}
u'_{0} =  \sum_{i=1}^{r} u_{i} \,, 
\quad
u'_{j} = u_{j} - u_r \;\; (i=1,\cdots,r-1) \,.
\label{relabelu}
\end{align}
We may discard the decoupled $U(1)$ vector multiplet at the classical level and compute the elliptic genus for the $G'=U(1)^{r-1}$ theory. However, we find it useful to maintain the decoupled $U(1)$ as elaborated below. 

A naive inclusion of the decoupled $U(1)$ makes the elliptic genus vanish.  From the path integral point of view, the vanishing is due to the gaugino zero modes. In the contour integral formula \eqref{jkformula}, we have $du'_0$ in the measure, but the integrand is independent of $u'_0$, leading to a vanishing result. 

In order to avoid the trivially vanishing result, we modify the definition of the index following the spirit of \cite{Cecotti:1992qh}. The key idea is to include the $R$-symmetry fugacity in the definition in \eqref{esa1e1} by inserting $b^R = e^{2\pi i \beta R}$ into the trace. Setting $\beta=0$ ($b=1$) gives the original index, which we call $\mathcal{I}_0(q;x_i)$. Because the $R$-charge does not commute with supercharges, for $\beta \neq 0$, the new twisted partition function is not protected by supersymmetry. However, we can consider its first derivative
\beal{esa1e200}
\mathcal{I}_1(q;x_i) \equiv \left.  \frac{\partial\mathcal{I}(q;x_i,b)}{2 \pi i \, \partial \beta} \right|_{\beta=0} \, ,
\eea
which under suitable conditions has a chance to become a supersymmetric index. 

Generally, checking whether $\mathcal{I}_1(q;x_i)$ qualifies as a supersymmetric index is challenging. In our case, however, the derivative in \eref{esa1e200} can be directly associated to the $U(1)$ decoupling. Since the free $U(1)$ decouples from the interacting part, any twisted partition function should factorize as follows
\begin{align}
    \mathcal{I}(q;x_i,b) = \mathcal{I}_\mathrm{free}(q;b) \times \mathcal{I}_\mathrm{int}(q;x_i,b) \, .
\end{align}
The free part, $\mathcal{I}_\mathrm{free}(q;b)$, is exact and $\bar{q}$-independent for arbitrary values of $\beta$, since the theory is {\em free}. This exactness does {\em not} rely on supersymmetry. Supersymmetry does imply $\mathcal{I}_\mathrm{free}(q;b=1)=0$. The interacting part, $\mathcal{I}_\mathrm{int}(q;x_i,b)$, becomes an index only if $b=1$. When $b \neq 1$, it is not protected by supersymmetry. Therefore, it may depend on $\bar{q}$ and get considerably renormalized. 

Going back to the first derivative in \eref{esa1e200}, we set $\beta =0$ to obtain
\beal{new-index}
&&
\mathcal{I}_1(q;x_i) =  \left. \frac{\partial \mathcal{I}(q;x_i,b)}{2 \pi i \, \partial \beta} \right|_{\beta=0} =
  \nn\\
  &&
  \hspace{0.5cm}
    \left. \frac{\partial \mathcal{I}_\mathrm{free}(q;b)}{2\pi i \, \partial \beta} \right|_{\beta=0} \times \mathcal{I}_\mathrm{int}(q;x_i,b=1) +   \mathcal{I}_\mathrm{free}(q;b=1) \times \left. \frac{\partial \mathcal{I}_\mathrm{int}(q;x_i,b)}{2 \pi i \, \partial \beta} \right|_{\beta=0} ~.~
\eea
Since $\mathcal{I}_\mathrm{free}(q;b=1) = 0$, the second term on the right-hand side vanishes. The remaining term qualifies as a supersymmetric index. In the rest of this paper, we will loosely call the first derivative $\mathcal{I}_1(q;x_i)$ the elliptic genus (or the index) and denote it by $\mathcal{I}(q;x_i)$ without any subscript. 

\begin{figure}[ht!!]
\begin{center}
\includegraphics[height=6cm]{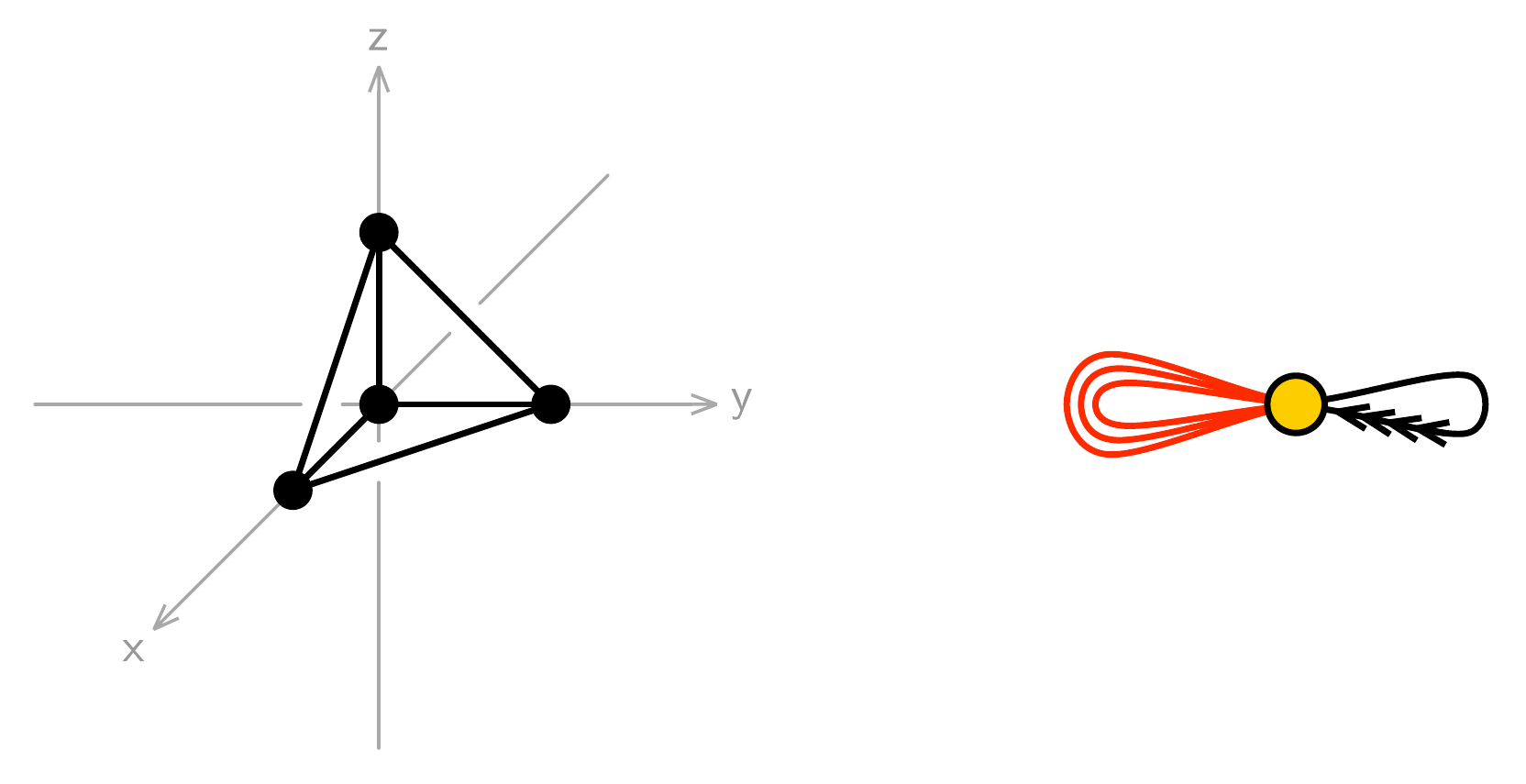} 
\caption{
Toric and quiver diagrams for $\mathbb{C}^4$. Black and red lines indicate chiral and Fermi fields, respectively.
\label{c4-toric}}
 \end{center}
 \end{figure}

\subsection*{Canonical Example: $\mathbb{C}^4$}

We present here the elliptic genus for the simplest abelian brane brick model corresponding to $\mathbb{C}^4$ \cite{GarciaCompean:1998kh, Franco:2015tna}.\footnote{The elliptic genera of its $SU(N)$ or $U(N)$ generalizations were thoroughly studied in \cite{Kologlu:2016aev}.} The theory has a single $U(1)$ gauge group. Its toric and quiver diagrams are shown in \fref{c4-toric}. The full global symmetry is $SU(3) \times U(1)_R$, where we assign fugacities $x$, $y$, $z$ to each of the $U(1)$ factors in the Cartan subalgebra of $SU(3)$.\footnote{It is important to note that the global symmetry of a $2d$ $(0,2)$ gauge theory depends not only on its quiver, but also on its $J$- and $E$-terms. For brevity, throughout the paper we will often provide only quiver diagrams, but the full theories are taken into account in our computations. Unless explicitly noted, the complete information about the theories we consider can be found in our earlier works \cite{Franco:2015tna,Franco:2015tya,Franco:2016nwv,Franco:2016qxh,Franco:2016fxm}.} \tref{tchc4} summarizes the global symmetry charges carried by the chiral and Fermi fields of the theory.

 \begin{table}[ht!!]
 \centering
 \begin{tabular}{c|cccc|ccc}
 field & 
 $X$ & $Y$ & $Z$ & $D$ &
 $\Lambda_1$ &  $\Lambda_2$ &  $\Lambda_3$ 
 \\
 \hline
 $U(1)_x$ & 
 $+1/2$ & $-1/2$ & $-1/2$ & $+1/2$ &
 $1$ & $0$ & $0$
 \\
 $U(1)_y$ &
$-1/2$ & $+1/2$ & $-1/2$ & $+1/2$ &
$0$ & $1$ & $0$
 \\
 $U(1)_z$ &
 $-1/2$ & $-1/2$ & $+1/2$ & $+1/2$ &
 $0$ & $0$ & $1$
 \\
 \hline
 $U(1)_R$ &
 $+1/2$ & $+1/2$ & $+1/2$ & $+1/2$ &
 $0$ & $0$ & $0$
 \\
 \end{tabular}
 \caption{
 Global charges of matter fields in the $\mathbb{C}^{4}$ theory. 
}
 \label{tchc4}
 \end{table}

The one-loop integrand from the matter sector is given by
\begin{align} \label{apparentc4}
\begin{split}
Z_\text{1-loop}= 
\frac{-i  \eta(q) 
 \theta_{1}(q,x) \theta_{1}(q,y)  \theta_{1}(q,z)
}{
\theta_{1}(q,\sqrt{b x y z}) 
\theta_{1}(q,\sqrt{b x/yz})
\theta_{1}(q,\sqrt{b y/xz})
\theta_{1}(q,\sqrt{b z/xy})
} \,.
\end{split}
\end{align}
The gaugino from the decoupled $U(1)$ contributes
\begin{align}
\left. \frac{\partial \mathcal{I}_\text{free} }{2\pi i \,\partial \beta} \right|_{\beta=0}  = \eta(q)^2 \,,
\end{align} 
since the elliptic genus of the free fermion reads
\begin{align} 
\mathcal{I}_\text{free}= i \frac{\theta_1(q,b)}{\eta(q)} \,.
\end{align}
Following the prescription in \eqref{new-index}, we have
\begin{align}
\begin{split}
\mathcal{I}_{\mathbb{C}^4}(q;x,y,z) &=
\frac{- i \eta(q)^3 
 \theta_{1}(q,x) \theta_{1}(q,y)  \theta_{1}(q,z)
}{
\theta_{1}(q,\sqrt{x/y z}) 
\theta_{1}(q,\sqrt{y/x z}) 
\theta_{1}(q,\sqrt{z/x y}) 
\theta_{1}(q,\sqrt{x y z}) 
}
\\
&= 
\frac{- i \eta(q)^3 
 \theta_{1}(q,x_1) \theta_{1}(q,x_2)  \theta_{1}(q,x_3)
}{
\theta_{1}(q,s_1) 
\theta_{1}(q,s_2) 
\theta_{1}(q,s_3) 
\theta_{1}(q,s_4) 
}
\,.
\end{split}
\label{C4-index}
\end{align}
In the second line, for later convenience, we introduced the shorthand notation, 
\begin{align}
\begin{split}
&\qquad \qquad \qquad \qquad
x_1 = x \,, 
\quad 
x_2 = y \,, 
\quad 
x_3 = z\,.
\\
&s_1 = \sqrt{x/yz} \,,
\quad 
s_2 = \sqrt{y/zx} \,,
\quad 
s_3 = \sqrt{z/xy} \,,
\quad 
s_4 = \sqrt{xyz} \,. 
\end{split}
\label{s-fugacity}
\end{align}

\subsubsection*{Poles in Fugacity Variables}

Since $\theta_1(q,y)$ has a simple zero at $y=1$, the index $\mathcal{I}_{\mathbb{C}^4}$ has simple zeroes at $x_a=1$ and and simple poles at $s_i=1$. Let us examine the $q$-expansion of the index: 
\begin{align}
\begin{split}
\mathcal{I}_{\mathbb{C}_4}(q;x,y,z) &= \frac{(1-x)(1-y)(1-z)}{\sqrt{x y z}} \mathcal{F}(q;x,y,z) \,,
\\
\mathcal{F}(q;x,y,z) &=  - \frac{1}{\prod_{i=1}^4 (1-s_i)} q^0 + 1\cdot q^1 + \left(1 + \sum_{i=1}^4 (s_i + s_i^{-1}) \right) q^2 + \mathcal{O}(q^3) 
\\
&\equiv \sum_{k=0}^\infty \mathcal{F}_k q^k \,.
\end{split}
\label{q-exp-1}
\end{align}
The $\mathcal{F}_0$ term inherits the poles at $s_i=1$. In contrast, $\mathcal{F}_1$ and $\mathcal{F}_2$ do not share the poles, and are Laurent polynomials in $s_i$. 
We can show that the poles are absent in $\mathcal{F}_k$ for all $k\ge 1$. Without loss of generality, we may focus on the pole at $s_4 =1$. Setting $z = \,e^{2\pi i \epsilon}/x y$ and taking the limit $\epsilon \rightarrow 0$, we obtain
\begin{align}
\begin{split}
\left. \mathcal{I}_{\mathbb{C}_4}\right|_{\epsilon \rightarrow 0} =
\frac{- i \eta(q)^3}{\theta_{1}(q,e^{+\pi i \epsilon})} 
\frac{\theta_{1}(q,x) \theta_{1}(q,y)  \theta_{1}(q,e^{2\pi i \epsilon}/x y)
}{
\theta_{1}(q,x e^{-\pi i \epsilon}) 
\theta_{1}(q,y e^{-\pi i \epsilon} ) 
\theta_{1}(q,e^{\pi i \epsilon}/x y)  
}
= \frac{1}{\pi i \epsilon}(1+\mathcal{O}(\epsilon)) \,,
\end{split}
\label{q-exp-2}
\end{align}
where we used the identity \eqref{theta-eta-iden}. 
Comparing \eqref{q-exp-1} and \eqref{q-exp-2}, we deduce that all $\mathcal{F}_{k \ge 1}$  belong to the $\mathcal{O}(\epsilon)$ part of \eqref{q-exp-2}, thereby proving the absence of the pole. 
Note that this proof relies crucially on the $\eta(q)^3$ factor in the numerator which originates from the definition of the modified elliptic genus \eqref{new-index}. 
 
In the next section, we will show that the absence of poles for $\mathcal{F}_{k\ge 1}$ generalizes to all toric CY$_4$.  

\section{Elliptic Genus from Geometry}
\label{sec:geo}

In this section, we propose a geometric formula for computing the elliptic genus. This formula only depends on the toric diagram of the Calabi-Yau 4-fold. The proposal for such a geometric formula is motivated by two relevant results known in the literature. The first comes from the equivariant localization approach to the computation of the elliptic genus for non-linear sigma models in \cite{Lerche:1988zy}. The second is based on the computation of the equivariant index, which counts holomorphic functions on a Calabi-Yau cone, in \cite{Martelli:2006yb}. This index is the Hilbert series of the coordinate ring formed by the holomorphic functions and has been shown to relate to the volume function of the base Sasaki-Einstein manifold. 

\subsubsection*{Martelli-Sparks-Yau Formula for Hilbert Series} 

We begin with a brief review of the geometric formula of \cite{Martelli:2006yb} derived by Martelli-Sparks-Yau (MSY), specialized to a CY$_4$; see also \cite{Benvenuti:2006qr}. We denote the CY cone by $X$ and the Sasaki-Einstein base by $Y$: $X=C(Y)$. 

The toric diagram of $X$ is defined by a collection of integer valued vectors $v_I = (v_I^1,v_I^2,v_I^3,v_I^4)  \in \mathbb{Z}^4$. The subscript $I$ runs from 1 to the number of external vertices of the toric diagram. The Calabi-Yau condition makes it possible to work in an $SL(4,\mathbb{Z})$ basis in which $v_I^4=1$ for all $I$. Projecting $v_I$ to the non-trivial components, we can visualize the toric diagram as a convex polytope in $\mathbb{Z}^3$. 

The toric diagram also defines a solid cone $\triangle_X \equiv \{y_i \in \mathbb{R}^4;(v_I\cdot y) \ge 0 \mbox{ for all I} \}$. Then $X$ is a $U(1)^4$ bundle over $\triangle_X$. Geometrically, the Hilbert series for $X$ enumerates lattice points on the solid cone $\triangle_X$, 
\begin{align}
H_X(t) = \sum_{\{m\}} \prod_{i=1}^4 t_i^{m^i} \quad 
(\{m \} \in \triangle_X \cap \mathbb{Z}^4 ) \,.
\end{align}
It was shown in \cite{Martelli:2006yb} that the normalized volume of the base $Y$ as a function of the Reeb vector $b^i$ can be derived from the Hilbert series via
\begin{align}
V_Y(b) \equiv \frac{\mathrm{Vol}(Y)}{\mathrm{Vol}(S^7)} 
= \lim_{\epsilon \rightarrow 0} \left[ \epsilon^4 H_X (t_i = e^{-\epsilon b^i} )\right]_{b^4 = 4}  \,.
\end{align}
The minimum of this function gives the volume of the Sasaki-Einstein base:
\begin{align}
\frac{\mathrm{Vol}(Y)}{\mathrm{Vol}(S^7)} = V_Y(b_*) \,.
\end{align}

Let us consider the simplest example, $X=\mathbb{C}^4$, for which $Y=S^7$. In this case, we have
\begin{align}
(v_I{}^i) = 
\begin{pmatrix}
1 & 0 & 0 & 1 \\
0 & 1 & 0 & 1 \\
0 & 0 & 1 & 1 \\
0 & 0 & 0 & 1 \\
\end{pmatrix} \,.
\end{align}  
The Hilbert series is 
\begin{align}
H_{\mathbb{C}^4}(t) = \frac{1}{(1-t_1)(1-t_2)(1-t_3)(1-t_4/t_1 t_2 t_3)} \,.
\label{HS-C4}
\end{align}
The volume function, 
\begin{align}
V_{S^7}(b) = \frac{1}{b_1 b_2 b_3 (4-b_1 -b_2 - b_3)} \,,
\end{align}
is minimized at
\begin{align}
b_* = (1,1,1)  \,.
\quad 
\Longrightarrow 
\quad 
V_{S^7}(b_*) = 1 \,.
\end{align}

Up to an $SL(4,\mathbb{Z})$ basis change, 
\begin{align}
t_1 = s_1 \,, 
\quad 
t_2 = s_2 \,,
\quad 
t_3 = s_3 \,,
\quad 
t_4 = s_1 s_2 s_3 s_4 \,, 
\label{C4-t-to-s}
\end{align}
the Hilbert series in \eqref{HS-C4} agrees with the standard formula for $\mathbb{C}^n$
\begin{align}
H_{\mathbb{C}^n}(s) = \prod_{i=1}^n \frac{1}{1-s_i} \,,
\end{align}
where each $s_i$ independently counts holomorphic monomials of $\mathbb{C}$. 
The key idea of the MSY formula for the Hilbert series is to triangulate the toric diagram by a set of minimal tetrahedra, treat each tetrahedra as a $\mathbb{C}^4$ and compute $H$ for it, and sum all these individual contributions.  
Concretely, consider a triangulation $\mathcal{T}_X$ consisting of minimal tetrahedra, $\{ a\} \in \mathcal{T}_X$, 
\begin{align}
\langle v_{a_1}, v_{a_2}, v_{a_3}, v_{a_4} \rangle = 
\epsilon_{ijkl} v_{a_1}^i v_{a_2}^j v_{a_3}^k v_{a_4}^l = 1 \,.
\end{align}
Introduce a dual vector for each face of a minimal tetrahedron:
\begin{align}
(w^{a_p})_i = \frac{1}{(3!)^2} \epsilon^{a_p a_q a_r a_s} \epsilon_{ijkl} v_{a_q}^j v_{a_r}^k v_{a_s}^l \,.
\end{align}
The set of dual vectors gives a formula for the Hilbert series:
\begin{align}
H_X(t) = \sum_{\{a\} \in \mathcal{T}_X} \prod_{p=1}^4 \frac{1}{1-\prod_i t_i^{(w^{a_p})_i}} \,.
\label{MSY-final}
\end{align}
The rigorous derivation of this formula, which is explained in \cite{Martelli:2006yb}, is an application of the Duijstermaat-Heckman localization formula \cite{MR674406}. 

\subsubsection*{Elliptic Genus from NLSM} 

On general grounds, we expect that the abelian GLSM's under consideration flow to NLSM's with CY$_4$ target spaces. As shown in \cite{Lerche:1988zy} in a similar but different context, it is possible to write down the NLSM and derive a formula for the elliptic genus from the path integral via localization. 
Let us sketch the derivation of the elliptic genus from the NLSM, leaving the details for a future work \cite{topub1}.

The field content of the NLSM of our interest is as follows. The bosonic fields $\phi^i$, $\bar{\phi}^{\bar{\imath}}$ $(i=1,2,3,4)$ represent complex coordinates on the CY$_4$. The right-moving $\psi^i$, $\bar{\psi}^{\bar{\imath}}$ describe the tangent bundle. The left-moving $\lambda_a$ $(1=1,\ldots,6)$ describe a vector bundle in the 6 (real) representation of the $SU(4)$ holonomy group of the CY$_4$. Finally, there are left-moving singlets $\chi$, $\bar{\chi}$, which are 
the NLSM counterpart of the decoupled $U(1)$ gaugini in the GLSM. 

The classical action of the NLSM contains suitably covariantized kinetic terms and a 4-Fermi $(\psi\bar{\psi} \lambda\lambda)$ curvature term. 
To compute the elliptic genus via path integral, one separates the zero modes and the quantum fluctuation around the zero modes. Supersymmetry ensures that the one-loop determinants, which capture the leading quantum fluctuations, become exact. The final result is a finite-dimensional integral over bosonic and fermionic zero modes, where the integrand is product of one-loop determinants over fluctuations. 

In computing the one-loop determinants, both the kinetic terms and the curvature term contribute. The dependence on the $\lambda$ zero-modes is cancelled in an intermediate step, so that the final result is a function of $R_{i\bar{\jmath}}\psi^i_0 \bar{\psi}^{\bar{\jmath}}_0$, where $R_{i\bar{\jmath}}$ is a contracted version of the curvature tensor. As usual, the Fermion zero modes $\psi_0$, $\bar{\psi}_0$ are interpreted as differential forms $d\phi$, $d\bar{\phi}$. Hence, the elliptic genus becomes a ($q$-dependent) characteristic class integrated over the manifold. 

So far, we have sketched how to compute the unflavored elliptic genus of the NLSM. To compute the flavored elliptic genus, we should deform the NLSM to include terms that depend on the Killing vectors for the $U(1)^3$ isometry of the toric CY$_4$. The path integral then localizes on the fixed points of the Killing vectors. The localization is similar to that used in the MSY formula we reviewed above. A crucial point is that the triangulation of the toric diagram amounts to dividing the CY$_4$ into a number of $\mathbb{C}^4$ patches, and that the localization simply collects contributions from each patch. 

\subsubsection*{Synthesis: Flavored Elliptic Genus from Triangulation} 

Combining the elements reviewed above, we are now ready to present the geometric formula for the elliptic genus of a CY 4-fold cone. For chiral fields, the key idea is 
to replace each $1/(1-t)$ in the Hilbert series by $i \eta(q)/\theta_1(q,t)$ for the elliptic genus. To account for Fermi fields, we supplement it by factors of $i \theta_1(q,z)/\eta(q)$. Finally, the non-zero modes of the singlet Fermis contribute $\eta(q)^2$.

The elliptic genus of $\mathbb{C}^4$ serves as the building block of the whole construction. For $\mathbb{C}^4$, the GLSM and the NLSM are equivalent and we can copy the result \eqref{C4-index}:
\begin{align}
\mathcal{I}_{\mathbb{C}^4}= 
\frac{
-i \eta(q)^3 \theta_1(q,x) \theta_1(q,y) \theta_1(q,z)}{
\theta_1(q,\sqrt{x/yz})
\theta_1(q,\sqrt{y/zx})
\theta_1(q,\sqrt{z/xy})
\theta_1(q,\sqrt{xyz})} \,.
\end{align}
To implement the $SL(3,\mathbb{Z})$ basis change in the triangulation, we rewrite this as 
\begin{align}
\mathcal{J}_{\mathbb{C}^4}(t) = \frac{-i \eta(q)^3 \theta_1(q,\sqrt{t_4}/t_2 t_3)\theta_1(q,\sqrt{t_4}/t_3 t_1)\theta_1(q,\sqrt{t_4}/t_1 t_2)}{\theta_1(q,t_1)\theta_1(q,t_2)\theta_1(q,t_3)\theta_1(q,t_4/t_1t_2t_3)} \,.
\end{align}
The relation between $t_i$ and $(x,y,z)$ for $X=\mathbb{C}^4$ is 
(see \eqref{C4-t-to-s} and \eqref{s-fugacity})
\begin{align}
t_1 = \sqrt{x/yz} \,, \quad
t_2 = \sqrt{y/zx} \,, \quad
t_3 = \sqrt{z/xy} \,, \quad
t_4 = 1\,.
\end{align}

Given a triangulation $\mathcal{T}(X)$ of the toric diagram for an arbitrary $X$, we first compute the ``pre-index"
\begin{align}
\mathcal{J}_X(t) =  \sum_{\{a\} \in \mathcal{T}(X)} \frac{
-i \eta(q)^3 \prod_{e=1}^3
\theta_1\left(q, z_e^{\{a\}}(t)\right)
}{
\prod_{p=1}^4 \theta_1\left(q, y_p^{\{a\}}(t)\right)
} \,.
\label{geo-pre}
\end{align}
The arguments in the denominator are the same as for the Hilbert series:
\begin{align}
y_p^{\{a\}}(t) = \prod_{i=1}^4 t_i^{(w^{a_p})_i} \,.
\end{align}
The arguments in the numerator are
\begin{align}
z_1^{\{a\}}(t) = \sqrt{\frac{y_1 y_4}{y_2 y_3}} \,,
\quad 
z_2^{\{a\}}(t) = \sqrt{\frac{y_2 y_4}{y_3 y_1}} \,,
\quad 
z_3^{\{a\}}(t) = \sqrt{\frac{y_3 y_4}{y_1 y_2}} \,.
\quad 
\end{align}
At the final stage, we translate $t_i$ into fugacities and turn off the R-symmetry fugacity: 
\begin{align}
\mathcal{I}_X(x,y,z) = \mathcal{J}_X\left(t_i = \prod_{a=1}^3 (x_a)^{m_{ia}}\right) \,,
\label{geo-R}
\end{align}
where $x_a = (x,y,z)$ as in \eqref{s-fugacity}. 

The exponents $m_{ia}$ in \eqref{geo-R} is determined by the requirement that the bosonic part of the chiral ring matches between the gauge theory and the geometry. For this purpose, we can go back to the Hilbert series reviewed earlier in this section. 

Recall that for $\mathbb{C}^4$, we expressed $t_i$ in terms of $s_i$ in \eqref{C4-t-to-s}. For orbifolds of $\mathbb{C}^4$, to be discussed in \sref{sec:orbi}, we can similarly rewrite $t_i$ in terms of $s_i$ by comparing the Hilbert series computed from triangulation with the Hilbert series computed from the Molien sum which implement the method of images. For example, for the orbifold $\mathbb{C}^4/\mathbb{Z}_2(0,0,1,1)$, the triangulation gives (see \fref{c4z2-0011-toric})
\begin{align}
H(t) =\frac{1}{(1-t_1) (1-t_2)} \left(
\frac{1}{(1-t_3)(1-t_4/t_1 t_2 t_3)}
+\frac{1}{(1-1/t_3)(1-t_3t_4/t_1t_2)} 
\right)\,,
\end{align}
while the Molien sum gives
\begin{align}
H(s) =\frac{1}{(1-s_1) (1-s_2)} \left(
\frac{1}{(1-s_3)(1-s_4)}
+\frac{1}{(1+s_3)(1+s_4)} 
\right)\,,
\end{align}
The two results agree if we change the variables as 
\begin{align}
t_1 = s_1  \,, \quad
t_2 = s_2  \,, \quad
t_3 = s_3/s_4 \,, \quad
t_4 = 1\,. 
\end{align}
The same principle applies to all orbifolds. 

For non-orbifolds, the Molien sum is not available, but we can still compare the chiral rings using the methods explained in \cite{Franco:2015tna,Franco:2015tya}. 

\subsubsection*{Index Theory and Fixed Point Formula}

Our discussion leading to the geometric formula \eqref{geo-R} relied heavily on 
the toric nature of the target space. Here we briefly digress to understand the formula from the standard index theory in a way less dependent on toric geometry. 

The elliptic genus of a general $(0,2)$ sigma model was derived in \cite{Kawai:1994np}. The sigma model consists of a $d$ dimensional K\"ahler target space $X$ equipped with a rank $r$ holomorphic vector bundle $E$. Anomaly cancellation gives restrictions on the first and second Chern classes of $E$ and those of the tangent bundle $T$. If we use the splitting principle to write formally, 
\begin{align}
c(E) = \prod_{i=1}^r (1+v_i) \,,
\quad 
c(T) = \prod_{j=1}^d (1+w_j) \,,
\end{align}
the elliptic genus turns out to be \cite{Kawai:1994np}
\begin{align}
Z_{X,E} = \int_X \prod_{i=1}^r P(\tau,v_i) \prod_{j=1}^d \frac{w_j}{P(\tau,-w_j)} \,, 
\quad 
P(\tau,z) = \frac{\theta(\tau|z)}{\eta(\tau)}\,.
\end{align}
Relating this to our formula takes a few steps. The most crucial step is to apply the standard fixed point formula to the characteristic class above by means of the toric isometry. Then, the integral over the target space is replaced by the sum over fixed points, and the curvature eigenvalues $v_i$ and $w_j$ are replaced by our fugacity variables $x_i$ and $s_j$. Additional care should be taken to incorporate the decoupled Fermi multiplets. A detailed derivation along this line will be given in \cite{topub1}.

\subsubsection*{Triangulation with Subtraction}

The geometric localization is based on a triangulation of the toric diagram. There are toric diagrams of CY$_4$ which do not admit simple triangulation, i.e. triangulations that only use the points in the toric diagram. For instance, the toric diagram of $\mathbb{C}^{4}/\mathbb{Z}_{2}(1,1,1,1)$ cannot be split into two unit tetrahedra anchored at integer lattice points. But, as we will see in \sref{sec:c4z21111}, it is possible to add up three tetrahedra and {\em subtract} one to construct the desired toric diagram. The geometric formula based on triangulation including subtraction was used in the computation of the Hilbert series in \cite{Lee:2012ft}. In this paper, we will apply the subtraction method to the geometric formula for the elliptic genus and find results compatible with other computations. 

\subsubsection*{Poles and Zero Modes}

Consider the $q$-expansion of the index obtained from the geometric formula,  
\begin{align}
\mathcal{I} = \mathcal{P} \sum_{k=0}^\infty \mathcal{F}_k  q^k \,.
\end{align}
where the prefactor $\mathcal{P}$ carries all zeroes of $\mathcal{I}$, but has no poles except at $x,y,z=0,\infty$. 

In the previous section, for the index of $\mathbb{C}^4$, we observed that $\mathcal{F}_0$ has codimension 1 simple poles in $(x,y,z)$ but all $\mathcal{F}_{k\ge 1}$ have no such poles and are Laurent polynomials in $(x,y,z)$. Combining that observation and the fact that the geometric formula sums up contributions from triangulation, we deduce that, for all toric CY$_4$, the codimension 1 poles are present only in $\mathcal{F}_0$ and absent from all $\mathcal{F}_{k\ge 1}$. 

Physically, from the NLSM point of view, 
the poles stem from the fact that the target space is non-compact. Without the fugacities, the ``center of momentum" degree of freedom moving in the non-compact direction will cause a divergence. The fugacities regulate the divergence. It is comforting to notice that all ``oscillator" degrees of freedom for $k\ge 1$ are not affected by the divergence. 

\section{Abelian Anomaly and its Cancellation}
\label{sec:anomaly}

\subsection{General Discussion} 

In $2d$ $(0,2)$ gauge theories, the difference in the spectrum of left-moving and right-moving fermions can potentially lead to anomalies. As explained in \cite{Franco:2015tna,Franco:2015tya,Franco:2016nwv}, gauge theories associated to brane brick models are automatically free of non-abelian gauge anomalies. But, depending on the particulars of matter multiplets, they may appear to suffer from abelian gauge anomalies. 

These gauge theories can be embedded in string theory \cite{Franco:2015tya,Franco:2016nwv,Franco:2016qxh}. So, there must exist an anomaly cancelling mechanism involving open string modes on branes and closed string modes away from branes. Although we have not identified the precise anomaly cancelling mechanism, we have found an ansatz for an anomaly cancelling factor in the contour integral formula. 

To set the stage for the anomaly cancelling factor, we should recall the relation between abelian gauge anomaly and modularity. 
In a theory with abelian gauge symmetry $U(1)^r$, the anomaly matrix 
is defined by 
\begin{align}
\mathcal{A}_{ij} = \Tr_\text{chiral}(Q_i Q_j) - \Tr_\text{Fermi}(Q_i Q_j) 
\quad (i,j=1,\cdots, r) \,. 
\end{align}
The same information is encoded in the anomaly polynomial defined by 
\begin{align}
\label{ano-poly}
\mathcal{A}(u) = \sum_{i,j=1}^{r} \mathcal{A}_{ij} u_i u_j \,,
\end{align}
where $u_i$ are the gauge holonomy variables. 

In the context of the elliptic genus, the abelian anomaly is tied to the modularity. We recall the modular properties of the $\theta$ and $\eta$ functions in the additive notation: 
\begin{align}
\frac{\theta_1(-1/\tau | z/\tau)}{\eta(-1/\tau)} = i e^{\pi i z^2/\tau} \frac{\theta_1(\tau |z)}{\eta(\tau)} \,.
\end{align}
For the gauge theories under consideration, the elliptic variable $z$ denotes gauge holonomy variables or flavor fugacities. It follows that the abelian gauge anomaly of the matter sector of the gauge theory is reflected in the modular property of the one-loop determinant $Z_{\Phi} Z_{\Lambda}$. Under the transformation $(\tau,u_i) \rightarrow (-1/\tau,u_i/\tau)$, the one-loop determinant picks up a multiplicative factor whose exponent is proportional to the anomaly polynomial \eqref{ano-poly}.

The contour integral for the elliptic genus \eqref{esa1e2} is well-defined only if the theory is anomaly-free. If we naively integrate an anomalous integrand, the result fails to exhibit definite modularity. So, we must ``cure" the anomaly before the integration. 

In the next subsection, we will present an anomaly cancelling factor that works for some class of gauge theories. Since one of our main results in this paper is to compare the gauge theory and the geometric computations, let us briefly digress to discuss the potential anomaly of the geometry formula

In the additive notation, each term in the geometric formula takes the form
\begin{align}
\frac{-i \eta(q)^3 \, \theta_1(\tau|z_1) \, \theta_1(\tau|z_2)\, \theta_1(\tau|z_3)}{
\theta_1(\tau|y_1) \, 
\theta_1(\tau|y_2) \,
\theta_1(\tau|y_3) \,
\theta_1(\tau|y_4)} \,, 
\end{align}
where the fugacities satisfy the relation, 
\begin{align}
z_a = \frac{1}{2}(y_4-y_1-y_2-y_3) +y_a \quad (a=1,2,3)\,.
\end{align}
The anomaly polynomial, reflected in the modular property, is 
\begin{align}
y_1^2+y_2^2+y_3^2+y_4^2 -z_1^2-z_2^2-z_3^2 = \frac{1}{4}(y_1+y_2+y_3+y_4)^2 \,. 
\label{geo-anomaly}
\end{align}
This factor vanishes as long as the triangulation of the toric diagram lives entirely in the CY hyperplane. We cannot compare the elliptic genus of the gauge theory with the geometric formula before determining how to cancel the anomaly of the gauge theory.

\subsection{Anomaly Cancelling Factor - An Ansatz}

Our ansatz works for theories in which the net contribution of chiral fields is greater than that of Fermi fields in such a way that the anomaly polynomial can be written as a sum of squares with unit positive coefficients. 

To illustrate the point, let us consider a one-parameter family of orbifolds denoted by $\mathbb{C}^{4}/\mathbb{Z}_{n}(1,1,-1,-1)$:
\begin{align}
(z_1, z_2, z_3, z_4) \sim (\omega_n z_1 , \omega_n z_2 , \omega_n^{-1} z_3, \omega_n^{-1} z_4) \,, 
\quad \omega_n \equiv e^{2\pi i/n} \,. 
\end{align}
The anomaly polynomial of an arbitrary $\mathbb{Z}_n$ orbifold was computed in \cite{Mohri:1997ef}. The result for the $\mathbb{C}^{4}/\mathbb{Z}_{n}(1,1,-1,-1)$ is 
\begin{align}
\begin{split}
\mathcal{A}(u) &= 4\sum_{i} (u_i - u_{i+1})^2 - \sum_{i} (u_i - u_{i+2})^2
\\
&= \sum_i (u_{i-1}+u_{i+1} - 2 u_i)^2 \equiv \sum_i \tilde{u}_i^2 \,.
\end{split}
\label{ano-orbi-simple}
\end{align}
Here, the sum runs from $1$ to $n$ and a cyclic identification mod $n$ is understood. The change of variables from $u_i$ to $\tilde{u}_i$ is not one-to-one, so rewriting $\mathcal{A}(u)$ in terms of $\tilde{u}$ is not equivalent to the standard diagonalization of a real symmetric matrix. 

Our ansatz for the anomaly cancelling factor for $\mathbb{C}^{4}/\mathbb{Z}_{n}(1,1,-1,-1)$ is 
\begin{align}
W_n(u_i;v) = \frac{\prod_i \theta_1(q,v \tilde{u}_i) + \prod_i  \theta_1(q,v/\tilde{u}_i)}{2 \theta_1(q,v)^n} \,.
\label{ano-ansatz}
\end{align}  
This factor has a few peculiar features. First, it has its own ``anomaly" which cancels precisely against the anomaly from the matter sector \eqref{ano-orbi-simple}. Second, it depends on an auxiliary variable $v$. Remarkably, once we integrate over the $u$ variables, the $v$-dependence completely disappears.  
Third, since the $u$-variables appear only in the numerator, the pole structure of the elliptic genus, which depends on the flavor fugacities, is not affected by the insertion of the anomaly cancelling factor. Fourth, the normalization of $W_n$ is such that when we expand the elliptic genus in a power series of $q$, the leading term is not affected by the insertion of $W_n$.

We discovered the factor $W_n$ in \eqref{ano-ansatz} ``experimentally" while working on the orbifold models $\mathbb{C}^{4}/\mathbb{Z}_{n}(1,1,-1,-1)$ with $n=2,3$. We will discuss these two examples in detail in \sref{sec:orbi}. But, further experiments revealed that it can be applied to a much larger class of theories. We conjecture that it works for all theories in which the anomaly polynomial admits the rewriting
\begin{align}
\mathcal{A}(u) = \sum_{i} \tilde{u}_i^2 \,,
\label{ano-rewriting}
\end{align}
where $\tilde{u}_i$ are linear combinations of $u_i$ with integer coefficients.

We can easily generalize the ansatz \eqref{ano-ansatz} to a larger class of orbifolds 
that include $\mathbb{C}^4/\mathbb{Z}_n(1, 1, -1, -1)$. According to \cite{Mohri:1997ef}, the anomaly matrix of the $\mathbb{Z}_n$ orbifold, whose action is labeled by integers $(a_1, a_2, a_3, a_4)$ satisfying $0\le a_i \le n-1$ and $\sum_i a_i \equiv 0$ $(\text{mod}\; n)$, is given by
\be
\mathcal{A}_{ij} = 2 \delta_{ij} - \sum_{\mu=1}^{4} \delta_{j,i+a_{\mu}} - \sum_{\mu=1}^{4} \delta_{i,j+a_{\mu}}+ \sum_{\mu=1}^{3} \delta_{j,i+a_4+a_{\mu}} + \sum_{\mu=1}^{3} \delta_{i,j+a_4+a_{\mu}} \,.
\ee
A large subset of these orbifolds, $\mathbb{C}^4/\mathbb{Z}_n (a,b,-b,-a)$, 
admit the rewriting \eqref{ano-rewriting},  
\begin{align}
\begin{split}
 \mathcal{A}(u) &= \sum_{i} \left( 4 u_{i}^2 - 4 u_{i} u_{i+a} -4 u_{i} u_{i+b} +2 u_{i}u_{i+b-a}  + 2  u_{i}u_{i-b-a} \right) \\
&= \sum_{i} \left( u_{i} - u_{i+a} - u_{i+b} + u_{i+a+b} \right)^2 
\equiv \sum_{i} \widetilde{u}_{i}^2 \,.
\end{split}
\end{align}
Thus the anomaly cancelling factor \eqref{ano-ansatz} is applicable to these orbifolds. Setting $a=b=1$ brings us back to the $\mathbb{C}^4/\mathbb{Z}_n (1,1,-1,-1)$ orbifolds considered earlier.

There is yet another large class of orbifolds to which the anomaly cancelling factor \eqref{ano-ansatz} applies: $\mathbb{C}^2/\mathbb{Z}_m(a,-a) \times \mathbb{C}^2/\mathbb{Z}_n(b,-b)$.  We can use a pair of indices $(i,j)$ ($i \in \{ 1,2, \cdots, m \} $, $j \in \{ 1, 2, \cdots, n \}$) to label gauge nodes and their holonomy variables. The anomaly matrix is given by \cite{Davey:2010px} 
\begin{align}
\begin{split}
\mathcal{A} = & \ 4 \mathbbm{1}_m \otimes \mathbbm{1}_n - \delta_{i_2,i_1+a} \otimes \mathbbm{1}_n - \delta_{i_2,i_1-a} \otimes \mathbbm{1}_n - \mathbbm{1}_m \otimes \delta_{j_2,j_1+b} -\mathbbm{1}_m \otimes \delta_{j_2,j_1-b} \\
& - \delta_{i_1,i_2+a} \otimes \mathbbm{1}_n - \delta_{i_1,i_2-a} \otimes \mathbbm{1}_n - \mathbbm{1}_m \otimes \delta_{j_1,j_2+b} -\mathbbm{1}_m \otimes \delta_{j_1,j_2-b} \\
& + \delta_{i_2,i_1+a} \otimes \delta_{j_2,j_1-b} + \delta_{i_1,i_2+a} \otimes \delta_{j_1,j_2-b} + \delta_{i_2,i_1-a} \otimes \delta_{j_2,j_1-b} + \delta_{i_1,i_2-a} \otimes \delta_{j_1,j_2-b} \,.
\end{split}
\end{align}
After multiplying by $mn$ holonomy variables $\{ u_{(i,j)} \}$, 
we can reorganize the anomaly polynomial as follows,
\begin{align}
\begin{split}
\mathcal{A}(u) &= 
\sum_{i,j}  \left[ 4 u_{(i,j)}^2 -2 u_{(i+a, j)} u_{(i,j)} - 2 u_{(i-a,j)} u_{(i,j)} - 2 u_{(i,j+b)} u_{(i,j)} -2 u_{i,j-b)} u_{(i,j)}  \right. 
\\
&\qquad\qquad \left. + u_{(i+a,j-b)} u_{(i,j)} + u_{(i-a,j+b)} u_{(i,j)} + u_{(i-a,j-b)} u_{(i,j)} + u_{(i+a,j+b)} u_{(i,j)}  \right] 
\\
&= \sum_{i,j}  (u_{(i,j)} - u_{(i+a,j)}-u_{(i,j-b)}+u_{(i+a,j-b)})^2 
\equiv \sum_{i,j} \widetilde{u}_{(i,j)}^2  \,.
\end{split}
\end{align}
In the next section, we will show how the anomaly cancelling factor works in concrete examples with small values of $m,n$.

\section{Orbifold Models}\label{sec:orbi}

In this section, we compute the elliptic genera of a few orbifold models. We find perfect agreement between the geometric computation and the gauge theory computation, even when the latter includes the anomaly cancelling factor. The results also agree with an independent computation using the standard orbifold CFT method. 

The orbifold CFT method expresses the elliptic genus in terms of a sum over twisted sectors.  
To be concrete, consider the $\mathbb{C}^4/\mathbb{Z}_n(a_1,a_2,a_3,a_4)$ orbifolds. The four integers $a_i$ satisfy 
$0 \le a_i \le n-1$ and $\sum_i a_i \equiv 0$ (mod $n$). 
It is useful to introduce the following notations, 
\begin{align}
&b_1 = a_1 + a_4 \,, 
\quad b_2 = a_2 + a_4 \,, 
\quad b_3 = a_3 + a_4 \,.
\end{align}
and recall the definition of $x_a$ and $s_i$ from \eqref{s-fugacity}. 

To incorporate the twisted boundary conditions, it is convenient to use the generalized theta functions:
\begin{align}
\theta[^{\alpha}_{\beta}](q,y) = \sum_{n\in \mathbb{Z}} q^{\frac{1}{2} (n+\alpha)^2} e^{2\pi i(n+\alpha)(z+\beta)} \,,
\quad 
q= e^{2\pi i\tau}\,, \;\; y = e^{2\pi iz} \,.
\label{twisted-theta}
\end{align}
For integer/half-integer values of $\alpha$, $\beta$, they reduce to the familiar $\theta_a(\tau,z)$ $(a=1,2,3,4)$:
\begin{align}
\theta_1 = - \theta[^{1/2}_{1/2}] \,, 
\quad 
\theta_2 = \theta[^{1/2}_{\ 0 }] \,, 
\quad 
\theta_3 = \theta[^{0}_{0}] \,, 
\quad
\theta_4 = \theta[^{\ 0 }_{1/2}] \,.
\end{align}
Additional information on the theta functions are collected in appendix \ref{sec:theta-conv}.

The orbifold form of the elliptic genus is given by
\begin{align}
\mathcal{I}_{\mathbb{C}^4/\mathbb{Z}_n(a_i)} = \frac{1}{n} \sum_{k,l=0}^{n-1} c_{k,l} \frac{N_{k,l}(b_a;x_a)}{D_{k,l}(a_i;s_i)} \,,
\label{orbi-formula}
\end{align}
where the numerator and denominator are 
\begin{align}
\begin{split}
N_{k,l}(b_a;x_a) &= i \eta(q)^3 \prod_{a=1}^3
\theta[^{1/2+b_a(k/n)}_{1/2+ b_a(l/n)}](q,x_a) \,,
\\
D_{k,l}(a_i;s_i) &= \prod_{i=1}^4 \theta[^{1/2+a_i(k/n)}_{1/2+ a_i(l/n)}](q,s_i) \,.
\end{split}
\end{align}
The phase factors $c_{k,l}$ in \eqref{orbi-formula} are fixed by requiring that the index should have a definite modular property and quasi-periodicity in shift of the fugacity variables according to the orbifold action. Barring the possibility of discrete torsion, these requirements should fix $c_{k,l}$ uniquely, as we verify in a number of examples. We will not discuss discrete torsion in this paper. 

\subsection{$\mathbb{C}^{4}/\mathbb{Z}_{2}(0,0,1,1)$}

\begin{figure}[ht!!]
\begin{center} 
\includegraphics[height=6cm]{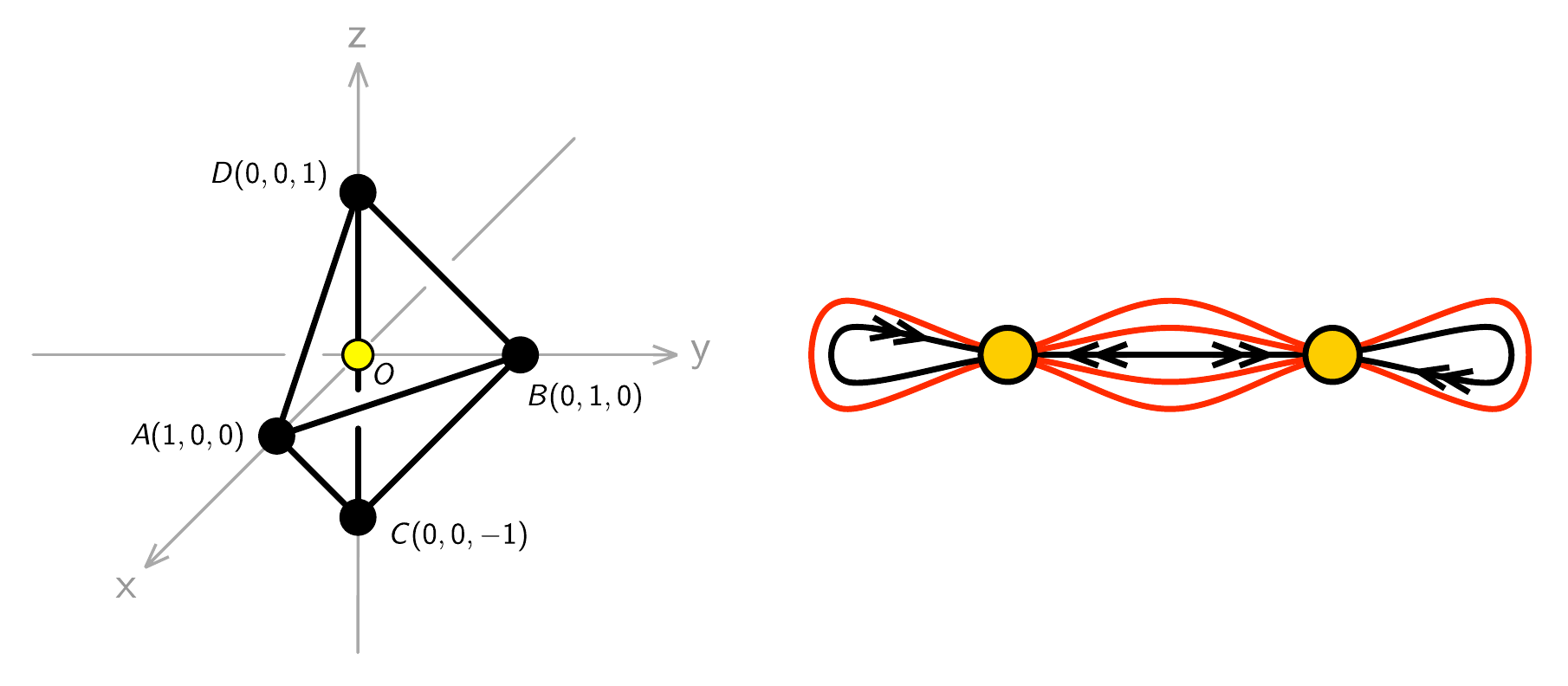} 
\caption{
Toric and quiver diagrams of the $\mathbb{C}^4/\mathbb{Z}_2(0,0,1,1)$ model.
\label{c4z2-0011-toric}}
 \end{center}
 \end{figure}

This is the simplest orbifold in the sense that the GLSM has two gauge nodes and that the gauge anomaly is absent. The toric diagram for the orbifold and the quiver diagram for the GLSM are shown in \fref{c4z2-0011-toric}.

\subsubsection*{Geometric Formula} 

To apply the geometric formula, we have to specify how to triangulate the toric diagram. For orbifold models, the toric diagram is a tetrahedron with a non-minimal volume. We assign labels A, B, C, D to the four external vertices and call the whole toric diagram $\mathcal{T}$(ABCD). The orientation is important here. Any even permutation of (ABCD) is equivalent to (ABCD), but an odd permutation implies an orientation reversal, which would flip the sign of the index.

Triangulation splits the toric diagram into a set of minimal tetrahedra. 
For the toric diagram in \fref{c4z2-0011-toric}, the result can be summarized as 
\begin{align}
\mathcal{T}\mbox{(ABCD)} = \triangle\mbox{(OABD)} +\triangle\mbox{(OACB)} \,, 
\end{align}
where the symbol $\triangle$ denotes a minimal tetrahedron.
Again, it is important to keep the orientation of all tetrahedra in a uniform way. The pre-index $J(t)$ explained in \sref{sec:geo} 
can be computed directly from the triangulation:
\begin{align}
\begin{split}
\mathcal{J}(t)
= &-\frac{i  \eta (q)^3 \theta_{1}(q,\sqrt{t_4}/t_1 t_2 )}{\theta_1(q,t_1) \theta_{1}(q, t_2) }
\times 
\\ 
& \biggl[ 
\frac{\theta_1(q,\sqrt{t_4}/t_2 t_3)\theta_1(q,\sqrt{t_4}/t_3t_1)}{\theta_1(q,t_3)\theta_1(q,t_4/t_1t_2t_3)} +
\frac{\theta_1(q,\sqrt{t_4} t_3/t_2)\theta_1(q,\sqrt{t_4} t_3/t_1)}{\theta_1(q,1/t_3)\theta_1(q,t_3t_4/t_1t_2)} 
\biggr]\,.
\end{split}
\label{pre-indexc4z2-0011}
\end{align}
The relation between $t_i$ and $(x,y,z)$ for this orbifold is 
\begin{align}
t_1 = s_1 = \sqrt{x/yz} \,, \quad
t_2 = s_2 = \sqrt{y/zx} \,, \quad
t_3 = s_3/s_4 = 1/xy \,, \quad
t_4 = 1\,, 
\end{align}
where $s_i$ are defined in \eqref{s-fugacity}.
The index is then 
\begin{align}
\begin{split}
\mathcal{I}
= \,& \frac{ i \eta (q)^3 \theta_{1}(q,z)}{\theta_1(q,\sqrt{x/yz}) \theta_{1}(q, \sqrt{y/xz}) }
\times 
\\ 
& \biggl[ 
\frac{\theta_1(q,\sqrt{x^3yz})\theta_1(q,\sqrt{xy^3z})}{\theta_1(q,xy)\theta_1(q,xyz)} +
\frac{\theta_1(q,\sqrt{x^3y/z})\theta_1(q,\sqrt{xy^3/z})}{\theta_1(q,xy)\theta_1(q,xy/z)} 
\biggr]\,.
\end{split}
\label{indexc4z2-0011-geo}
\end{align}

\subsubsection*{Orbifold CFT} 
The orbifold CFT computation gives
\begin{align}
\begin{split}
\mathcal{I}
=& - \frac{ i\eta (q)^3 \theta_{1}(q,z)}{\theta_1(q,\sqrt{x/yz}) \theta_{1}(q,\sqrt{y/xz}) }
\times 
\\ 
& \frac{1}{2} \left( \frac{\theta_1(q,x)\theta_1(q,y)}{\theta_1(q,\sqrt{z/xy})\theta_1(q,\sqrt{xyz})} - \sum_{a=2}^4 \frac{\theta_a(q,x)\theta_a(q,y)}{\theta_a(q,\sqrt{z/xy})\theta_a(q,\sqrt{xyz})} \right) \,.
\end{split}
\label{indexc4z2-0011-orbi}
\end{align}
Remarkably, the two results \eqref{indexc4z2-0011-geo} and \eqref{indexc4z2-0011-orbi} agree perfectly.

\subsubsection*{GLSM} 

We take the orbifold such that $X$, $Y$, $X'$, $Y'$ become adjoints while 
$Z$, $D$, $Z'$, $D'$ become (anti-)bifundamentals. Then the integrand becomes 
\begin{align}
\begin{split}
Z_\text{1-loop} = & \frac{ 2 \pi i  \eta(q)^4 \theta_{1}(q,z)^2}{\ \theta_{1}(q,\sqrt{x/yz})^2 \theta_{1}(q, \sqrt{y/xz})^2 } 
\\ 
& \times \frac{\theta_{1}(q,u x) \theta_{1}(q,x/u) \theta_{1}(q,uy) \theta_{1}(q,y/u) }{\ \theta_{1}(q,u\sqrt{z/xy}) \ \theta_{1}(q,u^{-1} \sqrt{z/xy})\theta_{1}(q,u \sqrt{xyz})  \theta_{1}(q,u^{-1}\sqrt{xyz})} \,,
\end{split}
\label{integrandc4z2-0011}
\end{align}
where $u = u_{12} = u_1/u_2$ is the non-trivial gauge holonomy variable after the overall $U(1)$ decoupling. 
Evaluating the JK residues with $\eta = +1$, we obtain 
\begin{align}
\begin{split}
\mathcal{I} 
= & \frac{ i \eta (q)^3 \theta_{1}(q,z)}{\theta_1(q,\sqrt{x/yz}) \theta_{1}(q, \sqrt{y/xz}) }
\times 
\\ 
& \biggl[ 
\frac{\theta_1(q,\sqrt{x^3yz})\theta_1(q,\sqrt{xy^3z})}{\theta_1(q,xy)\theta_1(q,xyz)} +
\frac{\theta_1(q,\sqrt{x^3y/z})\theta_1(q,\sqrt{xy^3/z})}{\theta_1(q,xy)\theta_1(q,xy/z)} 
\biggr]\,.
\end{split}
\label{indexc4z2-0011}
\end{align}
The two terms in \eqref{indexc4z2-0011} match precisely the two tetrahedron contributions in the geometric computation.

\subsection{$\mathbb{C}^{4}/\mathbb{Z}_{2}(1,1,1,1)$ 
\label{sec:c4z21111}}

\begin{figure}[ht!!]
\begin{center}
\includegraphics[height=5.5cm]{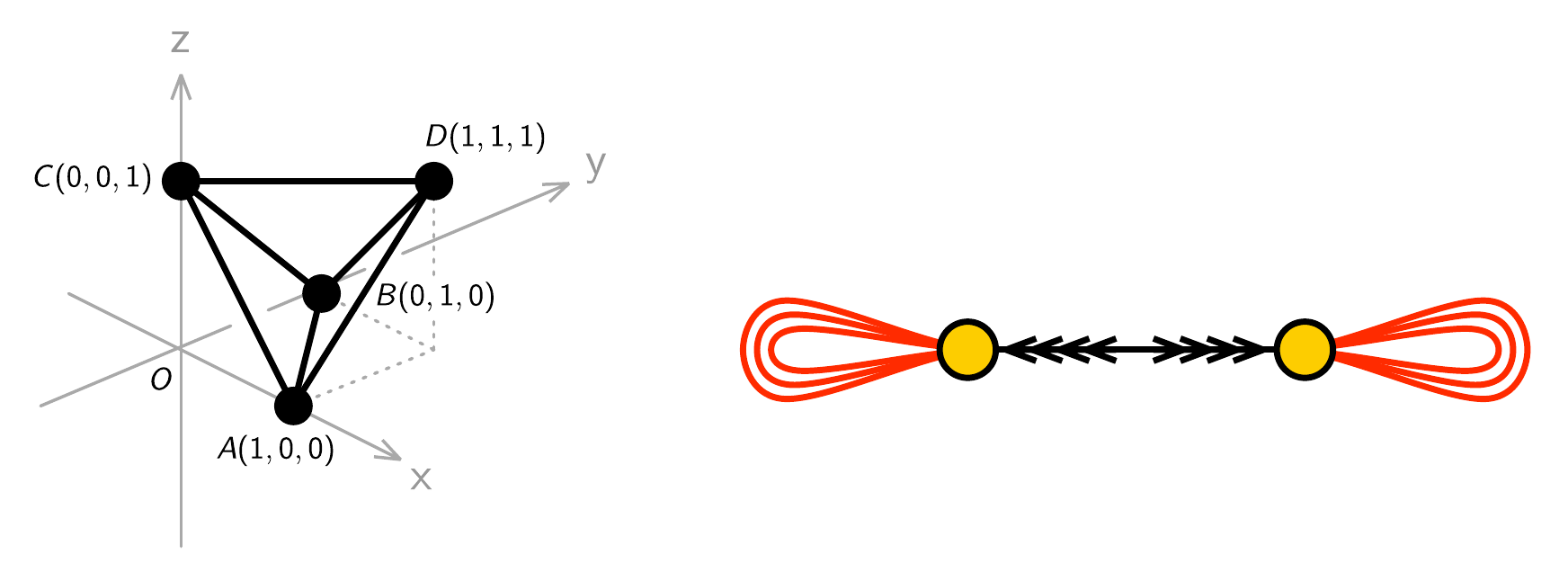} 
\caption{
Toric and quiver diagrams of the $\mathbb{C}^4/\mathbb{Z}_2(1,1,1,1)$ model.
\label{c4z2-1111-toric}}
 \end{center}
 \end{figure}

This is the simplest orbifold with non-vanishing gauge anomaly from the matter sector. Its Toric and quiver diagrams are shown in 
\fref{c4z2-1111-toric}.

\subsubsection*{Geometric formula (with subtraction)}

The triangulation needed for the subtraction method can be summarized as 
\begin{align}
\mathcal{T}(\mbox{ABCD}) = \triangle\mbox{(OABD)} +\triangle\mbox{(OBCD)}+\triangle\mbox{(OCAD)}-\triangle\mbox{(OABC)} \,.
\end{align}
The pre-index following from the triangulation is 
\begin{align}
\begin{split}
\mathcal{J}(t)=&\; \mbox{cyclic}\left[-
\frac{ i\eta(q)^3 \theta_1 (q,\sqrt{t_4}/t_2) \theta_1 (q,\sqrt{t_4}/t_1) \theta_1 (q,\sqrt{t_4}t_3^2/t_1t_2)}{\theta_1 (q,t_1/t_3) \theta_1 (q,t_2/t_3) \theta_1 (q,t_3) \theta_1 (q,t_3t_4/t_1t_2)}  \right]
\\
&  \qquad\qquad + \frac{ i\eta(q)^3 \theta_1 (q,\sqrt{t_4}/t_2t_3) \theta_1 (q,\sqrt{t_4}/t_3t_1) \theta_1 (q,\sqrt{t_4}/t_1t_2)}{\theta_1 (q,t_1) \theta_1 (q,t_2) \theta_1 (q,t_3) \theta_1 (q,t_4/t_1t_2t_3)} 
\,.
\end{split}
\label{c4z2-1111-pre}
\end{align}
Here, we defined a cyclic sum as 
\begin{align}
\mbox{cyclic}[f(t_1,t_2,t_3)] = f(t_1,t_2,t_3) + f(t_2,t_3,t_1) +f(t_3,t_1,t_2)\,.
\label{cyclic-sum}
\end{align}
The relation between $t_i$ and $(x,y,z)$ for this orbifold is 
\begin{align}
t_1 = s_1 s_4 = x \,, \quad
t_2 = s_2 s_4 = y \,, \quad
t_3 = s_3 s_4 = z \,, \quad
t_4 = 1\,.
\end{align}
The final result of the geometric computation is 
\begin{align}
\begin{split}
\mathcal{I}= 
-i &\left[ \frac{\eta(q)^3 \theta_1 (q,x) \theta_1 (q,y) \theta_1 (q,\frac{z^2}{x y})}{ \theta_1 (q,\frac{x}{z})  \theta_1 (q,\frac{y}{z})
\theta_1 (q,z) \theta_1 (q,\frac{z}{x y})} 
+\frac{\eta(q)^3 \theta_1 (q,y) \theta_1 (q,z) \theta_1 (q,\frac{x^2}{y z})}{ \theta_1 (q,\frac{y}{x}) \theta_1 (q,\frac{z}{x})
\theta_1 (q,x) \theta_1 (q,\frac{x}{y z})} \right.
\\
\quad &\left.  +\frac{\eta(q)^3 \theta_1 (q,z) \theta_1 (q,x) \theta_1 (q,\frac{y^2}{z x})}{ \theta_1 (q,\frac{z}{y}) \theta_1 (q,\frac{x}{y}) 
\theta_1 (q,y) \theta_1 (q,\frac{y}{x z})}
-\frac{\eta(q)^3 \theta_1 (q,x y) \theta_1 (q,y z) \theta_1 (q,z x)}{\theta_1 (q,x) \theta_1 (q,y) \theta_1 (q,z) \theta_1 (q,x y z)} \right]
\,.
\end{split}
\label{c4z2-1111-geo}
\end{align}
%

\subsubsection*{Single term formula}

It is possible to prove that \eqref{c4z2-1111-geo} is precisely equal to 
\begin{align}
\mathcal{I} = 
 \frac{ -i \eta(q)^3 \theta_1(q,x^2) \theta_1(q,y^2) \theta_1(q,z^2)}{
\theta_{1}(q,x/yz)
\theta_{1}(q,y/zx)
\theta_{1}(q,z/xy)
\theta_{1}(q,xyz)}
\,.
\label{c4z2-1111-curious}
\end{align}
It is interesting to observe that this is the same as the index for $\mathbb{C}^4$, 
except that all the fugacity variables have been ``squared".
The physical reason behind this expression is not clear to us at the moment.

\subsubsection*{Orbifold CFT} 

The orbifold CFT computation gives
\begin{align}
\mathcal{I} = - \frac{i}{2} \sum_{a=1}^4 
 \frac{ (-1)^{a+1} \eta(q)^3 \theta_1(q,x) \theta_1(q,y) \theta_1(q,z)}{
\theta_{a}(q,s_1)
\theta_{a}(q,s_2)
\theta_{a}(q,s_3)
\theta_{a}(q,s_4)}
\,.
\label{c4z2-1111-orbi}
\end{align}
It is straightforward to prove that \eqref{c4z2-1111-orbi} is 
equal to \eqref{c4z2-1111-geo}.

\subsubsection*{GLSM} 

The $\mathbb{C}^4/\mathbb{Z}_2$ $(1,1,1,1)$ model is the simplest model with non-vanishing gauge anomaly. The anomaly polynomial is \beal{es500a19}
\mathcal{A}(u) = 8(u_1-u_2)^2 = 2( 2 u_{12})^2 \equiv 2 (2u)^2 \,.
\eea
Following the general proposal of \sref{sec:anomaly},  
we insert the anomaly cancelling factor, 
\begin{align}
W(u^2;v)= 
\frac{
\theta_1 \left(q, v u^2 \right)
\theta_1 \left(v/u^2 \right)
}{
\theta_1 \left(q, v\right)^2
}~,~
\label{es500a20}
\end{align}
into the JK integral formula. 
The integrand of the JK integral is 
\begin{align}
Z_\text{1-loop} = \  \frac{2 \pi i \eta(q)^4 \theta_{1}(q,x)^2 \theta_{1}(q,y)^2 \theta_{1}(q,z)^2}{\prod_{i=1}^4 \theta_1(q,s_i u)\theta_1(s_i/u)} W(u^2;v)
\,.
\label{integrandc4z2}
\end{align}
With the choice $\eta=+1$, the elliptic genus is the collection of four residues, 
\begin{align}
\mathcal{I}
= i\eta (q)^3 \theta _1(q,x) \theta _1(q,y) \theta_1(q,z)\,
 \mbox{tetra}\left[
\frac{W(xyz;v)}{
\theta _1(q,x y) 
\theta _1(q,y z) 
\theta _1(q,z x) 
\theta_1(q,x y z)}
\right]\,,
\label{c4z2-1111-jk}
\end{align}
where we defined a sum reflecting the tetrahedral symmetry of the toric diagram, 
\begin{align}
\mbox{tetra}[g(x,y,z)] = g(x,y,z) + g(x,y^{-1},z^{-1}) + g(x^{-1},y,z^{-1}) + g(x^{-1},y^{-1},z)\,. 
\label{tetra-sum}
\end{align}
Remarkably, despite the appearance, \eqref{c4z2-1111-jk} is independent of the auxiliary variable $v$ and is equal to \eqref{c4z2-1111-geo} and \eqref{c4z2-1111-orbi}.
 
\subsubsection*{Comment on non-abelian global symmetry} 
The orbifold $\mathbb{C}^4/\mathbb{Z}_2(1,1,1,1)$ preserves the $SU(4)$ global symmetry of $\mathbb{C}^4$. It is interesting to see how the global symmetry is realized in each of the four different forms of the index. In the geometric form \eqref{c4z2-1111-geo}, 
the particular choice breaks manifest $SU(4)$ invariance.  
In the single term form \eqref{c4z2-1111-curious} or orbifold form \eqref{c4z2-1111-orbi}, each term is manifestly $SU(4)$-invariant. 
Finally, in the GLSM form \eqref{c4z2-1111-jk}, the four terms are related to each other by the Weyl group of $SU(4)$.

\subsubsection*{Bootstrap approach to the anomaly cancelling term}

The idea of a holomorphic anomaly cancelling factor was first conceived while studying this orbifold example. We now briefly review how a bootstrap approach, namely general consistency requirements, led us to the $W$ function.

It is reasonable to imagine that the correct anomaly cancelling mechanism will eventually contribute some extra theta function to the integrand of the JK residue integral. 
The new fields participating in the cancelling mechanism communicate directly to the anomalous $U(1)$ gauge field. Furthermore, the anomaly cancelling term should be invariant under the $SU(4)$ global symmetry. So the extra theta functions can only depend on the gauge fugacity $u$ but not on the global fugacities $x$, $y$, $z$. Let us denote the unknown combinations of extra theta functions by $W(q,u^2)$, 
where the square is introduced by convention to cancel square roots in the last line of \eqref{integrandc4z2}.

For simplicity, let us further assume that the extra factor does not introduce a new pole in the JK residue computation. The validity of this ansatz can be checked {\it a posteriori}. 
Comparing this ansatz with the simplest formula \eqref{c4z2-1111-curious} leads to the equation,
\begin{align}
\mbox{tetra}\left[\frac{\theta_1(xy/z) \theta_1(yz/x)\theta_1(zx/y)W(xyz)}{\theta_1(xy)\theta_1(yz)\theta_1(zx)} \right]
= 
 \frac{\theta_1(x^2) \theta_1(y^2)\theta_1(z^2)}{\theta_1(x)\theta_1(y)\theta_1(z)} \,.
\label{boot-eq}
\end{align}
To avoid clutter, in the remainder of this subsection, we suppress the $q$-dependence. 

It is not clear {\it a priori} whether a function $W$ satisfying this equation exists at all. Assuming its existence,
we can deduce several properties of the function $W(q,u^2)$. Switching to the additive notation momentarily, $W(\tau| \zeta)$, $q=e^{2\pi i \tau}$, $u^2 = e^{2\pi i \zeta}$, the crucial properties include:

\begin{enumerate}

\item Normalization and Parity:
$W(\tau | 0)=1$, $W(\tau | \zeta) = W( \tau | -\zeta)$.

\item 
Periodicity: 
$W(\tau | \zeta+1) = W(\tau | \zeta)$, $W(\tau | \zeta+\tau) = q^{-1}e^{-4\pi i \zeta} W(\tau | \zeta)$.

\item
Modularity: 
$W(\tau+1| \zeta) = W(\tau| \zeta)$, $W(-1/\tau| \zeta/\tau) = e^{2\pi i \zeta^2/\tau} W(\tau| \zeta)$. 

\end{enumerate}

\noindent
We can proceed further by taking the $z \rightarrow 1$ limit of \eqref{boot-eq}:
\begin{align}
\theta_1(xy)^2 W(x/y) - \theta_1(x/y)^2 W(xy) =  \theta_1(x^2)\theta_1(y^2) \,.
\end{align}
By a change of variables, $v=xy$, $w=x/y$, we can rewrite this as 
\begin{align}
\theta_1(v)^2 W(w) - \theta_1(w)^2 W(v) = \theta_1(vw)\theta_1(v/w) \,.
\label{boot-eq2}
\end{align}
The solution to this equation is far from unique. 
If $W_0(w)$ is a solution, then clearly
\begin{align}
W_k(w) = W_0(w) + k\, \theta_1(w)^2
\label{boot-gauge}
\end{align}
is another solution. Here, the factor $k$ is some modular form in $\tau$, independent of $z$. 

Suppose we use this ambiguity to choose a zero of $W$, i.e. we demand that $W(v_*)=0$ for some $v_*$ that is not equal to $m+n\tau$ $(m,n\in \mathbb{Z})$. Then \eqref{boot-eq2} gives 
\begin{align}
W(w;v_*) = \frac{ \theta_1(v_*w)\theta_1(v_*/w) }{\theta_1(v_*)^2} \,.
\label{boot-sol}
\end{align}
Conversely, all solutions of this form are equivalent in the sense of \eqref{boot-gauge}. 
To prove this claim, it suffices to use a well known identity, 
\begin{align}
\theta_1(vw)\theta_1(v/w)\theta_4(1)^2 = \theta_1(v)^2\theta_4(w)^2 - \theta_1(w)^2\theta_4(v)^2 \,, 
\label{4theta-identity}
\end{align}
to rewrite $W(w;v_*)$ as 
\begin{align}
W(w;v_*) = \left(\frac{\theta_4(w)}{\theta_4(1)}\right)^2  - \left(\frac{\theta_4(v_*)}{\theta_1(v_*)\theta_4(1)}\right)^2 \theta_1(w)^2  \,. 
\label{boot-sol2}
\end{align}
Clearly, the difference between two $W(w;v_*)$'s is proportional to $\theta_1(w)^2$. 

Finally, we should recall that we found the general solution \eqref{boot-sol} 
from the simplified equation \eqref{boot-eq2}. We should verify that it satisfies 
the original equation \eqref{boot-eq}. 
In terms of \eqref{boot-eq}, the ambiguity \eqref{boot-gauge} amounts to 
\begin{align}
\mbox{tetra}\left[\frac{\theta_1(xyz)}{\theta_1(xy)\theta_1(yz)\theta_1(zx)} 
\right] = 0 \,.
\label{boot-check1}
\end{align}
Inserting the solution \eqref{boot-sol2} with $\theta_4(v_*)=0$ to \eqref{boot-eq} gives
\begin{align}
\mbox{tetra}\left[\frac{\theta_1(xy/z) \theta_1(yz/x)\theta_1(zx/y)\theta_4(xyz)^2}{\theta_1(xy)\theta_1(yz)\theta_1(zx)\theta_4(1)^2} \right]
= 
\frac{\theta_1(x^2) \theta_1(y^2)\theta_1(z^2)}{\theta_1(x)\theta_1(y)\theta_1(z)} \,.
\label{boot-check2}
\end{align}
Both \eqref{boot-check1} and \eqref{boot-check2} can be proved using the argument given in appendix \ref{sec:theta-iden}.

\subsection{$\mathbb{C}^{4}/\mathbb{Z}_{3}(1,1,2,2)$}

\begin{figure}[ht!!]
\begin{center} 
\includegraphics[height=6cm]{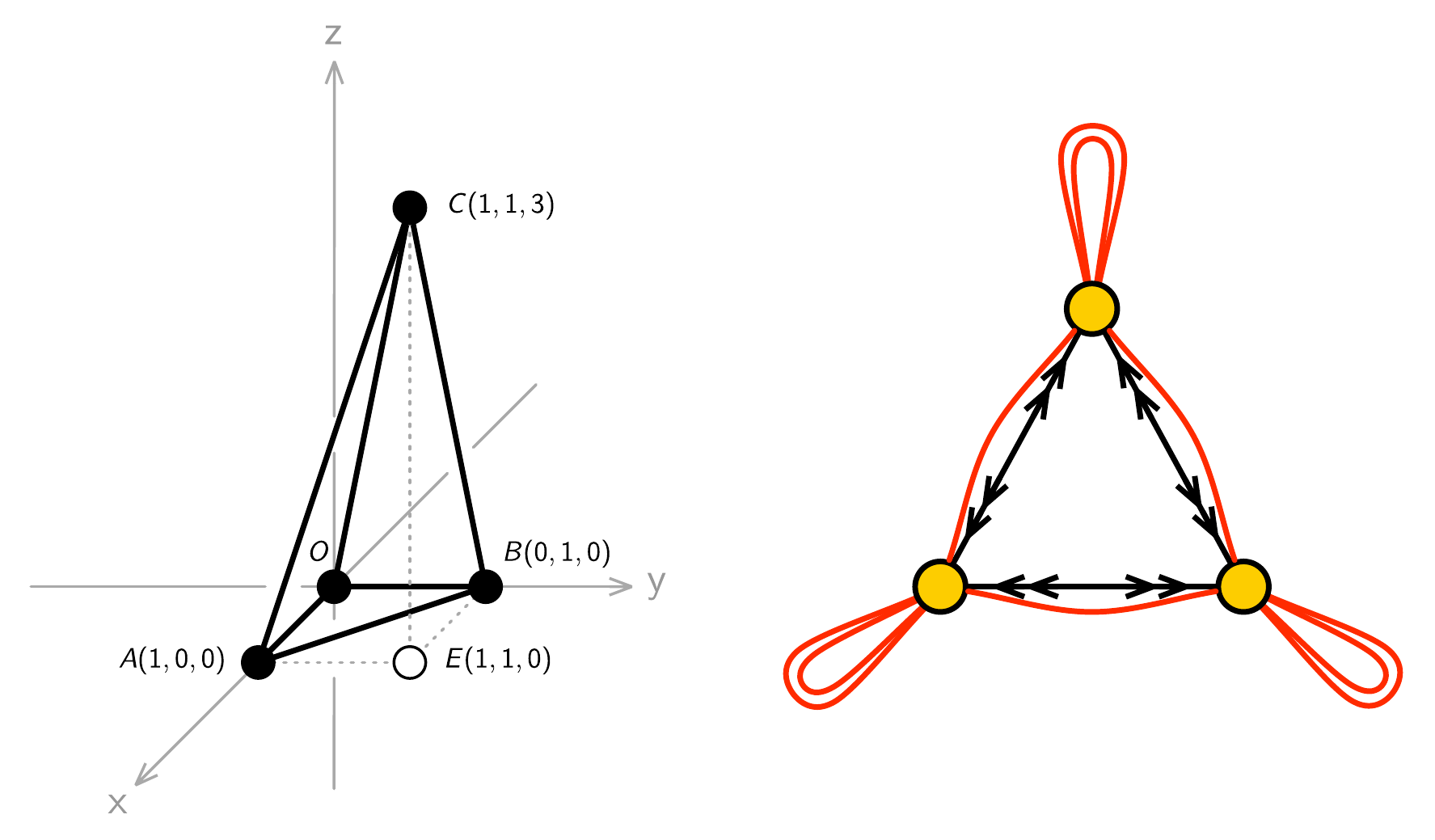} 
\caption{
Toric and quiver diagrams of the $\mathbb{C}^4/\mathbb{Z}_3(1,1,2,2)$ model. Point $E(1,1,0)$ is not part of the toric diagram of $\mathbb{C}^4/\mathbb{Z}_3(1,1,2,2)$, but plays a role as an external reference point for the geometric formula of the elliptic genus.
\label{c4z3-1122-toric}}
 \end{center}
 \end{figure}

\subsubsection*{Geometric formula (with subtraction)} 

The toric diagram for this orbifold, which is shown in \fref{c4z3-1122-toric}, can be triangulated with subtraction. To do so, it is convenient to introduce an external reference point, E$(1,1,0)$. The triangulation reads
\begin{align}
\mathcal{T}(\mbox{OABC}) = \mathcal{T}\mbox{(OAEC)} + \mathcal{T}\mbox{(OEBC)}- \mathcal{T}\mbox{(AEBC)} \,.
\label{c4z3-triangulation}
\end{align}
Each term on the right-hand side is a union of three minimal tetrahedra. Up to an $SL(3,\mathbb{Z})$ basis change, each term is isomorphic to the toric diagram of $\mathbb{C}^4/\mathbb{Z}_3(0,0,1,2)$. The pre-index $\mathcal{J}(t)$ computed from the triangulation \eqref{c4z3-triangulation} is a sum of nine terms. 
With a change of fugacity, 
\begin{align}
t_1 = s_1 s_4 = x \,, \quad
t_2 = s_2 s_4 = y \,, \quad
t_3 = (s_4)^3 = (x y z)^{3/2}  \,, \quad
t_4 = 1\,,
\label{c4z3-t-to-xyz}
\end{align}
we obtain a lengthy expression for the index $\mathcal{I}$ consisting of nine terms. 

Consider, instead, naively applying the geometric formula pretending as if the three terms on the right-hand side of \eqref{c4z3-triangulation} were minimal tetrahedra. 
The pre-index from this (unjustified) process is 
\begin{align}
\begin{split}
\widetilde{\mathcal{J}}(t) = -\frac{i \eta(q)^3}{\theta_1 (q,t_3) } & \left[ \frac{ \theta_1(q,t_3 t_4^{3/2}/t_1^3)  \theta_1(q,t_4^{3/2}/t_2^3)  \theta_1(q,t_2^3 t_4^{3/2}  /t_1^3 t_3)}{ \theta_1(q,t_1^3/t_2^3)  \theta_1(q,t_2^3/t_3)  \theta_1(q,t_4^3/t_1^3)} \right. 
\\
& \left. \quad + \frac{ \theta_1(q,t_3 t_4^{3/2}/t_2^3)  \theta_1(q,t_4^{3/2}/t_1^3)  \theta_1(q,t_1^3 t_4^{3/2}/t_2^3 t_3)}{ \theta_1(q,t_2^3/t_1^3)  \theta_1(q,t_1^3/t_3)  \theta_1(q,t_4^3/t_2^3)} \right. 
\\
& \left. \quad -\frac{ \theta_1(q,t_3 t_4^{3/2}/t_1^3) \theta_1(q,t_3 t_4^{3/2}/t_2^3) \theta_1(q,t_1^3 t_2^3/t_4^{9/2})}{ \theta_1(q,t_4^3/t_1^3) \theta_1(q,t_4^3/t_2^3) \theta_1(q,t_1^3 t_2^3/t_3 t_4^3)} \right] \,.
\label{c4z3-1122-pre}
\end{split}
\end{align}
Upon the change of the variable \eqref{c4z3-t-to-xyz}, the pre-index \eqref{c4z3-1122-pre} gives  
\begin{align}
\begin{split}
\widetilde{\mathcal{I}} = & \frac{-i \eta(q)^3}{\theta_1 (q,s_4^3)} 
\left[  
\frac{\theta_1(q, y^3) \theta_1(q, s_1^3) \theta_1(q, s_2^3/x^3) }{\theta_1(q, x^3) \theta_1(q, y^3/x^3) \theta_1(q,s_2^3)} 
+ \frac{\theta_1(q, x^3) \theta_1(q, s_1^3/y^3) \theta_1(q, s_2^3)}{\theta_1(q, y^3) \theta_1(q, x^3/y^3)  \theta_1(q, s_1^3)}
\right. 
\\
& \left.  
\qquad \qquad \qquad \qquad \qquad
+\frac{\theta_1(q, x^3 y^3) \theta_1(q, s_1^3) \theta_1(q, s_2^3)}{\theta_1(q, x^3) \theta_1(q, y^3) \theta_1(q, s_3^3)} 
\right] \,.
\end{split}
\label{c4z3-1122-geo-2}
\end{align}
Remarkably, although the intermediate step \eqref{c4z3-1122-pre} is not justified, the final result \eqref{c4z3-1122-geo-2} turns out to agree perfectly with the nine term expression we mentioned below \eqref{c4z3-triangulation}. 

While we do not fully understand why \eqref{c4z3-1122-geo-2} gives the correct result, we suspect that it has something to do with orbifold singularities. 
Despite the appearance, the three terms in \eqref{c4z3-1122-pre} do {\em not} match the three non-minimal tetrahedra in \eqref{c4z3-triangulation}. Presumably, 
the discrepancy is related to twisted sectors. Orbifold singularities away from the origin of $\mathbb{C}^4$ are encoded by the vertices along the edges of the toric diagram. In \fref{c4z3-1122-toric}, the vertices lie on the interval $\overline{CE}$. 
Now, the subtraction \eqref{c4z3-triangulation} is done in such a way that the 
resulting orbifold $\mathbb{C}^4/\mathbb{Z}_3(1,1,2,2)$ no longer contains orbifold singularity away from the origin. Hence, it is conceivable that the discrepancies associated with the twisted sectors have been cancelled out so as to produce the correct elliptic genus. 

\subsubsection*{Single term formula} Just as in \eqref{c4z2-1111-curious}, we find a remarkably simple single term expression for the index: 
\begin{align}
\mathcal{I} = 
 \frac{ -i \eta(q)^3 \theta_1(q,x^3) \theta_1(q,y^3) \theta_1(q,z^3)}{
\theta_{1}(q,s_1^3)
\theta_{1}(q,s_2^3)
\theta_{1}(q,s_3^2)
\theta_{1}(q,s_4^3)}
\,.
\label{c4z3-1122-curious}
\end{align}
We note that both $\mathbb{C}^4/\mathbb{Z}_2(1,1,1,1)$ and $\mathbb{C}^4/\mathbb{Z}_3(1,1,2,2)$ are free of singularities away from the origin. 

\subsubsection*{Orbifold CFT} 
The orbifold computation can be summarized as 
($\omega = e^{2\pi i/3}$)
\begin{align}
\begin{split}
\mathcal{I} = \; & \frac{ i \eta(q)^3 \theta_1 (q,x) \theta_1 (q,y)}{3} \times  
\\
& 
\sum_{k,l=0}^{2} 
\frac{\omega^k \, \theta[^{1/2-2k/3}_{1/2-2l/3}](q,z) }{\theta[^{1/2+k/3}_{1/2+l/3}](q,s_1) \theta[^{1/2+k/3}_{1/2+l/3}](q,s_2) \theta[^{1/2-k/3}_{1/2-l/3}](q,s_3)\theta[^{1/2-k/3}_{1/2-l/3}](q,s_4)  }  \,.
\label{c4z3-1122-orbifold}
\end{split}
\end{align}
It is straightforward to verify that \eqref{c4z3-1122-orbifold} agrees perfectly with \eqref{c4z3-1122-geo-2} and \eqref{c4z3-1122-curious}.

\subsubsection*{GLSM}

The quiver diagram for this orbifold model is given in \fref{c4z3-1122-toric}. When we decouple the overall $U(1)$, we choose a basis for the $u$ variables such that $u_3$ is decoupled, and $u_{13}$ and $u_{23}$ remain as independent variables. 

The anomaly polynomial for this theory is
\begin{align}
\begin{split}
\mathcal{A}(u) &= 3( u_{12}^2 + u_{23}^2 +u_{31}^2 ) 
\\
&= (u_{12} - u_{23})^2 + (u_{23} - u_{31})^2 + (u_{31} - u_{12})^2 
\equiv \tilde{u}_2^2 + \tilde{u}_3^2 + \tilde{u}_1^2 
\,.
\end{split}
\end{align}
To obtain the second expression, we used the fact that $u_{12} + u_{23} +u_{31} = 0$.

As discussed in \sref{sec:anomaly}, the anomaly cancelling factor for this orbifold is
\begin{align}
W_{(3)}(u_i;v) = \frac{\theta_1(q,v \tilde{u}_1)\theta_1(q,v \tilde{u}_2 )\theta_1(q,v \tilde{u}_3)+\theta_1(q,v/\tilde{u}_1)\theta_1(q,v/ \tilde{u}_2)\theta_1(q,v/\tilde{u}_3)}{2 \theta_1(q,v)^3} \,.
\label{boot-sol-3}
\end{align} 
In the multiplicative notation, the $\tilde{u}$ variables are defined as \begin{align}
\tilde{u}_1  = \frac{u_2 u_3}{u_1^2} \,, 
\quad 
\tilde{u}_2  = \frac{u_3 u_1}{u_2^2} \,,
\quad 
\tilde{u}_3  = \frac{u_1 u_2}{u_3^2} \,.
\end{align}
Since the overall $U(1)$ has decoupled, $W_{(3)}$ is a function of two independent variables, say, $u_{13}$ and $u_{23}$.  When the two fugacities satisfy $u_{13} = u = 1/u_{23}$, 
$W_{(3)}$ collapses to the simpler $W$ function we encountered earlier, 
\begin{align}
W_{(3)}(u_i;v)|_{u_{13} = u = 1/u_{23}} = W(u^3;v) \,.
\end{align}

Including $W_{(3)}$, 
the one-loop integrand is ($u_{ij}= u_i/u_j$) 
\begin{align}
Z_\text{1-loop}& =  \frac{-i (2 \pi)^2 \eta(q)^7  \theta_{1}(q,x)^3 \theta_{1}(q,y)^3 \theta_{1}(q,z u_{21}) \theta_{1}(q,z u_{32}) \theta_{1}(q,z u_{13}) W_{(3)}(u_i;v)}{\prod_{i=1}^2 \theta_1(q,s_i u_{21}) \theta_1(q,s_i u_{32}) \theta_1(q,s_i u_{13}) \prod_{i=3}^4 \theta_1(q,s_i u_{12}) \theta_1(q,s_i u_{23}) \theta_1(q,s_i u_{31}) }
\,.
\label{integrandc4z3}
\end{align}

\begin{figure}[ht!!]
\begin{center}
\resizebox{0.5\hsize}{!}{
\includegraphics[trim=0cm 0cm 0cm 0cm,totalheight=5 cm]{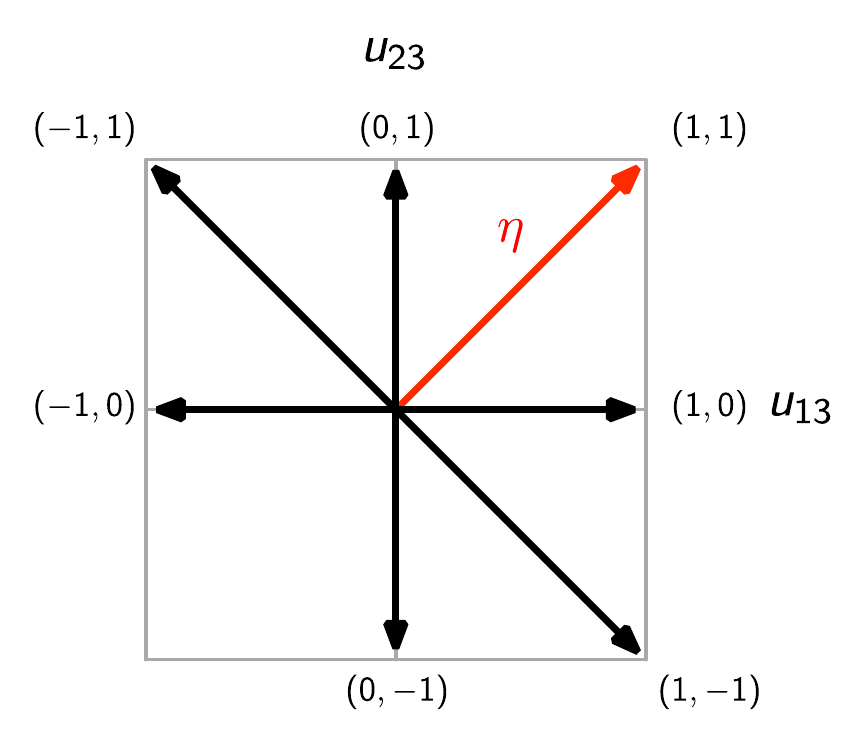}
}  
\caption{
Charge vectors of the $\mathbb{C}^4/\mathbb{Z}_3(1,1,2,2)$ theory 
in a basis where $u_{13}$ and $u_{23}$ are taken to be independent. 
For $\eta=(1,1)$, three cones participate in the JK residue calculus. Each cone contributes up to four residues, some of which may vanish. 
\label{qcovectordc4z3}}
 \end{center}
 \end{figure}
There are three pairs of charge vectors for the chiral multiplets. They split the charge plane into six regions as depicted in \fref{qcovectordc4z3}. Symmetries of the quiver diagram guarantees that all six regions are equivalent. 

In the notation of \fref{qcovectordc4z3}, we can paraphrase the general prescription of \sref{sec:JK-integral} as follows. Any choice of $\eta$ is equally good, and we choose $\eta=(1,1)$. Then we look for pairs of charge vectors, which we call ``cones", whose {\em positive} span contain $\eta$. For $\eta= (1,1)$, we find three cones, 
\begin{align}
A:\; (1,0)\&(0,1) \,,
\quad 
B:\; (1,0)\&(-1,1) \,,
\quad
C:\; (1,-1)\&(0,1) \,.
\label{c4z3-jk-pairs}
\end{align}
The cones determine the poles from which we take the residues. 
For example, cone $A$ picks up residues at $u_{12}s_i =1$ $(i=1,2)$ and $u_{23}s_j =1$ $(j=3,4)$. Cones $B$ and $C$ pick up residues in a similar way. 
In the end, the elliptic genus can be written as
\begin{align}
\mathcal{I} = \mathcal{R}_A + \mathcal{R}_B + \mathcal{R}_C \,,
\label{c4z3-jk-sum}
\end{align}
where $\mathcal{R}_{A,B,C}$ denote the partial sum of residues coming from the cones in \eqref{c4z3-jk-pairs}. 
Since every charge vector appears exactly twice in the denominator in \eqref{integrandc4z3}, each partial sum in \eqref{c4z3-jk-sum} can contribute up to four residues. 

It turns out that $\mathcal{R}_A$ contributes four terms, 
\begin{align}
\begin{split}
\mathcal{R}_A = & \frac{i \eta(q)^3 \theta_1(x) \theta_1(y)}{\theta_1(x/y) \theta_1(xy) } 
\left[ 
 \frac{\theta_1(x) W_{(3)}(s_1, s_3) }{\theta_1(s_1^3y) \theta_1( s_3^3y)}
 + \frac{\theta_1(y) W_{(3)}(s_1,s_4) }{ \theta_1(s_1^3/x) \theta_1(s_4^3/x)} 
\right. 
\\
& \left. 
\qquad \qquad \qquad \qquad
- \frac{\theta_1(y) W_{(3)}( s_2,s_3 ) }{ \theta_1(s_2^3 x)\theta_1(s_3^3 x) }  
- \frac{\theta_1(x) W_{(3)}(s_2,s_4 ) }{\theta_1(s_2^3/y) \theta_1(s_4^3/y)}
 \right] \,,
\label{c4z3-jk-A}
\end{split}
\end{align}
whereas $\mathcal{R}_B$ and $\mathcal{R}_C$ contribute two non-vanishing terms each, 
\begin{align}
\begin{split}
\mathcal{R}_B &= 
- \frac{i \eta(q)^3 \theta_1(x) \theta_1(y)}{\theta_1(x/y)} \left[  
 \frac{\theta_1(s_1/z) W_{(3)}(s_1,s_1^2)}{\theta_1(s_1^3/x)  \theta_1(s_1^3 y) \theta_1(s_1^3)}  
- \frac{ \theta_1( s_2/z) W_{(3)}(s_2,s_2^2)}{ \theta_1(s_2^3/y)\theta_1(s_2^3 x) \theta_1(s_2^3)}  \right] 
\,,
\\
\mathcal{R}_C &= -
\frac{i \eta(q)^3 \theta_1(x) \theta_1(y)}{\theta_1(xy)}
\left[  \frac{\theta_1(s_3z) W_{(3)}(s_3^2,s_3) }{\theta_1(s_3^3x)\theta_1(s_3^3y)  \theta_1(s_3^3)} 
- \frac{\theta_1(s_4z) W_{(3)}(s_4^2,s_4) }{ \theta_1(s_4^3/x)\theta_1(s_4^3/y) \theta_1(s_4^3)} \right]
\,.
\end{split}
\label{c4z3-jk-BC}
\end{align}
In \eqref{c4z3-jk-A} and \eqref{c4z3-jk-BC}, we suppressed the $q$-dependence of $\theta_1$ to simplify the expressions. We also wrote $W_{(3)}(u_1,u_2)$ in place of $W_{(3)}(u_1, u_2,u_3=1)$.
Remarkably, the sum \eqref{c4z3-jk-sum} of all residues  is independent of $v$ and equal to \eqref{c4z3-1122-geo-2}, \eqref{c4z3-1122-curious} and \eqref{c4z3-1122-orbifold}.

\subsection{$\mathbb{C}^{4}/\mathbb{Z}_{2}\times \mathbb{Z}_{2} (0,0,1,1)(1,1,0,0)$} \label{Z2Z2}

\begin{figure}[ht!!]
\begin{center}
\includegraphics[height=6cm]{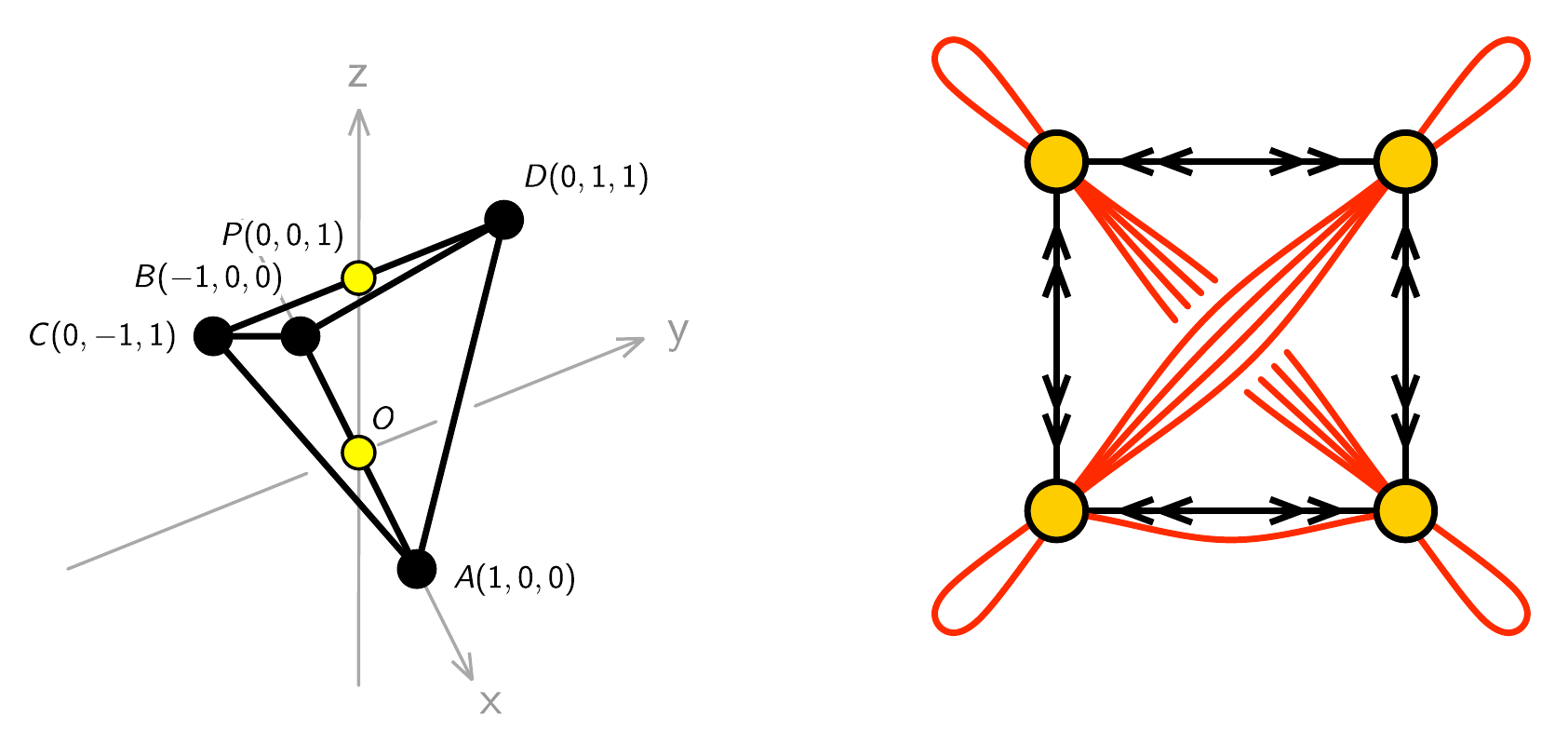} 
\caption{
Toric and quiver diagrams of the $\mathbb{C}^4/\mathbb{Z}_2\times \mathbb{Z}_2(0,0,1,1)(1,1,0,0)$ model.
\label{c4z2z2-toric}}
 \end{center}
 \end{figure}

\subsubsection*{Geometric formula} 
If we choose not to use any subtraction, 
there is a unique way to triangulate the toric diagram. 
In the notation of \fref{c4z2z2-toric}, the triangulation reads
\begin{align}
\mathcal{T}(\mbox{ABCD}) = \triangle\mbox{(OPAD)} + \triangle\mbox{(OBPD)} +\triangle\mbox{(OCAP)} + \triangle\mbox{(OBCP)} \,.
\end{align}
The pre-index (with $t_4=1$) is 
\begin{align}
\mathcal{J}(t) = \mbox{$xy$-parity}\left[ -\frac{ i\eta(q)^3 \theta_1 (q,1/t_3) \theta_1 (q,t_2/t_1t_3) \theta_1 (q,1/t_1t_2)}{\theta_1 (q,t_1) \theta_1 (q,t_2) \theta_1 (q,t_3/t_2) \theta_1 (q,1/t_1t_3)} 
\right]\,,
\end{align}
where we defined the ``$xy$-parity sum" as 
\begin{align}
\mbox{$xy$-parity}[h(t_1,t_2)] = h(t_1,t_2) + h(t_1^{-1},t_2) +h(t_1,t_2^{-1}) + h(t_1^{-1},t_2^{-1}) \,.
\label{xy-parity-sum}
\end{align}
Upon substitution,  
\begin{align}
t_1=s_4/s_3 = xy\,, \quad 
t_2= s_1/s_2 = x/y\,, \quad
t_3= s_1 s_2 = 1/z\,, 
\end{align}
we obtain
\begin{align}
\begin{split}
\mathcal{I} = - \frac{i \eta(q)^3 \theta_1(q,z)}{\theta_1(q,xy)\theta_1(q,x/y)}
\biggl[ 
&\frac{\theta_1(q,x^2)\theta_1(q,y^2/z)}{\theta_1(q,y/xz)\theta_1(q,z/xy)}  
- \frac{\theta_1(q,x^2 z)\theta_1(q,y^2)}{\theta_1(q,y/xz)\theta_1(q,xyz)} 
\\
&- \frac{\theta_1(q,x^2/z)\theta_1(q,y^2)}{\theta_1(q,x/yz)\theta_1(q,z/xy)} 
+ \frac{\theta_1(q,x^2)\theta_1(q,y^2z)}{\theta_1(q,x/yz)\theta_1(q,xyz)} 
\biggr]\,.
\end{split}
\label{c4z2z2-geo}
\end{align}
Some symmetries of the index are manifest from the formula:
\begin{align}
\mathcal{I}(x,y,z) = - \mathcal{I}(1/x,y,z) = -\mathcal{I}(x,1/y,z) = -\mathcal{I}(x,y,1/z) = 
+\mathcal{I}(y,x,z)\,.
\end{align}

\subsubsection*{Orbifold CFT} 
The orbifold CFT method gives the index in the form 
\begin{align}
\mathcal{I} = - \frac{i\eta(q)^3 \theta_1(q,z)}{4} \sum_{a,b=1}^4 
 \frac{ c_{a,b} \, \theta_{a*b}(q,x) \theta_{a*b}(q,y) }{
\theta_{a}(q,s_1)
\theta_{a}(q,s_2)
\theta_{b}(q,s_3)
\theta_{b}(q,s_4)}
\,.
\label{c4z2z2-orbi}
\end{align}
The phase factors $c_{a,b}$ and the labels $a*b$ for the theta functions in the numerator  are
\begin{align}
c_{a,b} = 
\begin{pmatrix}
+ & - & - & - \\
+ & - & + & + \\
+ & - & + & + \\
+ & - & - & - \\
\end{pmatrix} \,,
\quad 
a*b = 
\begin{pmatrix}
\ 1 \ & \ 2 \ & \ 3 \ & \ 4 \ \\
2 & 1 & 4 & 3 \\
3 & 4 & 1 & 2 \\
4 & 3 & 2 & 1 \\
\end{pmatrix} \,.
\label{c4z2z2-orbi-2}
\end{align}
We can regard the $\mathbb{C}^{4}/\mathbb{Z}_{2}\times \mathbb{Z}_{2}$ model 
as a $\mathbb{Z}_{2}$ orbifold of 
either the $\mathbb{C}^{4}/\mathbb{Z}_{2}(0,0,1,1)$ model or the $\mathbb{C}^{4}/\mathbb{Z}_{2}(1,1,1,1)$ model. In \eqref{c4z2z2-orbi-2}, 
the first rows of the matrices correspond to the $\mathbb{C}^{4}/\mathbb{Z}_{2}(0,0,1,1)$ model, 
whereas the diagonal elements correspond to the $\mathbb{C}^{4}/\mathbb{Z}_{2}(1,1,1,1)$ model.
See \eqref{indexc4z2-0011-orbi} and \eqref{c4z2-1111-orbi}.

It is straightforward (but tedious) to prove that \eqref{c4z2z2-geo} and \eqref{c4z2z2-orbi} agree perfectly. 

\subsubsection*{GLSM} 

The anomaly polynomial is
\begin{align}
\mathcal{A}(u) = 4(u_1-u_2+u_3-u_4)^2  \,.
\end{align}
The anomaly cancelling factor is given by
\begin{align}
V(u;v_1,v_2) = W(u;v_1) W(u;v_2) \,,
\quad 
W(u;v) = \frac{\theta_1(q,vu)\theta_1(q,v/u)}{\theta_1(q,u)^2}\,.
\label{boot-sol-z2z2}
\end{align}
The parameters $v_1$ and $v_2$ in \eqref{boot-sol-z2z2} are unconstrained except that $v_{1,2} \neq q^n$ $(n\in \mathbb{Z})$.

The one-loop integrand with the anomaly cancelling factor $V(u_i;v_j)$ is ($u_{ij} = u_i / u_j$)
\begin{align}
\begin{split}
Z_\text{1-loop} & = -i (2 \pi)^3 \eta(q)^{10} \ \theta_{1}(q,z)^4 \frac{\theta_{1}(q,x u_{24}) \theta_{1}(q,x u_{42}) \theta_{1}(q, y u_{24}) \theta_1(q,yu_{42})}{\prod_{i=1}^2 \theta_1(q,s_i u_{12}) \theta_1(q,s_i u_{21}) \theta_1(q,s_i u_{34}) \theta_1(q,s_i u_{43}) } \, \\
& \frac{\theta_1(q,x u_{13}) \theta_1(q,x u_{31}) \theta_1(q,y u_{13}) \theta_1(q,y u_{31})}{\prod_{i=3}^4 \theta_1(q,s_i u_{23}) \theta_1(q,s_i u_{32}) \theta_1(q,s_i u_{41}) \theta_1(q,s_i u_{14})} V(u_1u_3/u_2u_4;v_1,v_2) \,.
\end{split}
\label{integrandc4z2z2}
\end{align}

The charge plane contains eight vectors $\pm(u_{12}, u_{23}, u_{34}, u_{41})$ with the obvious constraint $u_{12}+u_{23}+u_{34}+u_{41}=0$. 
By taking any three linearly independent vectors out of the eight, we can make 32 cones. Some of the cones overlap with each other. 
\begin{figure}[ht!!]
\begin{center}
\resizebox{0.5\hsize}{!}{
\includegraphics[trim=0cm 2cm 0cm 0cm,totalheight=5 cm]{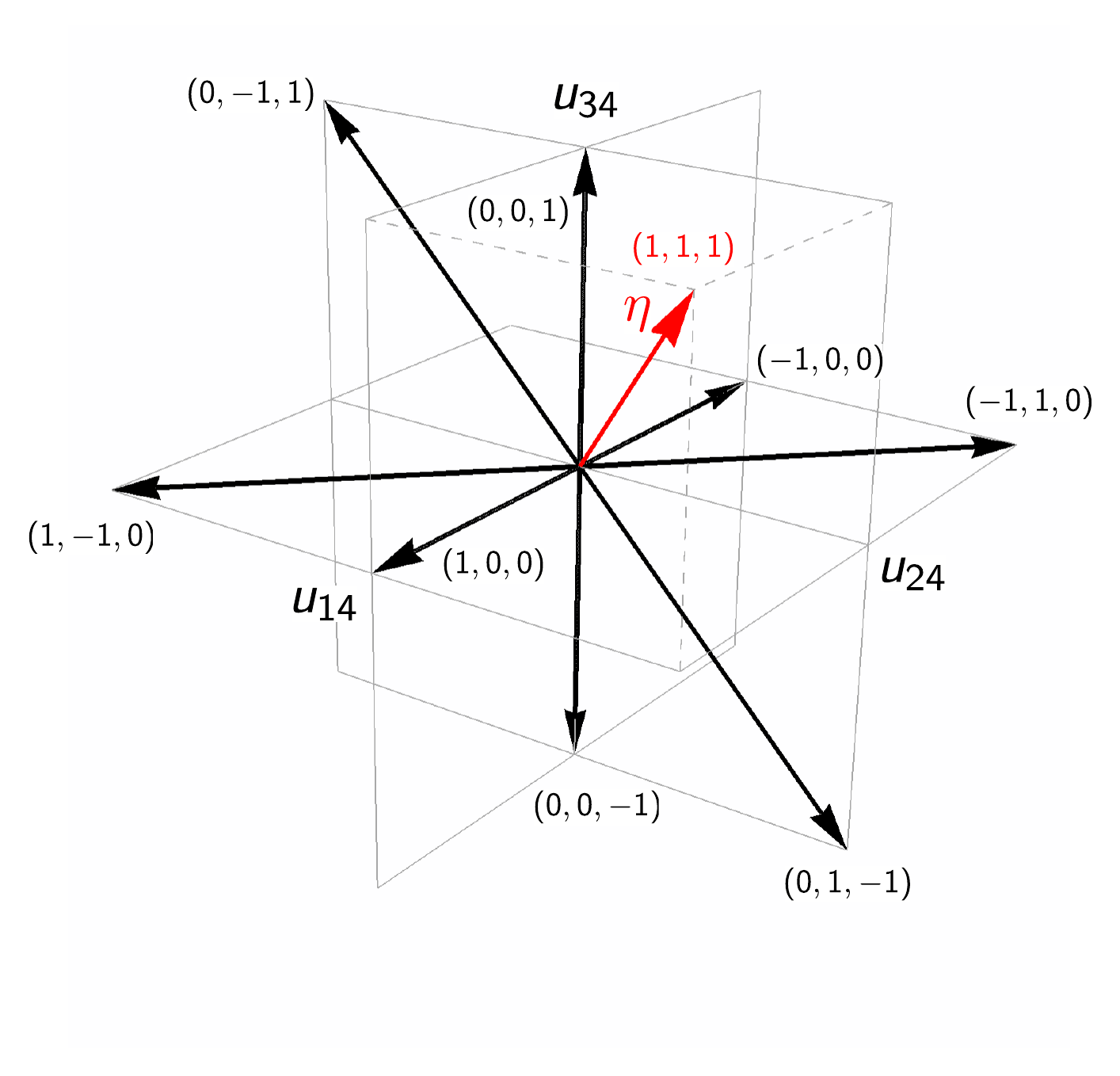}
}  
\caption{
Charge vectors of the $\mathbb{C}^4/\mathbb{Z}_2 \times \mathbb{Z}_2$ theory with $\eta=(1,1,1)$.
\label{qcovectordc4z2z2}}
 \end{center}
 \end{figure}

We will say that a choice of $\eta$ is {\it generic} if it does not lie on the boundary of any of the cones. For any generic choice, four cones contribute to the JK-calculus. 
Up to symmetries of the gauge theory, there are two inequivalent choices of $\eta$. Let us call them branch A and branch B. 
On both branches, two out of the four cones give vanishing contributions to the JK-integral. So the effective number of cones is two. 

On branch A, all non-vanishing residues come from degenerate poles with four planes intersecting. The two cones, which look different a priori, both land on the same degenerate poles. 
The end result is that there are four non-vanishing residues, which match the geometric formula \eqref{c4z2z2-geo} term by term. 
At the four poles, the $V$ gives a trivial contribution, $V(u=1;v_1,v_2)=1$. 

On branch B, there are 16 non-vanishing residues, all with non-trivial contributions from the $V$ function. 
The 16 residues can be divided into four groups of four terms. 
Each group matches a term in the geometric formula \eqref{c4z2z2-geo}.
A key step in proving the equality can be written as 
\begin{align}
\begin{split}
\frac{\theta_1(z)^2}{\theta_1(y/x)}\left[\frac{\theta_1(x)V(y)}{\theta_1(yz)\theta_1(y/z)\theta_1(y)}-\frac{\theta_1(y)V(x)}{\theta_1(xz)\theta_1(x/z)\theta_1(x)} \right] &
\\
+\frac{\theta_1(z)}{\theta_1(z^2)}V(z)\left[\frac{\theta_1(xyz)}{\theta_1(xz)\theta_1(yz)}+\frac{\theta_1(xy/z)}{\theta_1(x/z)\theta_1(y/z)}\right] &= \frac{\theta_1(xy)}{\theta_1(x)\theta_1(y)}\,.
\end{split}
\label{z2z2-boot-eq}
\end{align}
We can verify this identity by plugging in the proposed form of $V$ \eqref{boot-sol-z2z2} and using the simpler identity \eqref{4theta-identity} repeatedly.

\subsection{$\mathbb{C}^{4}/\mathbb{Z}_{4}(1,1,1,1)$}

\begin{figure}[ht!!]
\begin{center} 
\includegraphics[height=5cm]{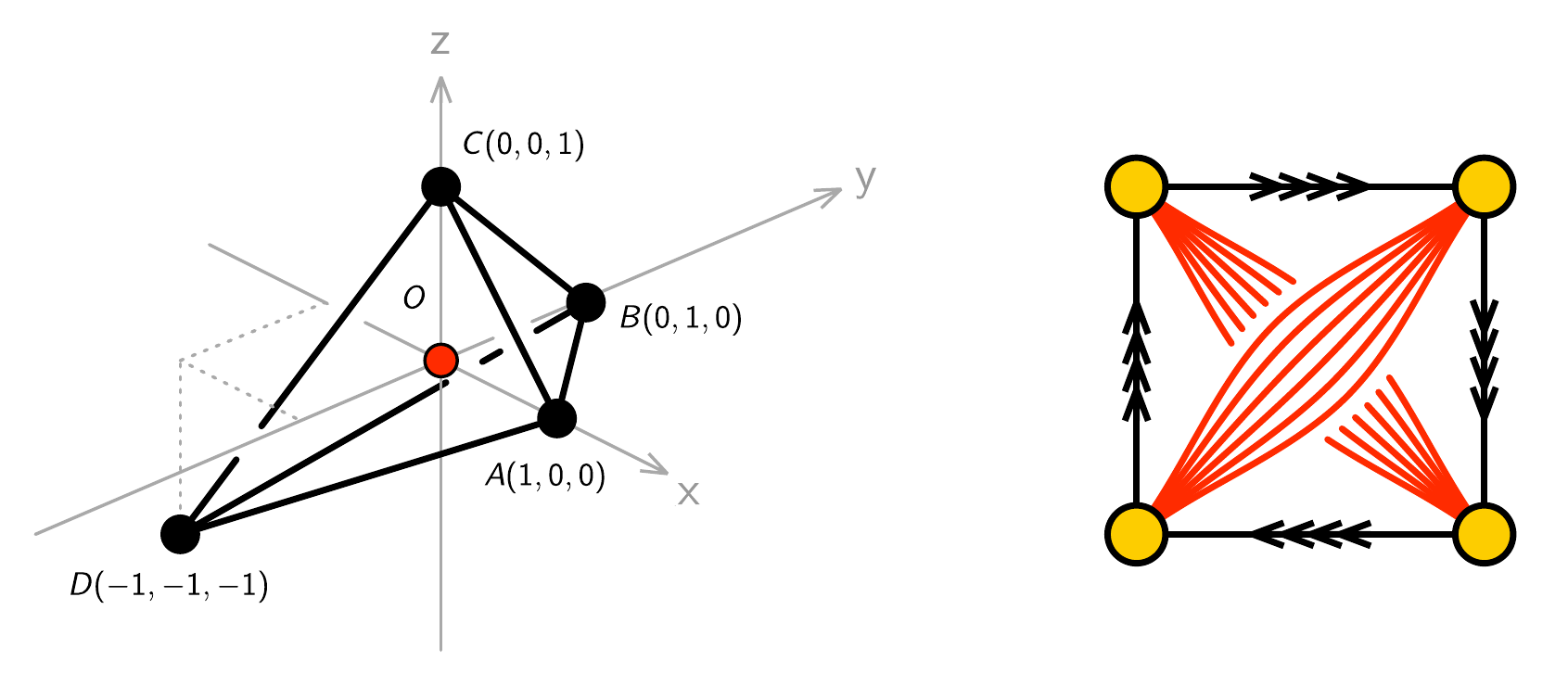} 
\caption{
Toric and quiver diagrams of the $\mathbb{C}^4/\mathbb{Z}_4(1,1,1,1)$ model.
\label{c4z4-1111-toric}}
 \end{center}
 \end{figure}

\subsubsection*{Geometric formula}
The toric diagram has four external vertices at $(1,0,0)$, $(0,1,0)$, $(0,0,1)$, $(-1,-1,-1)$, 
and an internal vertex at $(0,0,0)$. It can be triangulated in the usual way, 
\begin{align}
\mathcal{T}(\mbox{DABC}) = \triangle\mbox{(OABC)} + \triangle\mbox{(DABO)} +\triangle\mbox{(DBCO)} + \triangle\mbox{(DCAO)} \,.
\end{align}
so the geometric formula works well without using the subtraction method. 
The pre-index (with $t_4=1$ inserted) is 
\begin{align}
\begin{split}
\mathcal{J}(t)
= & - \frac{i  \eta (q)^3 \theta_{1}(q,1/t_1 t_2 )\theta_1(q,1/t_2 t_3)\theta_1(q,1/t_3t_1)}{\theta_1(q,t_1) \theta_{1}(q, t_2) \theta_1(q,t_3)\theta_1(q,1/t_1t_2t_3)} 
\\
& - \mbox{cyclic}\left[ \frac{i  \eta (q)^3 \theta_{1}(q,t_1^2/ t_2 t_3)\theta_1(q,t_1^2/t_2)\theta_1(q,t_1^2/t_3)}{\theta_1(q,1/t_1) \theta_{1}(q, t_2/t_1) \theta_1(q,t_3/t_1)\theta_1(q,t_1^3/t_2t_3)}\right]\,,
\end{split}
\label{pre-indexc4z4-1111}
\end{align}
where the cyclic sum was defined in \eqref{cyclic-sum}. 
The relation between $t_i$ and $(x,y,z)$ for this orbifold is 
\begin{align}
t_1 = s_1/s_4 = 1/yz \,, \quad
t_2 = s_2/s_4 = 1/zx \,, \quad
t_3 = s_3/s_4 = 1/xy \,.
\end{align}
The geometric index can be summarized as
\begin{align}
\mathcal{I} = 
\mbox{tetra}\left[\frac{i \eta(q)^3 \theta_1( q, x^2 yz)\theta_1( q, x y^2z)\theta_1( q, xyz^2)}{\theta_1( q, x y)\theta_1( q, yz)\theta_1( q, zx) \theta_1(q,x^2y^2z^2)}\right] \,, 
\label{c4z4-1111-geo}
\end{align}
where the tetrahedral sum was defined in \eqref{tetra-sum}.

\subsubsection*{Orbifold CFT and single term formula}

If we apply the orbifold CFT formula \eqref{orbi-formula} directly to this orbifold, we find a sum of $4^2=16$ terms. Since the shift parameters are multiples of $1/4$, the result is not particularly illuminating. 

We can take an alternative route to find something simpler. The key idea is that $\mathbb{C}^4/\mathbb{Z}_4(1,1,1,1)$ can be regarded as a $\mathbb{Z}_2$ orbifold of $\mathbb{C}^4/\mathbb{Z}_2(1,1,1,1)$. 
Recall that the index of the latter satisfies an interesting identity
(see \eqref{c4z2-1111-curious} and \eqref{c4z2-1111-orbi}), 
\begin{align}
\begin{split}
\mathcal{I}
&= 
- \frac{i}{2} \sum_{a=1}^4 
 \frac{ (-1)^{a+1} \eta(q)^3 \theta_1(q,x) \theta_1(q,y) \theta_1(q,z)}{
\theta_{a}(q,s_1)
\theta_{a}(q,s_2)
\theta_{a}(q,s_3)
\theta_{a}(q,s_4)} 
\\
& \qquad = 
 -\frac{ i \eta(q)^3 \theta_1(q,x^2) \theta_1(q,y^2) \theta_1(q,z^2)}{
\theta_{1}(q,s_1^2)
\theta_{1}(q,s_2^2)
\theta_{1}(q,s_3^2)
\theta_{1}(q,s_4^2)}
\,.
\end{split}
\end{align}
In words, we may say that $\mathbb{Z}_2(1,1,1,1)$ orbifolding resulted in squaring the fugacity variables. By applying the same orbifolding once again, we can deduce that 
\begin{align}
\begin{split}
\mathcal{I}
&= 
- \frac{i}{2} \sum_{a=1}^4 
 \frac{ (-1)^{a+1} \eta(q)^3 \theta_1(q,x^2) \theta_1(q,y^2) \theta_1(q,z^2)}{
\theta_{a}(q,s_1^2)
\theta_{a}(q,s_2^2)
\theta_{a}(q,s_3^2)
\theta_{a}(q,s_4^2)} 
\\
& \qquad = 
- \frac{ i \eta(q)^3 \theta_1(q,x^4) \theta_1(q,y^4) \theta_1(q,z^4)}{
\theta_{1}(q,s_1^4)
\theta_{1}(q,s_2^4)
\theta_{1}(q,s_3^4)
\theta_{1}(q,s_4^4)}
\,.
\end{split}
\label{c4z4-1111-curious}
\end{align}
To summarize, starting from the single term formula of the $\mathbb{C}^4/\mathbb{Z}_2(1,1,1,1)$ orbifold and applying further $\mathbb{Z}_2$ orbifolding, we derived a new single term formula \eqref{c4z4-1111-curious} for the $\mathbb{C}^4/\mathbb{Z}_4(1,1,1,1)$ orbifold. It is straightforward to prove that \eqref{c4z4-1111-curious} agrees perfectly with \eqref{c4z4-1111-geo}.

\subsubsection*{GLSM}

The anomaly polynomial is 
\begin{align}
\mathcal{A}(u) = 4(u_{12} + u_{34})^2 - 2 (u_{13})^2 - 2 (u_{24})^2 \,.
\end{align}
Since it has some negative terms, we cannot use the ansatz for the anomaly cancelling factor proposed in \sref{sec:anomaly}. 
Taking a leap of faith, we propose a new anomaly cancelling factor for this example, 
\begin{align}
U(u_i;v_1,v_2) = \frac{\theta_1(q,v_1 u_{12}u_{34})\theta_1(q,v_1/u_{12}u_{34})\theta_1(q,v_2 u_{12}u_{34})\theta_1(q,v_2/u_{12}u_{34})}{\theta_1(q,v_1 u_{13})\theta_1(q,v_1/u_{13})\theta_1(q,v_2 u_{24})\theta_1(q,v_2/u_{24})} \,.
\label{c4z4-ano-cancel}
\end{align}

\begin{figure}[ht!!]
\begin{center}
\resizebox{0.5\hsize}{!}{
\includegraphics[trim=0cm 2cm 0cm 0cm,totalheight=5 cm]{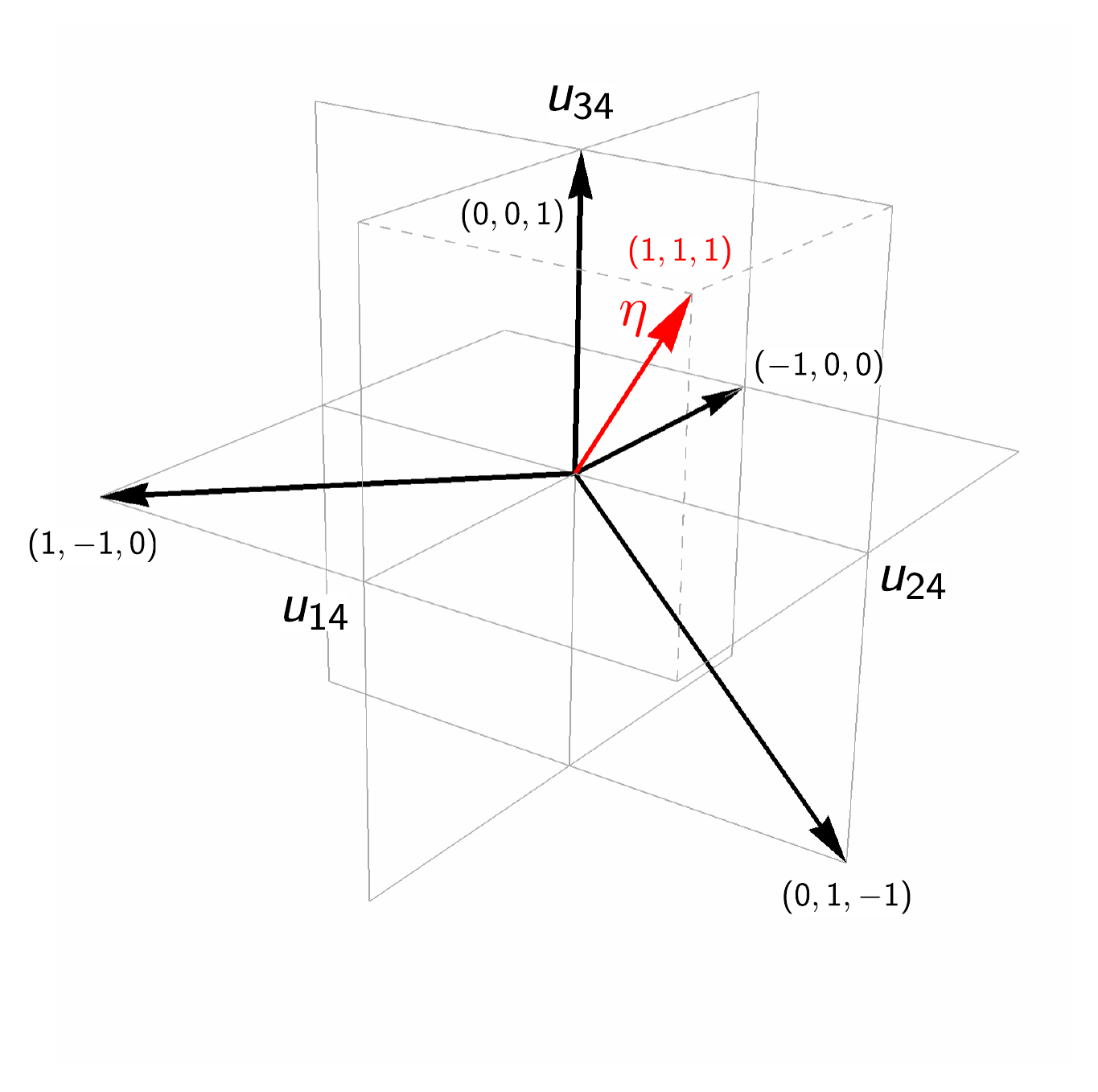}
}  
\caption{
Charge vectors of the $\mathbb{C}^4/\mathbb{Z}_4$ $(1,1,1,1)$ theory with $\eta=(1,1,1)$. \label{qcovectordc4z4}}
 \end{center}
 \end{figure}

The one-loop integrand is $(u_{ij} = u_i / u_j)$
\begin{align}
Z_\text{1-loop} &= \frac{-i (2\pi)^3 \eta(q)^{10} \prod_{a=1}^3 \theta_1(q,x_a u_{13}) \theta_1(q,x_a/u_{13}) \theta_1(q,x_a u_{24}) \theta_1(q,x_a/u_{24})}{\prod_{i=1}^4 \theta_1(q,s_i / u_{12}) \theta_1(q,s_i / u_{23}) \theta_1(q,s_i / u_{34}) \theta_1(q,s_i / u_{41}) } U(u_i)
\,.
\label{integrandc4z4}
\end{align}
The charge vectors from the chiral multiplets are depicted in \fref{qcovectordc4z4}. By choosing three out of four vectors, we can form four different cones. For any generic choice of $\eta$, only one cone contains $\eta$ inside it. The four cones are related by the tetrahedral symmetry of the toric diagram, so it is guaranteed that the result will be independent of the choice of $\eta$. 

Choosing $\eta = (1,1,1)$, the naive counting gives $4^3=64$ poles, among which 40 are normal and 24 are degenerate. A closer look reveals that the residues from the 24 degenerate poles all vanish, while 
all but four residues from the 40 normal poles also vanish. 
Interestingly, the anomaly cancelling factor \eqref{c4z4-ano-cancel} becomes 1 at the four surviving poles. The final result is a sum of four terms, which match the geometric formula \eqref{c4z4-1111-geo} term by term.

\section{Non-Orbifold Models}
\label{sec:non-orbi}

\subsection{$D_3$}\label{d3phase1index}

\begin{figure}[ht!!]
\begin{center}
\includegraphics[height=5cm]{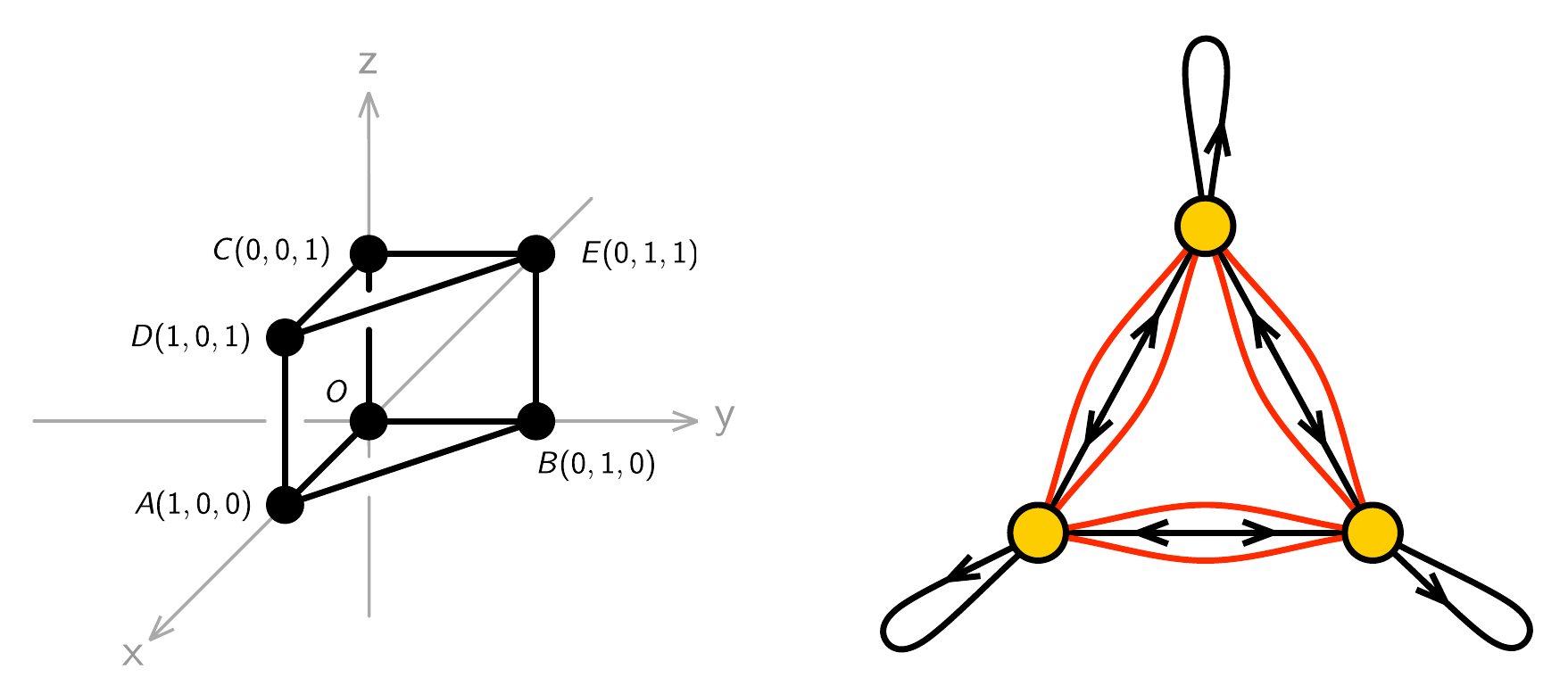}
\caption{
Toric and quiver diagrams for the $D_3$ model.
\label{D3-toric}}
 \end{center}
 \end{figure}

\subsubsection*{Geometric formula} 

The toric diagram of the $D_3$ theory is shown in \fref{D3-toric}. There are six different ways to triangulate it without subtraction. If we choose to triangulate it as 
\begin{align}
\mathcal{T}(D_3) = \triangle\mbox{(OABC)} + \triangle\mbox{(ABCE)} +\triangle\mbox{(ACDE)} \,,
\end{align}
the pre-index with $t_4 =1$ is
\begin{align}
\begin{split}
\mathcal{J}(t)
 = -i \eta(q)^3 &\left[
 \frac{\theta_1(q,t_1 t_2)  \theta_1(q,t_1 t_3)  \theta_1(q,t_2 t_3)}{ \theta_1(q,t_1)  \theta_1(q,t_2)  \theta_1(q,t_3)  \theta_1(q,t_1 t_2 t_3)} \right. 
 \\
& \quad + \frac{ \theta_1(q,t_2 ) \theta_1(q,t_3)  \theta_1(q,t_1^2 t_2 t_3)}{ \theta_1(q,t_1)  \theta_1(q,t_1 t_2)  \theta_1(q,t_1 t_3)  \theta_1(q,t_1 t_2 t_3)} 
\\
& \quad \left. + \frac{ \theta_1(q,t_1)  \theta_1(q,t_3/t_2)  \theta_1(q,t_1 t_2 t_3)}{ \theta_1(q,t_2)  \theta_1(q,t_3) \theta_1(q,t_1 t_2)  \theta_1(q,t_1 t_3) }\right]\,.
\end{split}
\label{pre-indexd3}
\end{align}
We set the other fugacities to be
\begin{align}
t_1 = x \,, \quad
t_2 = y \,, \quad
t_3 = z \,,
\end{align}
so that we have
\begin{align}
\begin{split}
\mathcal{I}
&=  -i \eta(q)^3 \left[ \frac{\theta_1(q,x y)  \theta_1(q,x z)  \theta_1(q,y z)}{ \theta_1(q,x)  \theta_1(q,y)  \theta_1(q,z)  \theta_1(q,x y z)} \right. \\
& \quad \left. + \frac{ \theta_1(q,y ) \theta_1(q,z)  \theta_1(q,x^2 y z)}{ \theta_1(q,x)  \theta_1(q,x y)  \theta_1(q,x z)  \theta_1(q,x y z)} + \frac{ \theta_1(q,x)  \theta_1(q,z/y)  \theta_1(q,x y z)}{ \theta_1(q,y)  \theta_1(q,z) \theta_1(q,x y)  \theta_1(q,x z) }\right]\,.
\end{split}
\label{d3-geo}
\end{align}
The symmetries of the toric diagram are reflected in the index. The $S_3$ subgroup is generated by the two elements:
\begin{align}
(x,y) \rightarrow (y,x)\,,
\quad 
(x,y) \rightarrow (y,1/xy)\,.
\label{D3-S3}
\end{align}
Another subgroup, $\mathbb{Z}_2$, acts as $z\rightarrow 1/z$. Not all symmetries are manifest in \eqref{d3-geo}.

The $q$-expansion of the index reads 
\begin{align}
\mathcal{I} = - \frac{1+x+ y+x^2 y + xy^2+x^2y^2 -6xy}{(1-x)(1-y)(1-xy)} - \frac{2(1-x)(1-y)(1-xy)}{xy} q + \mathcal{O}(q^2) \,.
\end{align}
As expected from the discussion in \sref{sec:geo}, the leading term contains codimension 1 poles, whereas the next term is a Laurent polynomial in $(x,y,z)$. 
Curiously, both terms are independent of $z$. 
In fact, we can prove that the index is completely $z$-independent. Then we may set $z$ to any convenient value 
to simplify the answer. For instance, setting $z=y$ and applying the symmetry \eqref{D3-S3}, we obtain
\begin{align}
\mathcal{I} =  \frac{-i \eta(q)^3}{\theta_1(q,xy)\theta_1(q,x/y)} \left[
 \frac{\theta_1(q,x)^2 \theta_1(q,y^2)}{\theta_1(q,y)^2} -\frac{\theta_1(q,y)^2 \theta_1(q,x^2)}{\theta_1(q,x)^2} \right] \,.
\end{align}

\subsubsection*{GLSM}
The quiver diagram of the $D_3$ theory is shown in \fref{D3-toric}.
The theory consists of three adjoint chirals, six bifundamental chirals and six bifundamental Fermis. Their charges under the global symmetry are summarized in \tref{tchd3}.

After the overall $U(1)$ decoupling, we have the integrand
\begin{align}
\begin{split}
Z_\text{1-loop} = & \  \frac{i (2 \pi)^2 \eta(q)^7}{\theta_{1}(q,x) \ \theta_{1}(q,y)\ \theta_{1}(q,1/xy)} 
\prod_{a=\pm 1} \left[ \frac{\theta_{1}(q. (u_2 z^{1/3})^a y \sqrt{x})}{\theta_{1}(q, (u_{2} z^{1/3})^a \sqrt{x})} \right.
\\
& \left. \times \frac{ \theta_{1}(q,  (u_1 z^{-1/3})^a / (x \sqrt{y}) ) \ \theta_{1}(q, (u_{12} z^{1/3})^a \sqrt{x /y})}{\theta_{1}(q, ( u_1 z^{-1/3})^a  \sqrt{y}) \ \theta_{1}(q, (u_{12} z^{1/3})^a / \sqrt{xy})} \right] \,.
\end{split}
\end{align}

 \begin{table}[ht!!]
 \centering
  \resizebox{1\hsize}{!}{
 \begin{tabular}{c|
ccc|ccc|ccc|ccc|ccc
 }
 field 
 & $X_{23}$ & $X_{31}$ & $X_{12}$ 
 & $X_{32}$ & $X_{13}$ & $X_{21}$ 
 & $Y_{11}$ & $Y_{22}$ & $Y_{33}$
 & $\Lambda_{23}$ & $\Lambda_{31}$ & $\Lambda_{12}$
 & $\Lambda_{32}$ & $\Lambda_{13}$  & $\Lambda_{21}$
\\
\hline
$U(1)_x$ & 
$+1/2$ & $0$ & $-1/2$ &
$+1/2$ & $0$ & $-1/2$ &
$+1$ & $0$ & $-1$ &
$+1/2$ & $-1$ & $+1/2$ &
$+1/2$ & $-1$ & $+1/2$ 
\\
$U(1)_y$ &
$0$ & $+1/2$ & $-1/2$ &
$0$ & $+1/2$ & $-1/2$ &
$0$ & $+1$ & $-1$ &
$+1$ & $-1/2$ & $-1/2$ &
$+1$ & $-1/2$ & $-1/2$ 
\\
$U(1)_z$ &
$+1/3$ & $+1/3$ & $+1/3$ &
$-1/3$ & $-1/3$ & $-1/3$ &
$0$ & $0$ & $0$ &
$+1/3$ & $+1/3$ & $+1/3$ &
$-1/3$ & $-1/3$ & $-1/3$ 
\\
\hline
 \end{tabular}
 }
 \caption{Global charges of matter fields in the $D_3$ theory.}
 \label{tchd3}
 \end{table}

\begin{figure}[ht!!]
\begin{center}
\resizebox{0.45\hsize}{!}{
\includegraphics[trim=0cm 0cm 0cm 0cm,totalheight=7 cm]{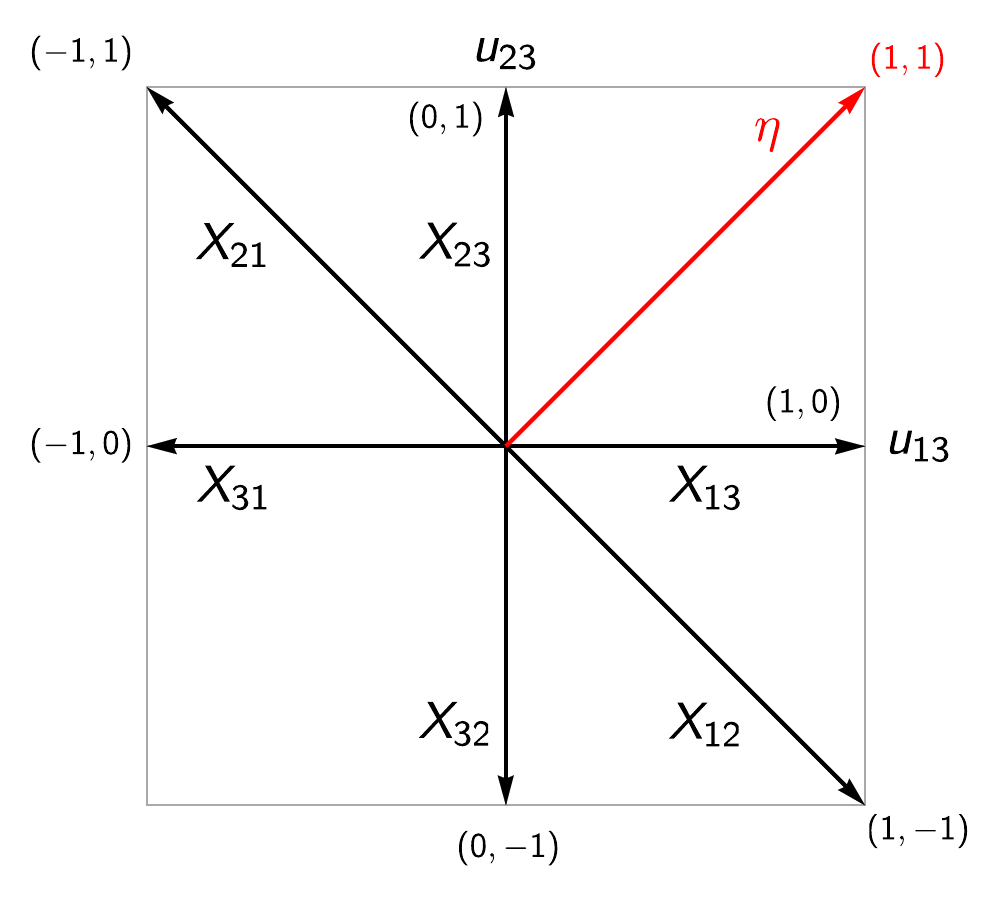}
}  
\caption{
Charge vectors of the $D_3$ theory. For $\eta=(1,1)$, residues from three cones contribute to the index: $\{X_{13}, X_{23} \}$, $\{X_{13}, X_{21}\}$ and $\{X_{12}, X_{23}\}$. 
\label{qcovectord3phase1}}
 \end{center}
 \end{figure}

\fref{qcovectord3phase1} shows the charge covectors of six singularity lines. When we take $\eta$ to be $(1,1)$, three out of twelve singularities contribute to the elliptic genus. 
By adding the residues from three cones $\{ X_{13}, X_{23}\}$, $\{X_{13}, X_{21}\}$ and $\{ X_{12} , X_{23} \}$, we obtain
\begin{align}
\begin{split}
\mathcal{I} &= i\eta(q)^3 \biggl[ \  
\frac{\theta_{1}(q,z)\ \theta_{1}(q,xy)\ \theta_1(q,xz/y)}{\theta_{1}(q,x)\ \theta_{1}(q,y) \ \theta_{1}(q,y/z) \ \theta_{1}(q,xz)} \\
&\qquad \qquad\quad  + 
\frac{\theta_{1}(q,x)\ \theta_{1}(q,y/z)\ \theta_1(q,xyz)}{\theta_{1}(q,y)\ \theta_{1}(q,z) \ \theta_{1}(q,xy) \ \theta_{1}(q,xz)} \\
&\qquad \qquad\quad  - 
\frac{\theta_{1}(q,y)\ \theta_{1}(q,xz)\ \theta_1(q,xy/z)}{\theta_{1}(q,x)\ \theta_{1}(q,z) \ \theta_{1}(q,xy) \ \theta_{1}(q,y/z)}
\ \biggr] \,.
\end{split}
\label{d3-JK}
\end{align}
It can be shown to be equal to \eqref{d3-geo}.

Note that the charge vectors in \fref{qcovectord3phase1} divide the plane into six regions. There is a one-to-one map between the six regions and six different ways to triangulate the toric diagram (without subtraction) in \fref{D3-toric}, such that the equality between the GLSM result and the geometric result become manifest without any change of variables or theta function identity.

\subsection{$H_4$ - Anomaly and Triality}\label{h4index}

\begin{figure}[ht!!]
\begin{center}
\resizebox{0.4\hsize}{!}{
\includegraphics[height=5cm]{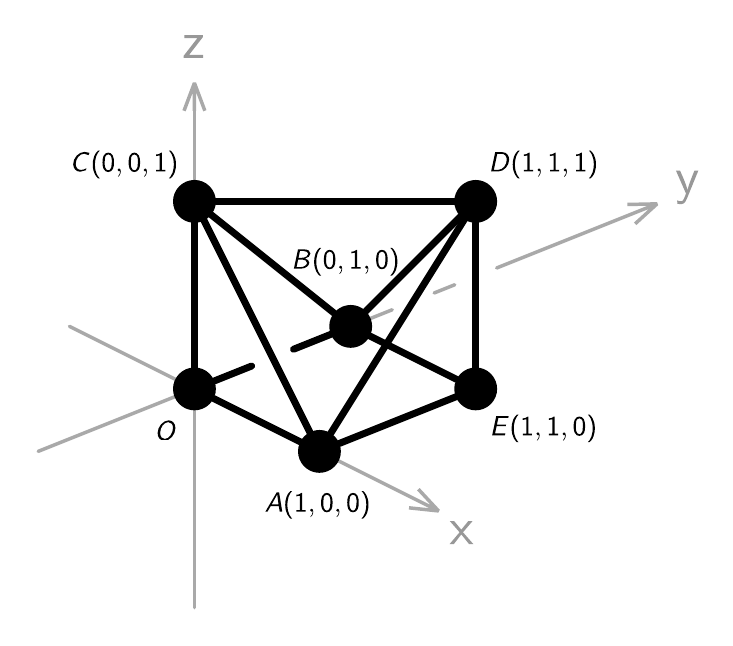}
}  
\caption{
Toric diagram for $H_4$.
\label{h4-toric}}
 \end{center}
 \end{figure}

Let us now consider an example that has not previously appeared in the literature. The toric diagram for this geometry is shown in \fref{h4-toric} and we refer to it as $H_4$.

\subsubsection*{Geometric Formula} 

In the notation of Figure \ref{h4-toric}, we use the triangulation
\begin{align}
\mathcal{T}(H_4) = \triangle\mbox{(OAEC)} +\triangle\mbox{(CAED)}+\triangle\mbox{(OEBC)}+\triangle\mbox{(CEBD)} \,.
\end{align}
With the change of variable, 
\begin{align}
t_1=x\,, 
\quad 
t_2=y\, 
\quad 
t_3=1/z\,, 
\quad
t_4=1 \,,
\end{align}
we obtain 
\begin{align} 
\label{IH4-geo}
\begin{split}
\mathcal{I} = \frac{-i \eta(q)^3}{\theta_1(x/y)} & \left[ \frac{\theta_1(x) \theta_1(x/yz) \theta_1(y/z)}{\theta_1(y) \theta_1(z) \theta_1(z/x)} - \frac{\theta_1(y) \theta_1(y/xz) \theta_1(x/z)}{\theta_1(x) \theta_1(z) \theta_1(z/y) } \right. \\
& \left. \quad + \frac{\theta_1(x) \theta_1(z) \theta_1(xz/y^2)}{\theta_1(y) \theta_1(y/z) \theta_1(xz/y)} 
-\frac{\theta_1(y) \theta_1(z) \theta_1(yz/x^2)}{\theta_1(x) \theta_1(x/z) \theta_1(yz/x)} \right] \,.
\end{split}
\end{align}
Its lowest terms in the $q$-expansion are
\begin{align}
\mathcal{I} = -\frac{(x-y)^2 z (1+z) }{(1-z)(xz-y)(yz-x)} + \frac{(x-y)^2 (1-z^2 )}{xyz} q+ \mathcal{O}(q^2) \,.
\end{align}

\subsubsection*{GLSM} 

\begin{figure}[ht!!]
\begin{center}
\includegraphics[height=7cm]{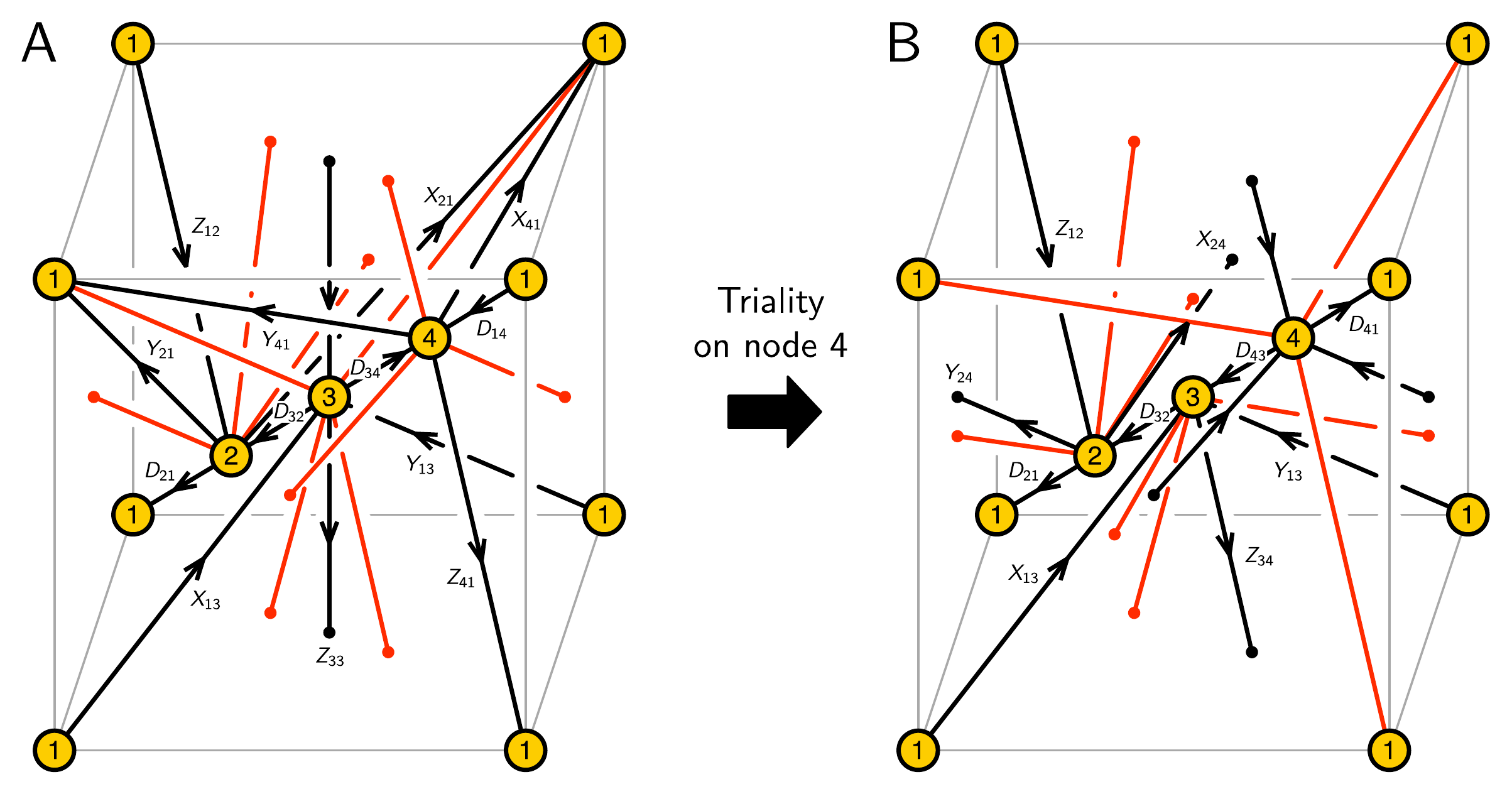}
\caption{
Periodic quivers for two toric phases of $H_4$.
\label{h4-quiver}}
 \end{center}
 \end{figure}

The $H_4$ model has several toric phases. \fref{h4-quiver} shows the periodic quivers for two of them and how they are related by triality. 
For a detailed discussion on how triality is realized on the periodic quiver, we refer the reader to \cite{Franco:2015tna,Franco:2015tya,Franco:2016nwv}.

\subsubsection*{Phase A} 

This phase has 13 chiral fields and 9 Fermi fields.
The plaquettes in the periodic quiver correspond to
\begin{align}
\begin{array}{rcccc}
 & & J & \ \ \ & E \\
\Lambda_{11}^x: \ & & D_{14} \cdot X_{41} - X_{13} \cdot D_{32} \cdot D_{21} 
& & Y_{13}\cdot D_{34} \cdot Z_{41}-Z_{12} \cdot Y_{21}  \\
\Lambda_{11}^y: \ & & D_{14}\cdot Y_{41} - Y_{13} \cdot D_{32} \cdot D_{21} 
& & Z_{12} \cdot X_{21}  - X_{13} \cdot D_{34} \cdot Z_{41}\\
\Lambda_{11}^z: \ & & D_{14} \cdot Z_{41} - Z_{12}\cdot D_{21} 
& & X_{13} \cdot D_{32} \cdot Y_{21} - Y_{13} \cdot D_{34} \cdot X_{41}\\
\Lambda_{13}^x: \ & & D_{32} \cdot X_{21}  - D_{34} \cdot X_{41}  
& & Y_{13} \cdot Z_{33} - D_{14} \cdot Z_{41} \cdot Y_{13} \\
\Lambda_{13}^y: \ & & D_{32} \cdot Y_{21}  - D_{34}\cdot Y_{41} 
& &  Z_{12} \cdot D_{21}\cdot X_{13} - X_{13} \cdot Z_{33}\\
\Lambda_{42}^x: \ & & X_{21} \cdot D_{14}  - D_{21} \cdot X_{13} \cdot D_{34} 
& & Z_{41} \cdot Y_{13} \cdot D_{32} -Y_{41} \cdot Z_{12} \\
\Lambda_{42}^y: \ & & Y_{21} \cdot D_{14}  - D_{21}\cdot Y_{13} \cdot D_{34} 
& & X_{41} \cdot Z_{12} - Z_{41}\cdot X_{13} \cdot D_{32}\\
\Lambda_{23}: \ & & Z_{33} \cdot D_{32}  - D_{32} \cdot D_{21} \cdot Z_{12} 
& & Y_{21}\cdot X_{13} -X_{21} \cdot Y_{13}  \\
\Lambda_{43}: \ & & Z_{33} \cdot D_{34}  - D_{34} \cdot Z_{41} \cdot D_{14} 
& & X_{41} \cdot Y_{13} - Y_{41}\cdot X_{13}
\end{array}
\end{align}
The global charges of the fields are given in Tables \ref{tchh4A-c} and \ref{tchh4A-f}.

 \begin{table}[ht!!]
 \centering
 \resizebox{1\hsize}{!}{
 \begin{tabular}{c|cccc|ccc|ccc|ccc}
 field & 
 $D_{21}$ & $D_{32}$ & $D_{34}$ & $D_{14}$ & 
 $X_{21}$ & $Y_{21}$ & $Z_{21}$ &
 $X_{13}$ & $Y_{13}$ & $Z_{33}$ & 
 $X_{41}$ & $Y_{41}$ & $Z_{41}$ 
 \\
 \hline
 $U(1)_x$ &
 $-1/4$ & $-1/4$ & $1/4$ & $-1/4$ &
 $-1/4$ & $3/4$ & $1/4$ &
 $-1/2$ & $1/2$ & $0$ &
 $-3/4$ & $1/4$ & $1/4$ 
 \\
 $U(1)_y$ &
 $-1/4$ & $-1/4$ & $1/4$ & $-1/4$ &
 $3/4$ & $-1/4$ &  $1/4$ & 
 $1/2$ & $-1/2$ & $0$ &
 $1/4$ &  $-3/4$ & $1/4$ 
 \\
 $U(1)_z$ &
 $-1/4$ & $-1/4$ & $1/4$ & $-1/4$ &
 $3/4$ & $3/4$ &  $-3/4$ & 
 $1/2$ & $1/2$ & $-1$ &
 $1/4$ &  $1/4$ & $-3/4$ 
 \\
 \hline
 \end{tabular}
 }
 \caption{Global charges of chiral fields in phase A of $H_4$.}
 \label{tchh4A-c}
 \end{table}

 \begin{table}[ht!!]
 \centering
\resizebox{0.65\hsize}{!}{
 \begin{tabular}{c|ccc|cc|cc|cc}
 field & 
 $\Lambda_{11}^{x}$ & $\Lambda_{11}^{y}$ & $\Lambda_{11}^{z}$ &
 $\Lambda_{13}^{x}$ & $\Lambda_{13}^{y}$ &
 $\Lambda_{42}^{x}$ & $\Lambda_{42}^{y}$ & 
 $\Lambda_{23}$ & $\Lambda_{43}$
 \\
 \hline
 $U(1)_x$ &
 $1$ & $0$ & $0$ & 
 $1/2$ & $-1/2$ &
 $1/2$ & $-1/2$ &
 $1/4$ & $-1/4$ 
 \\
 $U(1)_y$ &
 $0$ & $1$ & $0$ & 
 $-1/2$ & $1/2$ &
 $-1/2$ & $1/2$ &
 $1/4$ & $-1/4$ 
 \\
 $U(1)_z$ &
 $0$ & $0$ & $1$ & 
 $-1/2$ & $-1/2$ &
 $-1/2$ & $-1/2$ &
 $5/4$ & $3/4$ 
 \\
 \hline
 \end{tabular}
}
 \caption{Global charges of Fermi fields in phase A of $H_4$.}
 \label{tchh4A-f}
 \end{table}

The anomaly polynomial is
\begin{align}
\mathcal{A}(u) = 2(2u_1 -u_2 -u_4)^2 \,,
\end{align}
so we can use the ansatz \eqref{ano-ansatz} for the anomaly cancelling factor at $n=2$.  

\begin{figure}[ht!!]
\begin{center}
\resizebox{0.45\hsize}{!}{
\includegraphics[trim=0cm 0cm 1cm 0cm,totalheight=5 cm]{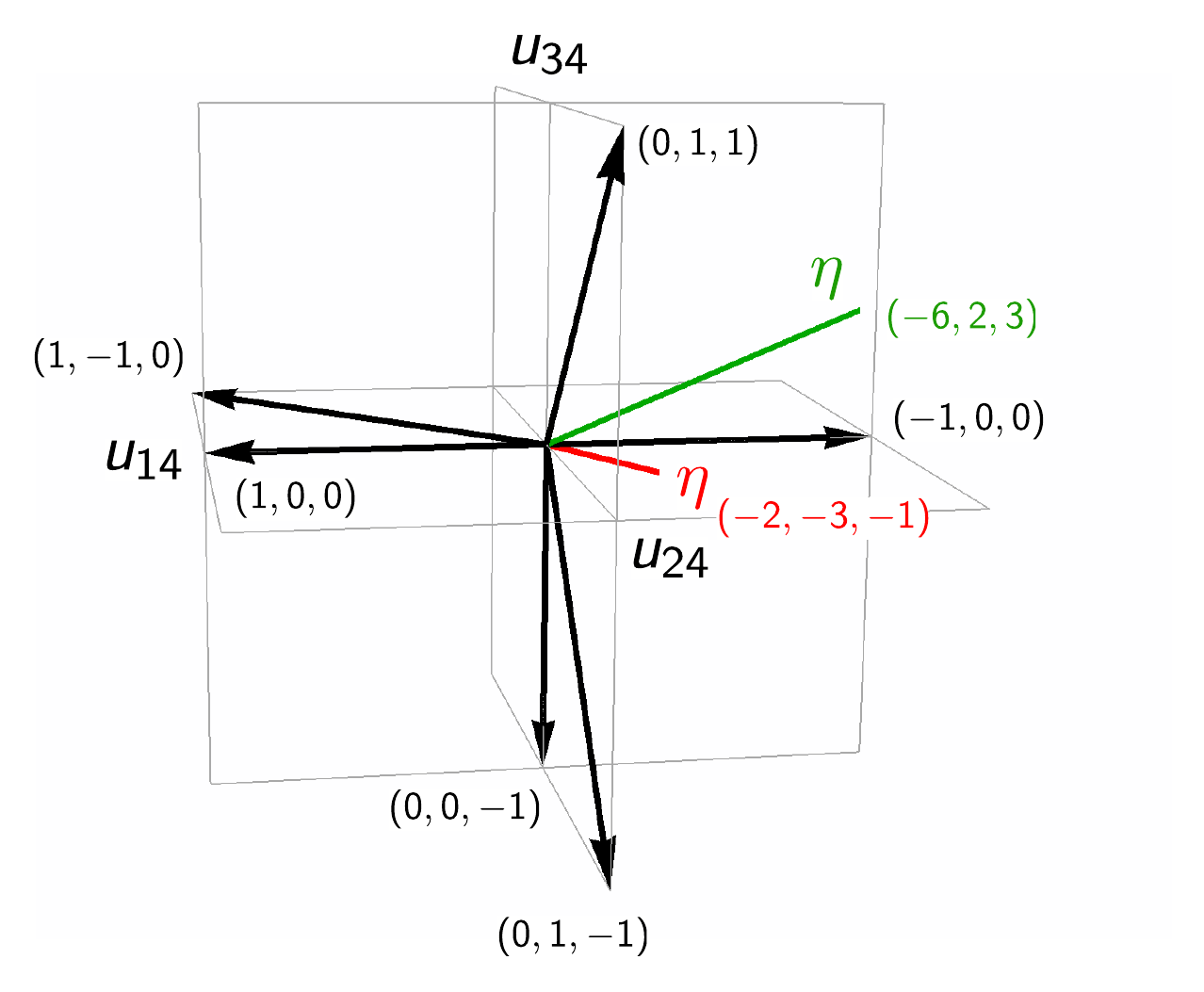}
}  
\caption{
Charge vectors in phase A of $H_4$ with $\eta=(-2,-3,-1)$ and $\eta=(-6,2,3)$. 
\label{qcovectorh4p1}}
 \end{center}
 \end{figure}

The JK residue computation with $\eta=(-2,-3,-1)$ leads to
\begin{align} 
\label{IH4A}
\begin{split}
\mathcal{I}_A = \frac{-i \eta(q)^3}{\theta_1(x/y)} & \left[ \frac{\theta_1(x) \theta_1(x/yz) \theta_1(y/z)}{\theta_1(y) \theta_1(z) \theta_1(z/x)} - \frac{\theta_1(y) \theta_1(y/xz) \theta_1(x/z)}{\theta_1(x) \theta_1(z) \theta_1(z/y) } \right. \\
& \left. \quad + \frac{\theta_1(x) \theta_1(z) \theta_1(xz/y^2)}{\theta_1(y) \theta_1(y/z) \theta_1(xz/y)} 
-\frac{\theta_1(y) \theta_1(z) \theta_1(yz/x^2)}{\theta_1(x) \theta_1(x/z) \theta_1(yz/x)} \right] \,,
\end{split}
\end{align}
which agrees with \eqref{IH4-geo} term by term. 
For this particular choice of $\eta$, the $W$-function becomes trivial 
at each of the four contributing poles. 

For a different choice, say, $\eta = (-6,2,3)$, we obtain a more complicated expression, 
\begin{align}
\label{IH4A-2}
\begin{split}
\mathcal{I}_A = \; & \frac{-i \eta(q)^3\theta_1(x) \theta_1(y) \theta_1(z)}{\theta_1(x/y) \theta_{1}(v)^2} \times  
\\
&
\Biggl[ 
\left( \frac{\theta_1(vx) \theta_1(v/x) }{\theta_1(x)^2 } 
 +\frac{\theta_1(vy) \theta_1(v/y) }{\theta_1(y)^2} 
  - \frac{\theta_1(vz) \theta_1(v/z)}{\theta_1(z)^2} \right)
  \\
  & \quad \times 
 \left( \frac{ \theta_1(y/xz) }{ \theta_1(z/y) \theta_1(xz)} -\frac{ \theta_1(x/yz)}{\theta_1(z/x) \theta_1(yz)} \right) 
\\
& \; + \frac{\theta_1(vx/yz) \theta_1(vyz/x) }{ \theta_1(x/yz) \theta_1(z/x) \theta_1(yz)} 
- 
\frac{\theta_1(vy/xz) \theta_1(vxz/y) }{\theta_1(y/xz)\theta_1(z/y) \theta_1(xz) }  \Biggr] \,.
\end{split}
\end{align}
Despite its appearance, it can be shown to be independent of $v$ and agree with \eqref{IH4-geo}.

\subsubsection*{Phase B} 

This phase has 10 chiral fields and 6 Fermi fields. 
The plaquettes in the periodic quiver correspond to
\begin{align}
\begin{array}{rcccc}
 & & J & \ \ \ & E \\
\Lambda_{23}^x: \ & & D_{32} \cdot Y_{24} \cdot D_{41} \cdot Z_{12} -Z_{34} \cdot D_{41} \cdot Y_{13} \cdot D_{32}   
& & D_{21} \cdot X_{13} - X_{24}\cdot D_{43}   \\
\Lambda_{23}^y: \ & & Z_{34} \cdot D_{41} \cdot X_{13} \cdot D_{32} - D_{32} \cdot X_{24} \cdot D_{41} \cdot Z_{12}
& & D_{21} \cdot Y_{13} - Y_{24} \cdot D_{43}  \\
\Lambda_{23}^z: \ & & D_{32} \cdot D_{21} \cdot Z_{12}  - Z_{34} \cdot D_{43} \cdot D_{32}  
& & X_{24} \cdot D_{41} \cdot Y_{13} - Y_{24}\cdot D_{41} \cdot X_{13}   \\
\Lambda_{14}^x: \ & & D_{43} \cdot D_{32}\cdot X_{24} \cdot D_{41} - D_{41} \cdot X_{13} \cdot D_{32} \cdot D_{21} 
& & Y_{13} \cdot Z_{34} - Z_{12} \cdot Y_{24}   \\
\Lambda_{14}^y: \ & & D_{43} \cdot D_{32}\cdot Y_{24} \cdot D_{41} - D_{41} \cdot Y_{13} \cdot D_{32} \cdot D_{21}  
& & Z_{12} \cdot X_{24} - X_{13}\cdot Z_{34}  \\
\Lambda_{14}^z: \ & & D_{43} \cdot Z_{34} \cdot D_{41} - D_{41} \cdot Z_{12} \cdot D_{21}
& & X_{13} \cdot D_{32} \cdot Y_{24} - Y_{13} \cdot D_{32}\cdot X_{24} \\
\end{array}
\end{align} 
The global charges of the fields are given in \tref{tchh4B}.

 \begin{table}[ht!!]
 \centering
\resizebox{1\hsize}{!}{
 \begin{tabular}{c|cccc|cc|cc|cc|ccc|ccc}
 field & 
 $D_{21}$ & $D_{32}$ & $D_{43}$ & $D_{41}$ &
 $X_{13}$ & $Y_{13}$ &
 $X_{24}$ & $Y_{24}$ &
 $Z_{12}$ & $Z_{34}$ &
 $\Lambda_{23}^{x}$ & $\Lambda_{23}^{y}$ & $\Lambda_{23}^{z}$ &
 $\Lambda_{14}^{x}$ & $\Lambda_{14}^{y}$ & $\Lambda_{14}^{z}$
 \\
 \hline
 $U(1)_x$ &
 $-1/4$ & $-1/4$ & $-1/4$ & $1/4$ &
 $-1/2$ & $1/2$ &
 $-1/2$ & $1/2$ &
 $1/4$ & $1/4$ &
 $-3/4$ & $1/4$ & $1/4$ &
 $3/4$ & $-1/4$ & $-1/4$
 \\
 $U(1)_y$ &
 $-1/4$ & $-1/4$ & $-1/4$ & $1/4$ &
 $1/2$ & $-1/2$ &
 $1/2$ & $-1/2$ &
 $1/4$ & $1/4$ &
 $1/4$ & $-3/4$ & $1/4$ &
 $-1/4$ & $3/4$ & $-1/4$
 \\
 $U(1)_z$ &
 $-1/4$ & $-1/4$ & $-1/4$ & $1/4$ &
 $1/2$ & $1/2$ &
 $1/2$ & $1/2$ &
 $-3/4$ & $-3/4$ &
 $1/4$ & $1/4$ & $5/4$ &
 $-1/4$ & $-1/4$ & $3/4$
 \\
\hline
 \end{tabular}
}
 \caption{Global charges of fields in phase B of $H_4$.}
 \label{tchh4B}
 \end{table}

The anomaly polynomial is
\begin{align}
\mathcal{A}(u) = 2(u_1 -u_2 -u_3 + u_4)^2 
\,,
\end{align}
so we can use the $W$-function as the anomaly cancelling factor. 

\begin{figure}[ht!!]
\begin{center}
\resizebox{0.53\hsize}{!}{
\includegraphics[trim=0cm 0cm 0cm 0cm,totalheight=5 cm]{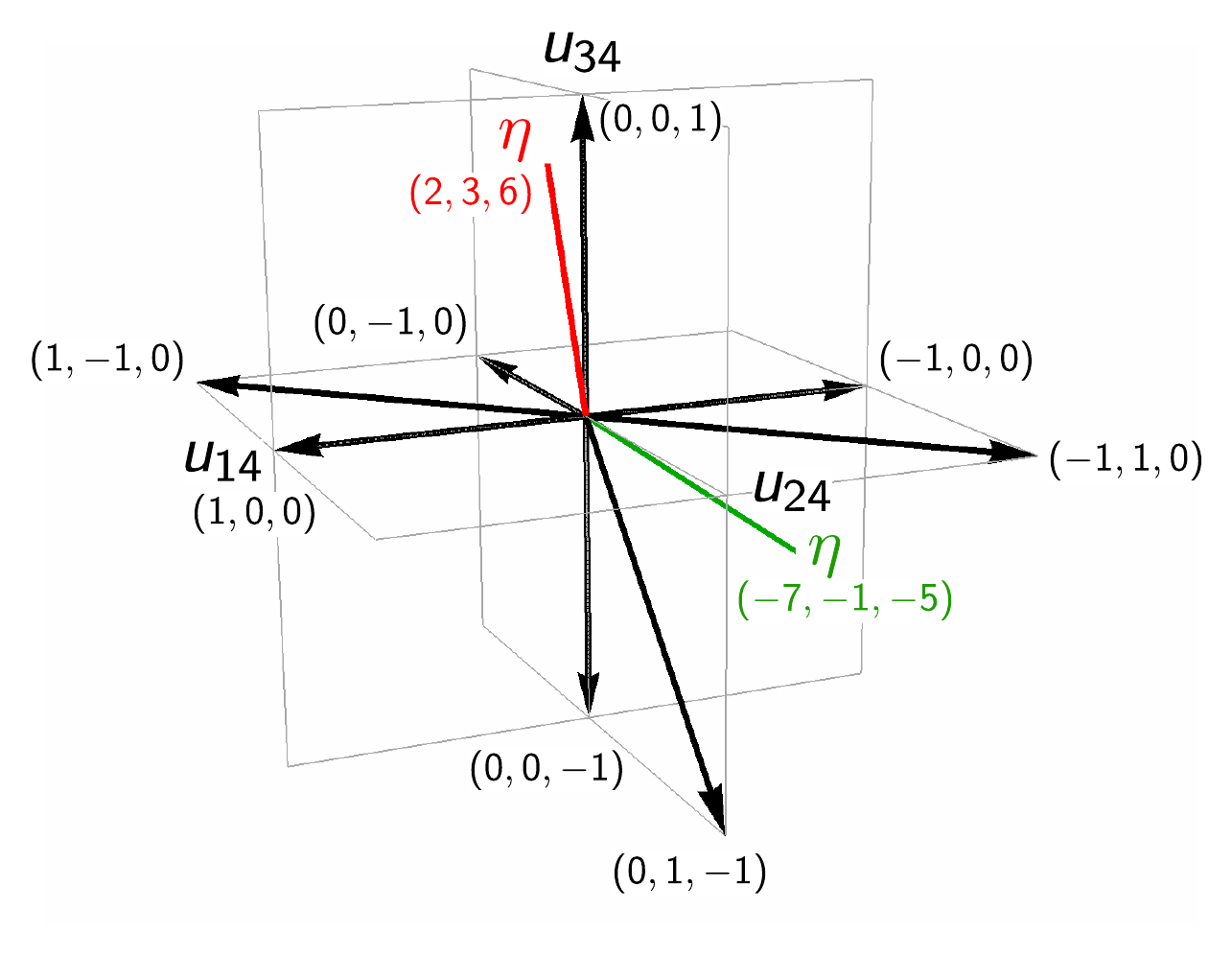}
}  
\caption{
Charge vectors in phase B of $H_4$ with $\eta=(2,3,6)$ and $\eta=(-7,-1,-5)$.
\label{qcovectorh4p2}}
 \end{center}
 \end{figure}

The JK residue computation with the choice $\eta=(2,3,6)$ 
gives 
\begin{align}
\label{IH4B2}
\begin{split}
\mathcal{I}_B = \frac{-i \eta(q)^3}{\theta_1(x/y)} & \left[ \frac{\theta_1(x) \theta_1(x/yz) \theta_1(y/z)}{\theta_1(y) \theta_1(z) \theta_1(z/x)} + \frac{\theta_1(y) \theta_1(z/x) \theta_1(xz/y)}{\theta_1(x) \theta_1(z) \theta_1(y/z) } \right. \\
& \left. + \frac{\theta_1(x) \theta_1(z) \theta_1(xz/y^2)}{\theta_1(y) \theta_1(y/z) \theta_1(xz/y)} +  \frac{\theta_1(y) \theta_1(z) \theta_1(x^2/yz)}{\theta_1(x) \theta_1(z/x) \theta_1(x/yz)} \right] \,,
\end{split}
\end{align}
which agrees with \eqref{IH4-geo} term by term. 
Again, the particular choice of $\eta$ makes the $W$ function trivial at each of the four contributing poles. 

If we instead choose $\eta = (-7,-1,-5)$, we obtain
\begin{align}
\label{IH4B}
\begin{split}
\mathcal{I}_B =\; & \frac{-i \eta(q)^3 \theta_1(x) \theta_1(y) \theta_1(z)}{\theta_1(x/y)\theta_1(v)^2} \times 
\\
&\left[ 
\frac{\theta_1(vx) \theta_1(v/x) \theta_1(y/xz)}{ \theta_1(x)^2  \theta_1(z/y) \theta_1(xz)} 
-
\frac{\theta_1(vy) \theta_1(v/y)  \theta_1(x/yz)}{\theta_1(y)^2 \theta_1(z/x) \theta_1(yz)} \right. 
\\
& 
\quad + 
\frac{\theta_1(vx/yz) \theta_1(vyz/x)}{\theta_1(x/yz)\theta_1(z/x) \theta_1(yz)  } 
-  
\frac{\theta_1(vy/xz) \theta_1(vxz/y) }{\theta_1(y/xz)\theta_1(z/y) \theta_1(xz) } 
\\
& \left.
\quad -\frac{\theta_1(x/y)\theta_1(xyz) }{\theta_1(xy)  \theta_1(xz) \theta_1(yz)}
\left( 
\frac{\theta_1(vz) \theta_1(v/z) }{\theta_1(z)^2 } 
- 
\frac{\theta_1(vxyz) \theta_1(v/xyz)}{ \theta_1(xyz)^2} 
\right)
\right]
\\
& 
+ 
\frac{i\eta(q)^3  \theta_1(xy) \theta_1(xz) \theta_1(yz)}{\theta_1(x) \theta_1(y) \theta_1(z) \theta_1(xyz)} 
 \,.
\end{split}
\end{align}
Again, it can be shown to be independent of $v$ and agree with \eqref{IH4-geo}.

\section{Discussion} 
\label{sec:discussion}

We computed the elliptic genera for a class of $2d$ $(0,2)$ gauge theories that arise on the worldvolume of D1-branes probing toric $\text{CY}_4$ singularities. These quiver gauge theories are efficiently described by T-dual Type IIA brane configurations of D4-branes suspended from NS5-branes known as brane brick models. 
The elliptic genera were computed using two independent methods.
First, we calculated the elliptic genus using an integral formula based on the GLSM quiver description known from our previous work on brane brick models. 
Then, we proposed a new formula for the elliptic genus based on the $\text{CY}_4$ target space geometry that the NLSM is expected to describe in the IR. 
For both methods, we regulated divergences arising from the non-compactness of the target space by introducing appropriate $U(1)^3$ global symmetry fugacities. 
It was shown in several examples that the two methods lead to the same result. 
We hence conclude that quantum effects, which are expected in the IR, do not drastically alter the correspondence between the target space geometry and this class of $2d$ $(0,2)$ gauge theories.

For $2d$ $(0,2)$ gauge theories that naively suffer from abelian gauge anomalies, we introduced an anomaly cancelling term in the integral formula for the elliptic genus. 
Using this extra term, we verified the agreement with the computation based on the target space geometry.

We also provided further evidence for triality of $2d$ $(0,2)$ theories by matching the elliptic genera of brane brick models with the same target space geometry. 
For brane brick models, the target space of the NLSM is the classical mesonic moduli space of the corresponding $2d$ $(0,2)$ quiver gauge theory. 
Our results suggest that the mesonic moduli space is a good IR observable for the examples considered in this paper. This is far from being the case for more general $2d$ $(0,2)$ theories unrelated to brane brick models.
It would be desirable to expand our investigation to more examples.

A natural question, which we leave for future work, is whether the elliptic genus formula proposed in this paper based on the triangulation of the toric diagram can be extended to more general geometries.
For instance, non-toric $\text{CY}_4$ form a rich class of geometries yet to be studied in the context of $2d$ $(0,2)$ theories. 

In addition, it would be desirable to obtain a stringy explanation for the anomaly cancelling factor in the GLSM computation of the elliptic genus. 
We expect that there is a stringy anomaly inflow mechanism analogous to the Green-Schwarz mechanism that cancels the anomaly. 
This would be in line with our derivation of the anomaly cancelling term in the integral formula for the elliptic genus, which relies on its modular properties and holomorphy.

Given that throughout this paper we have restricted ourselves to abelian brane brick models, it would be interesting to study their non-abelian extensions. 
Following \cite{Dijkgraaf:1996xw}, one can expect that the elliptic genus of such a non-abelian theory relies on a symmetric product of the target space of the corresponding abelian theory. 
However, a recent study of $2d$ maximal super-Yang-Mills with $\mathbb{C}^4$ as its target space showed that the naive symmetric product of $\mathbb{C}^4$ is not enough to obtain the elliptic genus of the non-abelian theory \cite{Kologlu:2016aev}.
We hope to return to non-abelian brane brick models in the near future. 

Finally, it would be interesting to find observables that are more refined than the elliptic genus, such as the ones proposed in \cite{Closset:2015ohf}, for brane brick models.

\acknowledgments{
We would like to thank M. Romo, P. Putrov, P. Yi and C. Vafa for enjoyable and helpful discussions. The work of S. F. is supported by the U.S. National Science Foundation grant PHY-1518967 and by a PSC-CUNY award. The work of D.G. is supported by POSCO TJ Park Science Foundation. The work of S. L. is supported by Samsung Science and Technology Foundation under Project Number SSTF-BA1402-08 and by the IBM Einstein Fellowship of the Institute for Advanced Study. The work of S. L. was also performed in part at the Aspen Center for Physics, which is supported by National Science Foundation grant PHY-1066293. R.-K. S. is supported by the ERC STG grant 639220 ``Curved SUSY''.
}
\\


\appendix 
\section{Theta Functions and their Identities } 

\subsection{Theta Functions \label{sec:theta-conv}}

To make the paper self-contained, we collect some well-known properties of eta and theta functions. We follow the standard conventions shared by many physics textbooks, including \cite{hori2003mirror,blumenhagen2009introduction} and consistent with \cite{Benini:2013xpa}.

The Dedekind eta function is 
\begin{align}
\eta(\tau) = q^{1/24} \prod_{n=1}^\infty (1-q^n) \,, \quad 
q= e^{2\pi i\tau}\,, \quad 
\text{Im}(\tau) > 0 \,.
\end{align}
Its modular properties are
\begin{align}
\eta(\tau+1) = e^{\pi i/12} \eta(\tau) \,,
\quad 
\eta\left(-\frac{1}{\tau}\right) = \sqrt{-i\tau} \,\eta(\tau) \,.
\end{align}

The Jacobi theta functions with arbitrary twists have a sum representation, 
\begin{align}
\theta[^{\alpha}_{\beta}](\tau|z) &= \sum_{n\in \mathbb{Z}} q^{\frac{1}{2} (n+\alpha)^2} e^{2\pi i(n+\alpha)(z+\beta)} \,, 
\quad \alpha,\beta \in \mathbb{R} \,,
\label{twisted-theta-sum}
\end{align}
and a product representation, 
\begin{align}
\frac{\theta[^{\alpha}_{\beta}](\tau|z)}{\eta(\tau)} &= e^{2\pi i\alpha(z+\beta)} q^{\frac{\alpha^2}{2}-\frac{1}{24}} \prod_{n=1}^\infty \left(1 + q^{n+\alpha-\frac{1}{2}} e^{2\pi i(z+\beta)}\right) \left( 1 +q^{n-\alpha -\frac{1}{2}} e^{-2\pi i(z+\beta)} \right) \,.
\label{twisted-theta-prod}
\end{align}
They have quasi-periodicity in $\alpha$ and $\beta$,
\begin{align}
\theta[^{\alpha+1}_{\ \beta}] (\tau|z) = \theta[^\alpha_\beta] (\tau|z) = e^{-2 \pi i \alpha} \theta[^{\ \alpha}_{\beta+1}] (\tau|z)  \,,
\end{align}
as well as quasi-periodicity in $z$,
\begin{align}
\theta[^\alpha_\beta](\tau|z+1) = e^{2\pi i \alpha}\theta[^\alpha_\beta](\tau|z) 
\,, 
\quad 
\theta[^\alpha_\beta](\tau|z+\tau) =  e^{-2\pi i (z+ \beta)}\theta[^\alpha_\beta](\tau|z) 
\,.
\end{align}
Their modular properties are
\begin{align}
\begin{split}
\theta[^\alpha_\beta](\tau+1|z) &= e^{\pi i \alpha(1-\alpha)} \theta[^{\quad \alpha}_{\alpha+\beta-\frac{1}{2}}](\tau|z) 
\,, 
\\
\theta[^\alpha_\beta]\left(-\frac{1}{\tau}\right| \left.\frac{z}{\tau}\right) &= \sqrt{-i \tau} \,e^{2 \pi i \alpha \beta + \pi i \frac{z^2}{\tau} } \theta[^{\ \beta}_{-\alpha}](\tau|z) 
\,.
\end{split}
\end{align}
When $\alpha, \beta$ are restricted to $0$ or $1/2$, they reduce to the more familiar $\theta_a$ $(a=1,2,3,4)$ functions:
\begin{align}
\begin{split}
\theta_1(\tau|z) &= - \theta[^{1/2}_{1/2}] (\tau|z) 
= - i q^\frac{1}{8} (y^{\frac{1}{2}}-y^{-\frac{1}{2}})\prod_{k=1}^{\infty} (1-q^k) (1-y q^k) (1-y^{-1} q^{k}) \,, 
\\
\theta_2(\tau|z) &= \; \theta[^{1/2}_{\ 0}] (\tau|z) \;
= q^\frac{1}{8} (y^{\frac{1}{2}}+y^{-\frac{1}{2}}) \prod_{k=1}^{\infty} (1-q^k) (1+y q^k) (1+y^{-1} q^{k}) \,, 
\\
\theta_3(\tau|z) &= \; \theta[^{\ 0 \ }_{\ 0 \ }] (\tau|z) \;
= \prod_{k=1}^{\infty} (1-q^k) (1+y q^{k-\frac{1}{2}}) (1+y^{-1} q^{k-\frac{1}{2}}) \,, 
\\
\theta_4(\tau|z) &= \; \theta[^{\ 0}_{1/2}] (\tau|z) \;
= \prod_{k=1}^{\infty}(1-q^k) (1-y q^{k-\frac{1}{2}}) (1-y^{-1} q^{k-\frac{1}{2}}) \,,
\end{split}
\end{align}
where $y=e^{2 \pi i z}$. 

As mentioned in \sref{sec:JK-integral}, the definition and computation of the (modified) elliptic genus often makes use of the following identity, 
\begin{align}
\left. \frac{\partial}{\partial z} \theta_1(\tau|z) \right|_{z=0} = 2\pi\, \eta(\tau)^3 \,.
\label{theta-eta-iden}
\end{align}

Throughout the main body of this paper, we mostly use the multiplicative (exponential) notation with variables $q$ and $y$. 
Sometimes, when we discuss the anomaly polynomial, it is more convenient to use the additive (logarithmic) notation with variables $\tau$ and $z$. Switching between the two notations is straightforward:
\begin{align}
\theta_a(\tau|z) = \theta_a(q = e^{2\pi i\tau} , y = e^{2 \pi i z}) \,.
\end{align}
Hopefully, which notation we are using is always clear from the context.

\subsection{Proving Theta Function Identities \label{sec:theta-iden}}

We have encountered a number of theta function identities in the main text. Here we sketch a proof for them. Consider a function $F(\tau|z)$ holomorphic in $\tau$ and meromorphic in $z$. 
Suppose $F(\tau|z)$ is quasi-periodic in $z$ in the following sense:
\begin{align}
F(\tau|z+1) = F(\tau|z) \,,
\quad F(\tau|z) = e^{i \gamma(\tau) - 2\pi i \delta z} F(\tau|z) \,, 
\label{quasi-periodic}
\end{align}
where $\gamma(\tau)$ is some linear function of $\tau$, and $\delta$ is an integer. For example, 
$\theta_a(\tau|z)^2$, $\theta_a(\tau|2 z)$ $(a=1,2,3,4)$ are quasi periodic holomorphic functions. Their ratio is in general a quasi-periodic meromorphic function. When two or more functions have the same periodicity, their linear combinations are also quasi-periodic.  

Suppose we want to prove an identity of the type
\begin{align}
F_L(\tau|z) = F_R(\tau|z) \,,
\end{align}
where $F_L$ and $F_R$ are known to have the same periodicity. 

A standard way to prove this is to show that $F_L$ and $F_R$ have the same sets of zeros and poles, counted with multiplicity. Then $F_L/F_R$ would be a holomorphic function that is strictly periodic ($\gamma = 0 = \delta$) on the torus $z\sim z+1 \sim z+\tau$. An elementary theorem in complex analysis asserts that 
a holomorphic function on a compact connected manifold must be a constant. 
Thus, to complete the proof, it suffices to show that $F_L/F_R=1$ at some convenient value of $z$. 

For the identities in this paper, the poles and residues of $F_{L,R}$ are easy to find,
but the zeros are not. Suppose we have shown that the poles and residues match. Then we know that $\Delta F = F_L - F_R$ is holomorphic. If $\Delta F$ is strictly periodic ($\gamma=0=\delta$), then the theorem mentioned above guarantees that $\Delta F$ is a constant, which can be determined by evaluating $\Delta F$ at some $z$. 

Let us make a slight digression. When $\delta \neq 0$, $F(\tau|z)$ is not really a function but a section of a holomorphic line bundle over the torus. 
The integer $\delta$ is the first Chern number.  
Intuitively, $\delta$ counts the number of zeros minus the number of poles of $F(\tau|z)$ on the torus, with multiplicity.

Returning to the main problem, if $F_L$ and $F_R$ have $\delta < 0$, it follows that $\Delta F$ must be a holomorphic section which has more poles than zeros. To avoid a contradiction, it must be that $\Delta F =0$, completing the proof. 
All the identities in the main text have either $\gamma = 0 = \delta$ or $\delta < 0$. Hence this proof applies to all of them. 

The following identity between $\theta_1$ and $\theta_4$, in the multiplicative notation, is used many times in the main text:
\begin{align}
\theta_1 (q, ab) \theta_1 (q, a/b) \theta_4 (q,1)^2 = \theta_1(q,a)^2 \theta_4(q,b)^2 - \theta_4(q,a)^2 \theta_1(q,b)^2 \,.
\end{align}

\section{Degenerate Poles and the Jeffrey-Kirwan Residue \label{sdegpoles}}

Additional ideas are necessary for computing the JK residue in \eqref{JKdefinition} in the presence of degenerate poles. Such poles appear when, for gauge group $U(1)^r$, more than $r$ hyperplanes corresponding to the charge vectors of the JK integral meet at a point. Degenerate poles can be regulated using flags as we review in this section. The calculation of the JK residue using the flag method was explained in detail in \cite{Benini:2013xpa}.

A degenerate pole $u_*$ is associated to $l \geq r$ singular hyperplanes identified by the charge vectors $\mathbf{Q_{u_*}} = \{ Q_{1} , \cdots , Q_{r}, \cdots , Q_{l}\} $.
We first choose a subset of $r$ vectors from $\mathbf{Q_{u_*}}$ and order them to form 
\beal{es100e1}
\mathbf{B}(\mathcal{F}) = \{Q_{i_1} \cdots Q_{i_r} \} \subset \mathbf{Q_{u_*}} ~.~
\eea
Let $\mathcal{F}_k$ be the vector space whose basis is given by the first $k$ vectors in $\mathbf{B}(\mathcal{F})$. The vector spaces $\mathcal{F}_k$, with $k=1,\dots,r$, form a set 
\beal{es100e2}
\mathcal{F}= \left[ \mathcal{F}_0 =\{0\} \subset \mathcal{F}_1 \cdots \subset \mathcal{F}_r = \mathbb{C}^r \right] \,, \text{  dim}(\mathcal{F}_k) = k \, .
\eea
This is known as a flag $\mathcal{F}$, whose basis is given by the ordered set $\mathbf{B}(\mathcal{F})$. $\mathbf{B}(\mathcal{F})$ is not necessarily unique for a degenerate pole associated to $\mathbf{Q_{u_*}}$. Hence, 
we have a finite set of all possible flags $\mathcal{F}$ for a given $\mathbf{Q_{u_*}}$, which we call $\mathcal{FL}(\mathbf{Q_{u_*}})$.

Once we construct $\mathcal{FL}(\mathbf{Q_{u_*}})$, 
we define an ordered set $\kappa^\mathcal{F}$ for each flag $\mathcal{F}\in\mathcal{FL}(\mathbf{Q_{u_*}})$ as
\beal{es100e3}
\kappa^\mathcal{F} = \left \{ \sum_{Q_{i} \in \mathbf{Q_{u_*}} \cap \mathcal{F}_{k}} Q_{i} \ \Big | \quad k=1,\cdots,r \right \} ~.~
\eea
After imposing the strong regularity condition on a trigger $\eta$ \cite{Benini:2013xpa}, we finally determine the JK residue.

For flags satisfying the positivity condition with respect to $\kappa^\mathcal{F}$, the JK residue reads
\beal{es100a5}
\JK residue_{u=u_*} = \sum_{\mathcal{F} \in \mathcal{FL}^+(\mathbf{Q_{u_*}})} \nu(\mathcal{F}) \residue_{\mathcal{F}} ~,~
\eea
where $\mathcal{FL}^{+} (\mathbf{Q_{u_*}},\eta) = \{ \mathcal{F} \in \mathcal{FL}(\mathbf{Q_{u_*}}) | \eta \in \text{Cone}( \kappa_1^\mathcal{F}, \cdots, \kappa_r^\mathcal{F})  \}$ and $\nu$ is a sign factor associated to the flag $\mathcal{F}$; $\nu (\mathcal{F}) = \text{sgn} \left( \text{det} ( \kappa_{1}^\mathcal{F} , \cdots , \kappa_{r}^\mathcal{F} ) \right)$. The functional $\residue_\mathcal{F}$ on the right hand side of \eref{es100a5} is known as the iterative residue, which is simply the residue computed in the ordered basis $\mathbf{B}(\mathcal{F})$. The iterative residue is defined under the following coordinate transformation
\beal{es100a6}
\tilde{u}_{k} = Q_{i_{k}} \cdot u ~,~
\eea
for $k=1,\cdots,r$. The $r$-form integrand $\omega$ is transformed into $\omega = \tilde{\omega}_{1\cdots r} \, d\tilde{u}_{1} \wedge \cdots \wedge d\tilde{u}_r $. Of course, we should not forget the Jacobian factor from the coordinate transformation. The iterative residue becomes straightforward in the $\tilde{u}$ coordinates
\beal{es100a7}
\residue_\mathcal{F} \omega = \residue_{\tilde{u}_r=\tilde{u}_{*r}} \cdots \residue_{\tilde{u}_{1}=\tilde{u}_{*1}} \tilde{\omega}_{1\cdots r} ~,~
\eea
where the $\residue$ on the right hand side is the single variable residue obtained by assuming that all the other variables are generic.

\bibliographystyle{JHEP}
\bibliography{mybib}


\end{document}